\documentclass[traditabstract]{aa} 
\usepackage{graphicx}
\usepackage{color}
\usepackage{url}
\usepackage{amssymb}
\usepackage{amsmath}
\usepackage{txfonts}
\usepackage{multirow}
\usepackage{lscape}
\usepackage{float}
\usepackage{longtable}
\usepackage{natbib}
\bibpunct{(}{)}{;}{a}{}{,} 

\def\xmm{XMM-{\it~Newton}}
\newfont{\gwpfont}{cmssq8 scaled 1000}

\newcommand{\excpres}{{\gwpfont EXCPRES}}

\begin{document}
   \title{The matter distribution in $z \sim 0.5$ redshift clusters
of galaxies}
   \subtitle{II : The link between dark and visible matter
   \thanks{Based on observations obtained with MegaPrime/MegaCam, a
joint project of CFHT and CEA/DAPNIA, at the Canada-France-Hawaii
Telescope (CFHT) which is operated by the National Research Council
(NRC) of Canada, the Institut National des Science de l'Univers of the
Centre National de la Recherche Scientifique (CNRS) of France, and
the University of Hawaii. This research also used the facilities of
the Canadian Astronomy Data Centre operated by the National Research
Council of Canada with the support of the Canadian Space Agency. Also
based on observations obtained with XMM-Newton, an ESA science mission
with instruments and contributions directly funded by ESA Member States
and NASA'. } }

   \author{
          G. Soucail\inst{1,2}
          \and
          G. Fo\"ex\inst{1,2,4}
          \and
          E. Pointecouteau\inst{1,3}
	  \and
          M. Arnaud\inst{5}
	  \and
          M. Limousin\inst{6,7}
          }
   \institute{
          Universit\'e de Toulouse; UPS-Observatoire Midi-Pyr\'en\'ees; IRAP; Toulouse, France
          \and
          CNRS; Institut de Recherche en Astrophysique et
Plan\'etologie (IRAP); 14 Avenue Edouard Belin, F--31400 Toulouse,  France
         \and
             CNRS; Institut de Recherche en Astrophysique et
Plan\'etologie (IRAP); 9 avenue Colonel Roche, F--31028 Toulouse cedex 4, France
         \and
           Departamento de F\'isica y Astronom\'ia; Universidad de
Valpara\'iso; Avda. Gran Bretana 1111; Valpara\'iso; Chile
         \and
             Laboratoire AIM; IRFU/Service d'Astrophysique; CEA/DSM; CNRS and 
Universit\'{e} Paris Diderot; B\^{a}t. 709, CEA-Saclay, F-91191 Gif-sur-Yvette Cedex, France
         \and
	      Laboratoire d'Astrophysique de Marseille (LAM); Universit\'e d'Aix-Marseille \& 
CNRS; UMR7326; 38 rue Fr\'ed\'eric Joliot-Curie, F-13388 Marseille Cedex 13, France
         \and
	      Dark Cosmology Centre, Niels Bohr Institute, University of Copenhagen; 
               Juliane Maries Vej 30, DK-2100 Copenhagen, Denmark
             }

   \date{Received Sept. 17, 2013; accepted  May 5, 2015}

  \abstract  
{ We present an optical analysis of a sample of 11 clusters built from the
\excpres\ sample of X-ray selected clusters at intermediate redshift ($z
\sim 0.5$). With a careful selection of the background galaxies we provide
the mass maps reconstructed from the weak lensing by the clusters. We
compare them with the light distribution traced by the early-type
galaxies selected along the red sequence for each cluster. The strong
correlations between dark matter and galaxy distributions are confirmed,
although some discrepancies arise, mostly for merging or perturbed
clusters. The average M/L ratio of the clusters is found to be: $M/L_r =
160 \pm 60$ in solar units (with no evolutionary correction), in excellent
agreement with similar previous studies. No strong evolutionary effects
are identified even if the small sample size reduces the significance
of the result. We also provide a individual analysis of each cluster in
the sample with a comparison between the dark matter, the galaxies and
the gas distributions. Some of the clusters are studied for the first
time in the optical. }


   \maketitle

%

\section{Introduction}
Although the origin and evolution of linear-scale clustering is well
described by the concordance model \citep{spergel07}, gravitational
clustering of matter on smaller scales (galaxy clusters and groups)
belongs to a non-linear regime of structure formation. This regime is more
difficult to understand and to simulate because its evolution must include
the role of baryons, driven by complex physics. Clusters of galaxies
which are the most massive gravitationally bounded structures have been
widely used over the past years to probe the cosmic evolution of the large
scale structures in the Universe \citep{voit05,allen11}. In the standard
model of structure formation driven by gravitation only, clusters form a
self-similar population characterized only by their mass and redshift.
Including baryon physics introduces some distortions in the scaling
relations between the mass and other physical quantities like temperature,
X-ray or optical luminosity, etc ... \citep{kaiser86,giodini13}. Most
research works have recently focused on the relationship between the
dominant dark matter and the baryonic matter that forms gas and stars
\citep{lin03,giodini09}. Both the mass-to-light (M/L) ratio of structures
and the halo occupation number (HON, or the number of satellite galaxies
per halo) correspond to observables easy to compare to predictions
from numerical simulations \citep{cooray02, tinker05}. They are both
representative of the way stellar formation occurred in the early
stages of halo formation \citep{marinoni02,borgani11}. Recent progress
on numerical simulations \citep{murante07,conroy07,aghanim09} have also
stressed the role of hierarchical building of structures in enriching
the intra-cluster medium (ICM) with stars in a consistent way with the
observed amount of ICM globular clusters and ICM light. This ICM light,
although hardly detectable, can be considered as the extension of the
diffuse envelope often seen in the central galaxy in rich clusters of
galaxies. It is an important component, although not the only one, which
explains the formation of the Brightest Cluster Galaxies (hereafter BCG)
in the centre of clusters of galaxies \citep{dubinsky98,presotto14}.

To quantify these processes and to compare them with those included
in numerical simulations, it is of prime importance to get reliable
masses and mass distribution in clusters of galaxies. But getting
accurate mass estimates is a difficult task and large uncertainties
reduce the validity of the relation between the light and total
mass distribution \citep{vale04,rozo14}. The determination of
the mass distribution using the weak gravitational lensing of the
background galaxies by clusters of galaxies is a powerful approach
to address this question \citep{schneider06,hoekstra13}. Lensing
is able to trace directly the dark matter component in rich
clusters of galaxies down to massive groups: see a few examples in
\citet{broadhurst05,gavazzi07,bradac08,umetsu12,zitrin13,gastaldello14}.
Recent analysis of cluster samples have shown strong improvements in the accuracy of the mass measurements \citep{vonderlinden14,applegate14,kettula14}, thanks to the powerful capacities of ground based wide field imaging. Finally, spectacular results were obtained in the case of merging clusters like
the ``Bullet'' cluster for which the dark matter distribution traces
closely the galaxy distribution while the intra-cluster gas traces by
X-ray emission is rather uncorrelated from the non-collisional components
\citep{clowe06}.

Another way to characterize the relation between the mass distribution
and the stellar light is to compare the M/L ratio with cluster
properties. In particular, recent results obtained with weak lensing
masses seem to show a slight scaling dependence of the M/L ratio
with mass: this is demonstrated from the MaxBCG sample built from the
SDSS, with structures ranging from small groups to massive clusters
\citep{sheldon09} and also from samples of clusters of galaxies
\citep{muzzin07,popesso07,bardeau07}. All these clusters are at low
redshift ($<0.2$ typically) because there are severe observational limits
at larger redshifts.

The purpose of this work is to present a detailed view of the relations
between dark matter and stellar light from a sample of 11 clusters at
intermediate redshift ($\sim 0.5$). This sample is part of the \excpres\
sample of cluster, built as an un-biased sample of clusters of luminous
X-ray clusters at redshift around 0.5, covering a wide range of dynamical
mass and X-ray temperature. The whole sample of 29 clusters was observed
in X-ray with \xmm\ (Arnaud et al. in preparation) to test the evolution
of clusters properties with redshift. The 11 brightest ones were observed
at CFHT for optical follow-up. A weak lensing analysis was proposed to
provide a mass estimate for these clusters. Its practical implementation,
as well as the global mass analysis were presented in \citet[hereafter
Paper 1]{foex12}. In the present paper, we focus on the comparison between
the optical properties of the clusters and their mass distribution and
we present the characteristics of each individual cluster. The paper is
organized as follow : Section 2 presents the data used in the analysis
and the selections of the different catalogs. Section 3 presents the
global optical properties of the clusters while Section 4 is dedicated
to the dark matter bi-dimensional distribution from the weak lensing
maps. In Section 5 we discuss the properties of the sample and the links
between the stellar light distribution and the total mass. Conclusions
are given in Section 6. The individual properties of the 11 clusters of
the sample are detailed in the Appendix.

Throughout this paper, we use a standard $\Lambda$-CDM cosmology with
$\Omega_\mathrm{M}=0.3, \Omega_\Lambda=0.7$ and a Hubble constant $H_0 = 70$ km
s$^{-1}$ Mpc$^{-1}$ or $h=H_0/100 = 0.7$.


\section{Observational data}
\begin{table*}
\caption{General properties of the clusters and summary of the
observations done in the $r'$ band, {\it i.e.} the data used for the
weak lensing analysis. The last columns give the mean galaxy number densities
before and after removal of foreground and cluster contamination (in units
of arcmin$^{-2}$).
}
\label{table:clusters}
\centering 
\begin{tabular}{l c c c c c c c c}
\hline\hline\noalign{\smallskip}
Cluster & RA & Dec & Redshift & Exp. time & Seeing & Completeness 
& Galaxy & Background \\
   & (J2000) & (J2000) &  $z$ & (sec.) & (\arcsec ) & in $r'$  (50\%)
& density & galaxy density\\
\noalign{\smallskip}\hline\noalign{\smallskip}
MS 0015.9+1609 & $00^{h}18^{m}33.26^{s}$ & $+16^{\circ}26{'}12.9{''}$ 
& 0.541 &  5600 & 0.82 & 24.50 &  26.0 & 18.0 \\
MS 0451.6--0305 & $04^{h}54^{m}10.85^{s}$ & $-03^{\circ}00{'}57.0{''}$ 
& 0.537 & 7200  & 0.77 & 24.75 & 30.6 & 23.5 \\
RXC J0856.1+3756 & $08^{h}56^{m}12.69^{s}$ & $+37^{\circ}56{'}15.0{''}$
& 0.411 & 7200  & 0.66 & 24.90 & 32.7 & 23.3 \\
RX J0943.0+4659 & $09^{h}42^{m}56.60^{s}$ & $+46^{\circ}59{'}22.0{''}$
& 0.407 & 7200  & 0.87 & 24.60 & 23.7 & 16.5 \\
RXC J1003.0+3254 & $10^{h}03^{m}04.62^{s}$ & $+32^{\circ}53{'}40.6{''}$ 
& 0.416 & 7200 & 0.79 & 24.55 & 26.2 & 19.2 \\
RX J1120.1+4318 & $11^{h}20^{m}07.47^{s}$ & $+43^{\circ}18{'}06.0{''}$ & 
0.612 & 7200 & 0.60 & 24.85 & 29.6 & 23.5 \\
RXC J1206.2--0848 & $12^{h}06^{m}12.13^{s}$ & $-08^{\circ}48{'}03.6{''}$ 
& 0.441 & 7200 & 0.85 & 24.90 & 32.0 & 26.5 \\
MS 1241.5+1710 & $12^{h}44^{m}01.46^{s}$ & $+16^{\circ}53{'}43.9{''}$
& 0.549 & 7200 & 0.72 & 24.85 & 33.0 & 16.3 \\
RX J1347.5--1145 & $13^{h}47^{m}32.00^{s}$ & $-11^{\circ}45{'}42.0{''}$
& 0.451 & 7200 & 0.77 & 24.95 & 29.6 & 25.8 \\
MS 1621.5+2640 & $16^{h}23^{m}35.16^{s}$ & $+26^{\circ}34{'}28.2{''}$ 
& 0.426 &  7200 & 0.60 & 25.05 & 37.1 & 28.9 \\
RX J2228.5+2036 & $22^{h}28^{m}33.73^{s}$ & $+20^{\circ}37{'}15.9{''}$ 
& 0.412 & 7200 & 0.69 & 24.85 & 33.5 & 25.6 \\
\noalign{\smallskip}\hline
\end{tabular}
\end{table*}

\subsection{Observations and data reduction}
Data were obtained for the whole cluster set with the MegaPrime instrument at
the Canada-France-Hawaii Telescope, during the 3 observing periods in 2006 and
2007 (RunIDs: 06AF26, 06BF26 and 07AF8; PI: G. Soucail). The camera MegaCam is a
wide field CCD mosaic covering 1 square degree, with a pixel size of
0.186\arcsec. Multi-color imaging was obtained with the 4 photometric filters
$g', r', i'$ and $z'$ with integration times of 1600 sec., 7200 sec., 1200 sec.
and 1800 sec. respectively. In practice, some clusters were observed with
slightly longer integration times due to the re-observation of some images in
poorer observing conditions. All data obtained in $r'$ were done in good seeing
conditions (IQ less than 0.8\arcsec ) and during photometric periods. The
integration time in $r'$ was defined to obtain a limiting magnitude for weak
lensing studies $r' \simeq 26$: we consider that at this magnitude limit, the
bulk of the background sources is at redshift higher than 1 and that the lensing
strength of the clusters is maximum. For the 3 other colors, the strategy was
defined to detect cluster galaxies up to $m^\star + 4$ (with $m_r^\star \simeq
20$ at $z=0.5$) in reasonable seeing conditions (IQ $< 1\arcsec$). The summary
of the observations is presented in Table \ref{table:clusters}.

Data reduction was done in a standard way for large CCD mosaics. After the
on-line preprocessing done at CFHT (correction of instrumental pixel-to-pixel
effects) with the Elixir
pipeline\footnote{\url{http://www.cfht.hawaii.edu/Instruments/Elixir/}}, the
global processing was done either by the Terapix team or by the authors, using
the Terapix tools\footnote{\url{http://www.terapix.fr/}} locally. A global
astrometric solution was found with {\sc Scamp} using the star references of the
USNO-B1 catalog \citep{monet03}, as well as the photometric alignment of the
different images. The final stacking of each image set was done with {\sc
Swarp}, producing a single wide field image and its associated weight map. A
$\chi^2$-image was built from $g', r', i'$ images and was used for objects
detection. Note that efficient flat-fielding in the far-red was difficult so the
$z'$-images were not included in the $\chi^2$-images.

The cluster RXJ1347.5--1145 was retrieved directly from the CFHT-CADC
archives\footnote{The Canadian Astronomy Data Centre is operated by the National
Research Council of Canada with the support of the Canadian Space Agency.}: it
was observed in $g'$ and $r'$ (PI: H. Hoekstra, runID: 05AC10) and extensively
analyzed previously in weak and strong lensing \citep{bradac08, halkola08}. Note
also that the field of view of RXJ2228.5+2036, being at low galactic latitude,
is highly contaminated by bright stars. It is necessary to mask large areas
around all the stars and this prevents an efficient weak lensing analysis.
Results obtained with this cluster will have to be taken with caution.


\subsection{Multi-color photometry}
Photometric catalogs were built with {\sc Sextractor} \citep{bertin96} in
dual mode, the detection of the objects being made on the $\chi^{2}$-image
for each cluster. For the magnitudes of the objects, we used the MAG\_AUTO
parameters while for color indices, we used the MAG\_APER magnitudes
measured in an constant circular aperture of 3\arcsec\ in diameter. We also
corrected uncertainties in the zero point calibration, a critical step for
further estimates of photometric redshifts. We used the color-color
distributions of the stars detected in each field and we compared these
distributions with the expected ones computed by convolving a well
calibrated spectral stellar library \citep{pickles98} with the filter and
instrumental transmissions. The position of the ``knee'' seen in the stellar
color-color diagrams was also matched to the observed colors. We finally
considered the $r'$-band photometry as a reference and computed the best
corrections to apply for the other filters, using a $\chi^2$ minimization
between both distributions.

Separation between stars and galaxies were obtained by following the
methodology developed by \citet{bardeau05} and all stellar-like objects
were removed from the catalogs. The 50\% completeness limit of the galaxy
catalogs in the $r'$-band are given in Table \ref{table:clusters}.

\subsection{Photometric redshifts and cluster member selection}
To identify cluster members in the photometric catalogs,
we used an updated version of the public code HyperZ
\footnote{\url{http://webast.ast.obs-mip.fr/hyperz/}} (Version 11, June
2009) and computed photometric redshifts. HyperZ is based on a template
fitting method \citep{bolzonella00, pello09}: the measured spectral
energy distribution (SED) is fitted with a library of templates built
from different spectral types, star formation histories and redshifts. In
the present case, photometric redshifts were computed in the redshift
range $[0,4]$.  We did not try to compute Bayesian redshifts with a
prior on luminosity but we used a simple cut in the permitted range of
absolute magnitudes: $ -25 < M < -14$.

\begin{figure}
\centering
\includegraphics[width=\hsize]{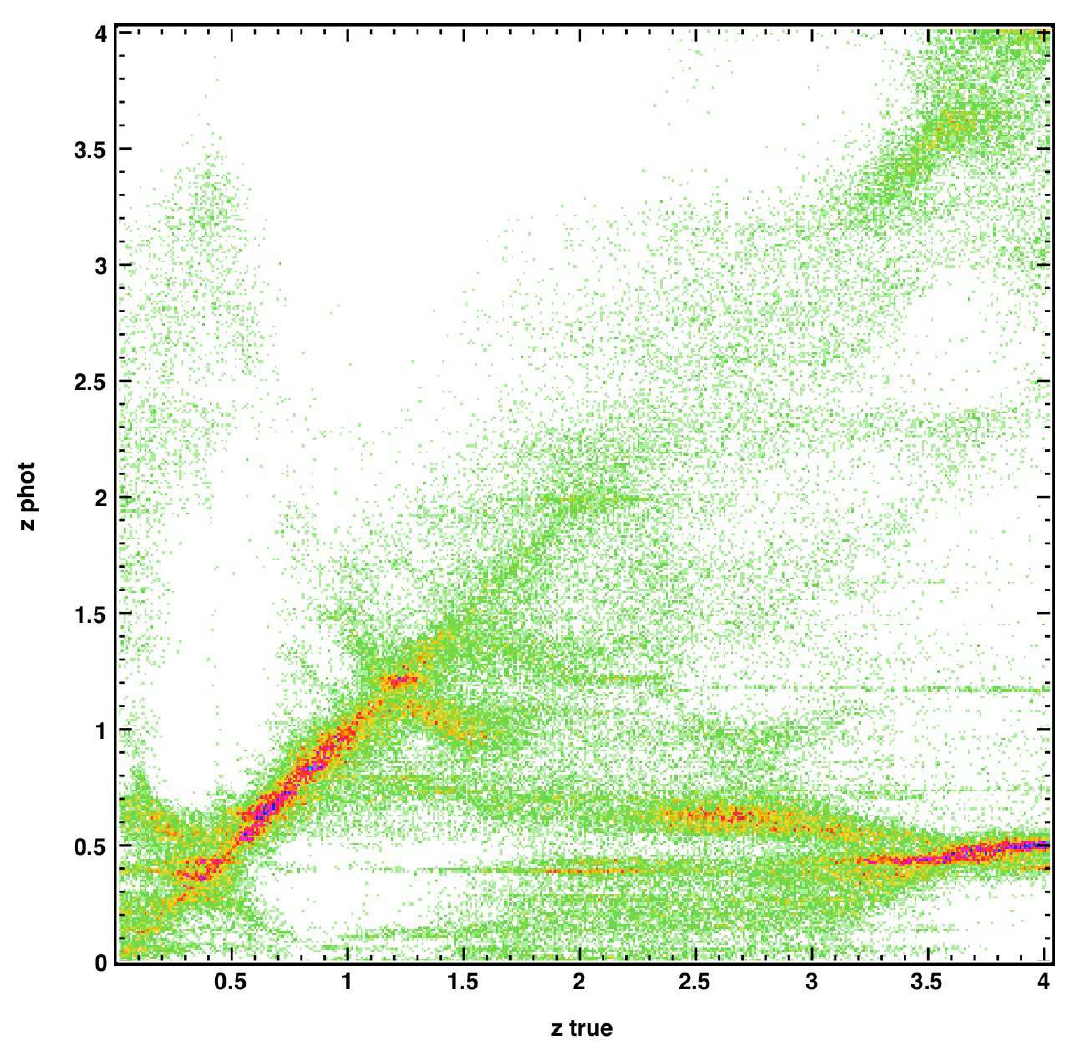}
\caption{ 
Photometric redshifts estimated by HyperZ for a simulated flat
distribution of galaxies. The color scale
shows the density of points, from black to yellow. 
}
\label{fig:simu}
\end{figure}

\begin{figure*}
\centering
\includegraphics[width=0.9\hsize]{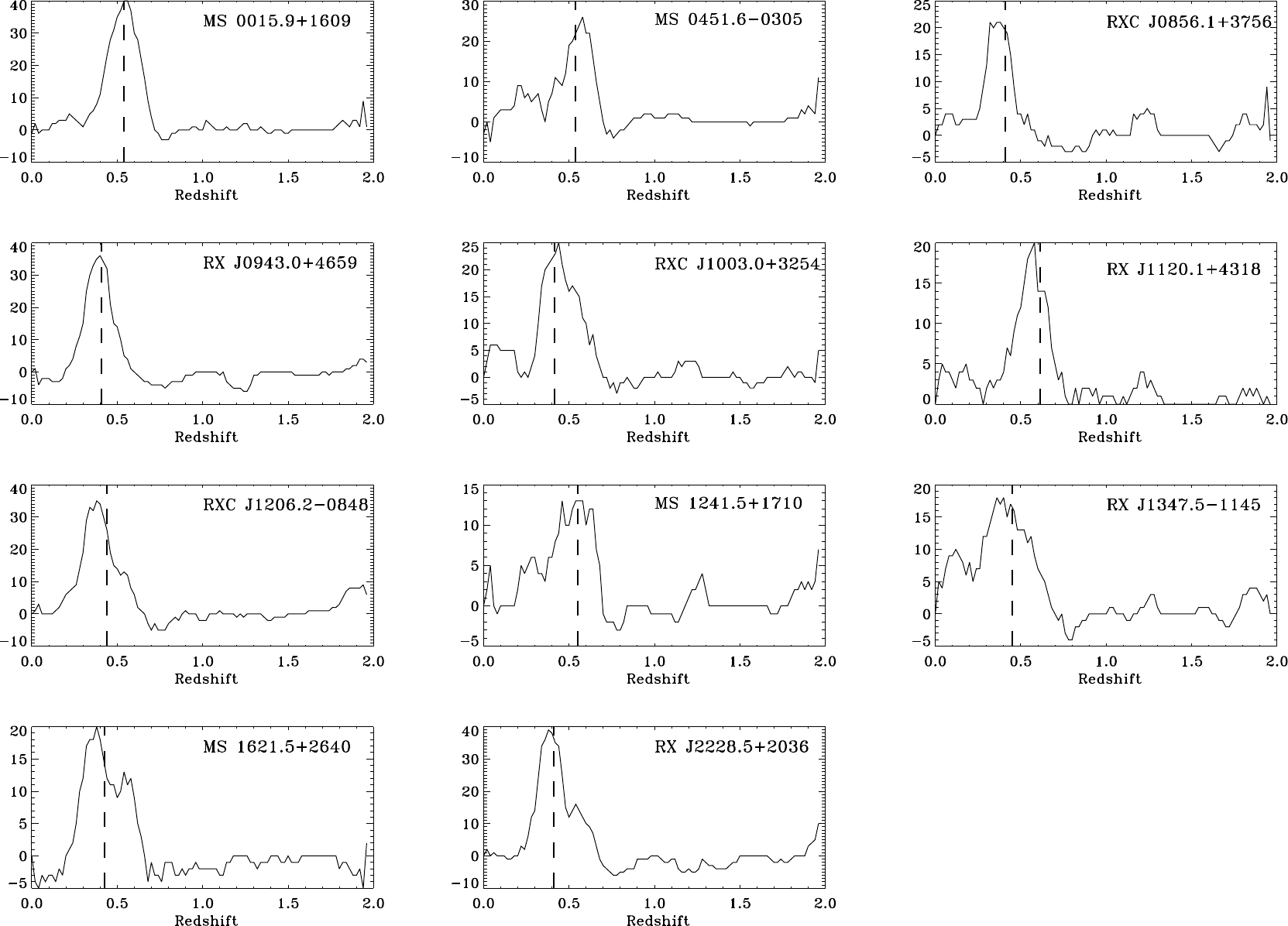}
\caption{ 
Over-densities in the photometric redshifts distribution for
each cluster. In each panel the vertical line shows the spectroscopic
redshift of the cluster. The plot represents the redshift distribution
of the central area defined as $R <5\arcmin$ minus the
distribution of an annulus of same area starting at $R = 10\arcmin$. 
}
\label{fig:cluster_photoz}
\end{figure*}

HyperZ can be very efficient with a sufficient number of photometric
bands but in this work we only have magnitudes in 4 filters, with
different limiting magnitudes so it is quite challenging to perform good
photometric redshifts in all redshift ranges. So we did not try to assign
a photometric redshift to each individual galaxy and we focused mainly
on the selection of cluster members and the detection of the cluster
over-density.  To have a more quantitative estimate of the reliability
of these photometric redshifts, we simulated a photometric catalog
with similar properties as our present observations. For simplicity,
we used a flat redshift distribution, which is sufficient to test our
redshift ranges of interest. We generated simulated magnitudes for
100,000 galaxies in the 4 MegaCam filters adding noise according to the
average signal-to-noise ratio observed in our data.  Then we ran HyperZ
on this simulated catalog and compared the photometric redshifts to the
expected values (Fig. \ref{fig:simu}).  Because of several color-color
degeneracies, the reliability of HyperZ is not constant across the whole
redshift range and many high redshifts galaxies with $z_{true}>1.5$ have
an under-estimated photometric redshift $z_{phot} \sim 0.5$.  But for
galaxies with $0.4<z_{true}<0.6$, the results are quite satisfactory:
most of the galaxies have a ``correct'' photometric redshift ($\pm
0.1$) and only a small fraction of them have ``catastrophic'' values.
Therefore, we consider that the photometric redshifts estimated by HyperZ
can be used safely to pre-select cluster members.  A consistency
check was done by looking at the clusters over-density of galaxies:
we determined the redshift distribution far from the cluster centre and
subtracted it from the one obtained in a central region covering the same area. The
resulting redshift distribution shows clearly a peak located close to
the cluster spectroscopic redshift (Figure \ref{fig:cluster_photoz}).

\subsection{The cluster color-magnitude diagram}
\begin{figure}
\center
\includegraphics[angle=-90,width=\hsize]{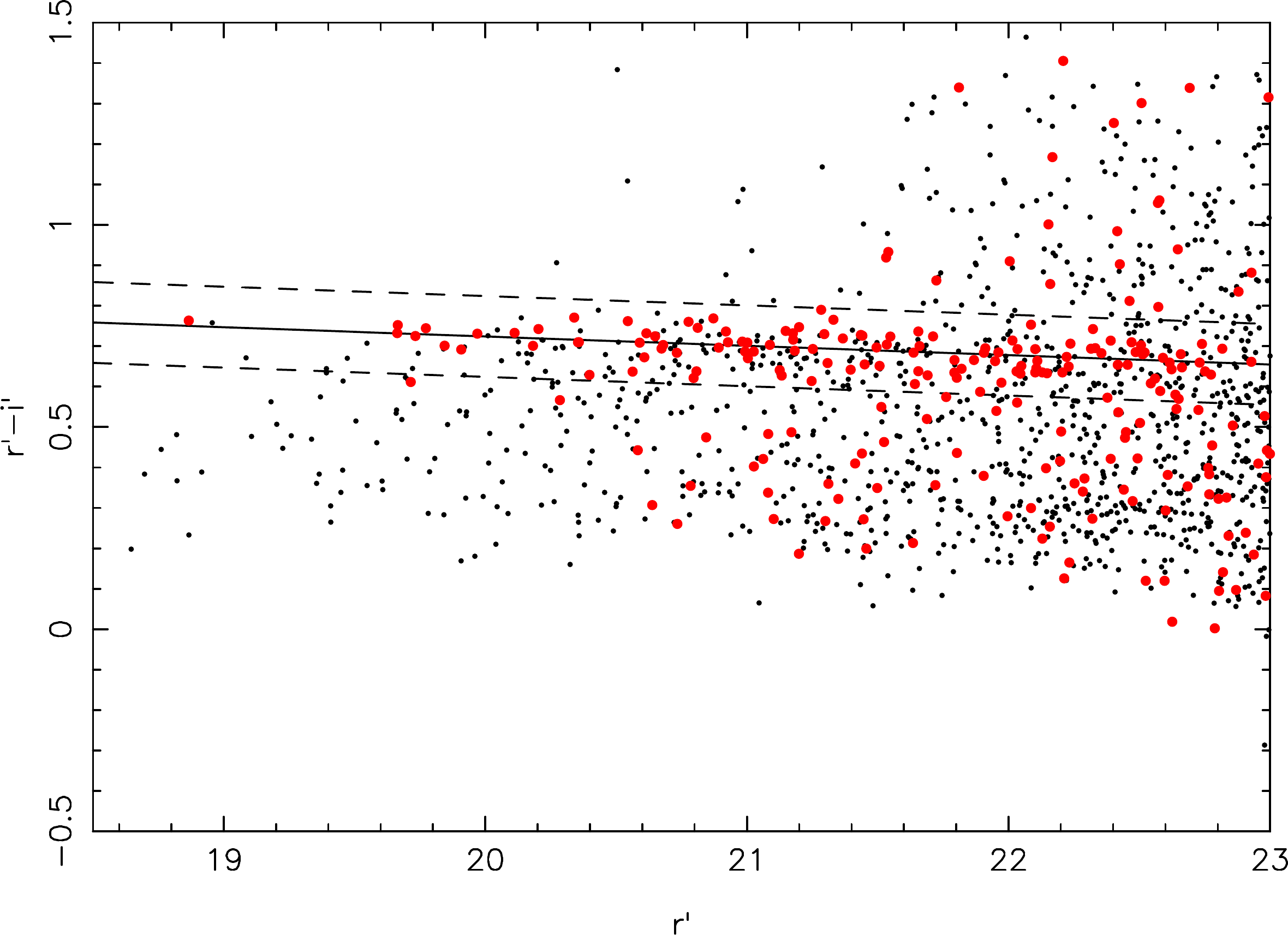}
\caption{Color-magnitude diagram in the field of MS1621.5+2640. The
black points are the galaxies located at $r<500''$ from the cluster
centre. The red dots are those with $r<200''$ and a $z_{phot}$ compatible
with the cluster redshift. The straight line is the best fit to the
red sequence and the dashed lines are the $3\sigma$ limits of the red
sequence. At the cluster redshift ($z=0.43$), the expected color,
computed with the synthetic evolutionary code of \citet{bruzual03}
is $r'-i' = 0.77$.} 
\label{fig:cmd} 
\end{figure}

Early-type galaxies form a homogeneous population whose spectral energy
distribution is dominated by red and old stars. For a given redshift
these galaxies are distributed along a well defined ``red sequence''
in a color-magnitude diagram and this sequence extends over several
magnitudes with a small scatter (smaller than 0.1 mag typically). This
characteristics has long been used as a powerful tool to detect clusters
of galaxies in large photometric surveys \citep{gladders00}. In the
present study we used the red sequence of the galaxies to clean the
lensing catalogs from cluster members.  In order to identify this red
sequence we followed the method described by \citet{stott09}: we first
selected a sub-sample of objects located in the central area of the images
($R< 200\arcsec$ from the cluster centre) and we only kept galaxies with
a photometric redshift compatible with the cluster redshift. We then
performed a linear fit of the red sequence with a $3\sigma$ clipping to
determine analytically the color-magnitude relation in the $(r',r'-i')$
diagram (Fig. \ref{fig:cmd}). On average, the dispersion $\sigma$ around
the red sequence is $0.07 \pm 0.02$. Finally we excluded all the galaxies in
the whole catalog along this relation and within $\pm 3\sigma$ of the
Gaussian fit. We also applied a magnitude cut $18<m_{r}<23$ because
fainter galaxies are no longer dominated by cluster members and it is
of prime importance to keep a background density as large as possible
for the weak lensing analysis.

In summary, the background galaxies catalogs were built for the present
work with the following rules: a magnitude cut $22<r'<26$ and a color
cut outside the red sequence $\pm 3\sigma$, up to $r'=23$.  Thus, most of
the cluster members were removed. We checked that the galaxies density
profile of the remaining galaxies is flat, except very near the cluster centre 
where some residual contamination remains (see Paper 1). However,
this has no strong impact of the global morphology of the clusters
that is described in this paper. The values of the average background
galaxies density are given in Table \ref{table:clusters}, before and
after cleaning the catalogs.


\section{Stellar light distribution}

\begin{table*}
\caption{Global properties of the cluster sample: the radius
$R_{200}$ and the 2D projected mass $M_{200}^{2D}$ are derived from
the weak lensing analysis. The optical luminosity $L_{200}$ and the
galaxy number $N_{200}$ are measured inside the radius $R_{200}$ and
are corrected from the background contamination. The total luminosity
has been corrected from the incompleteness at faint magnitudes. For
comparison we also give the number of cluster members inside a physical
radius of 1 Mpc,  N (1Mpc). The global M/L ratio is given in solar units
in the $r'$-band. }
\label{table:opt_prop}
\begin{center}
\begin{tabular}{l c c c c c c}
\hline\hline
\noalign{\smallskip}
Cluster & $R_{200}$ & $M_{200}^{2D}$ & $L_{200}$ & $N_{200}$ & N(1Mpc) & M/L \\
  & ($h_{70}^{-1}$ Mpc) & ($10^{15} h_{70}^{-1} M\odot$) & ($10^{12}
h_{70}^{-2} L\odot$) & & & ($h_{70}
M_\odot/L\odot$) \\ 
\noalign{\smallskip}\hline\noalign{\smallskip}
MS 0015.9+1609 & $2.33 \pm 0.13$ & $3.27 \pm 0.58$ & $17.2 \pm 1.6$ & $135\pm 12$ & $69 \pm 8$ & $190 \pm 51$\\
MS 0451.6--0305 & $1.92 \pm 0.12$ & $1.84 \pm 0.35$ & $10.8\pm 1.2$ & $81 \pm 9$ & $48 \pm 7$ & $170 \pm 52$\\ 
RXC J0856.1+3756 & $1.65 \pm 0.10$ & $1.0 \pm 0.175$ & $5.1 \pm 0.7$ & $48 \pm 7$ & $28 \pm 5$ & $197 \pm 63$\\
RX J0943.0+4659 & $1.77 \pm 0.80$ & $1.18 \pm 0.25$ & $13.6\pm 1.4$ & $100 \pm 11$ & $64 \pm 8$ & $ 87 \pm 28$\\
RXC J1003.0+3254 & $1.69 \pm 0.11$ & $0.94 \pm 0.20$ & $4.8 \pm 0.9$ & $28 \pm 5$ & $29 \pm 5$ & $198 \pm 81$\\
RX J1120.1+4318 & $1.46 \pm 0.14$ & $0.75 \pm 0.26$ & $8.2 \pm 1.2$ & $52 \pm 7$ & $46 \pm 7$ & $ 91 \pm 45$\\
RXC J1206.2--0848 & $2.03 \pm 0.10$ & $1.93 \pm 0.29$ & $17.4 \pm 1.7$ & $111 \pm 11$ & $54 \pm 7$ & $ 111\pm 27$\\
MS 1241.5+1710 & $1.88 \pm 0.13$ & $1.78 \pm 0.38$ & $8.0 \pm 1.1$ & $55 \pm 8$ & $30 \pm 6$ & $224 \pm 79$\\
RX J1347.5--1145 & $2.40 \pm 0.10$ & $3.27 \pm 0.40$ & $13.9\pm 1.4$ & $99 \pm 10$ & $42 \pm 7$ & $235 \pm 53$\\
MS 1621.5+2640 & $1.90 \pm 0.11$ & $1.57 \pm 0.26$ & $7.9 \pm 1.0$ & $63 \pm 8$ & $37 \pm 6$ & $199 \pm 59$\\
RX J2228.5+2036 & $1.68 \pm 0.12$ & $1.09 \pm 0.24$ & $16.0\pm 1.9$ & $117 \pm 13$ & $58 \pm 8$ & $ 68 \pm 23$\\
\noalign{\smallskip}\hline
\end{tabular}
\end{center}
\end{table*}

\subsection{Selection of cluster galaxies and global cluster properties}
As stated in Section 2, we specifically built galaxy cluster catalogs
by selecting galaxies for which the color $r'-i'$ falls within $\pm
3\sigma$ of the cluster red sequence. We also limited the sample to galaxies
brighter than $0.4 L^\star$, {\it i.e.} $m^\star +1$ to avoid too much
contamination in the faint magnitude bins. The $m^\star$ magnitude was
computed for each cluster, assuming an absolute magnitude $M^\star - 5
\log h = - 20.44$ \citep{blanton03} and adding the appropriate distance
modulus, the $k$-correction for an early-type galaxy for each cluster
and the galactic reddening correction. Adding an evolution correction
in the luminosity of the elliptical galaxies would amount to $\sim
0.65$ mag.
This would decrease the intrinsic luminosity at a given apparent
magnitude by a factor 1.8 and it would also change the magnitude cut
in the galaxy catalogues and reduce the number of cluster galaxies.
All in all the expected change in the total luminosity of the clusters
is a factor of 2 to 2.2. But it is highly uncertain because it 
strongly depends on the galaxy evolutionary scheme used in the
modelling of the evolution correction and it is generally not included
in the global studies of clusters. To remain consistent with previous works
(see \citet{popesso04,bardeau07} for example), we chose not to include it
in the present work.  

\begin{table*}
\caption{Morphological properties of the luminous component of the clusters. The
ellipticity $\epsilon=1-b/a$ and the position angle PA are given for the
clusters luminosity density map as well as the semi-major axis of the measured ellipse. For the Brightest
Cluster Galaxy (BCG), only $\epsilon$ and PA are given. PA are in degrees, counted 
counter-clockwise with respect to the North-South axis.}
\label{table:optMorpho}
\begin{center}
\begin{tabular}{l c c c c c}
\hline\hline
\noalign{\smallskip}
Cluster & \multicolumn{3}{c}{Cluster light} & \multicolumn{2}{c}{BCG} \\
  & $\epsilon$ & PA (deg.) & radius (kpc and \arcsec) & $\epsilon$ & PA (deg.) \\ 
\noalign{\smallskip}\hline\noalign{\smallskip}
MS 0015.9+1609 &  $0.20\pm0.10$ & $36\pm9$ & 520 (80\arcsec) & $0.26\pm0.03$ & $75\pm8$ \\
MS 0451.6--0305 &  $0.24\pm0.02$ & $150\pm3$ & 510 (89\arcsec) & $0.28\pm0.02$ & $101\pm8$ \\ 
RXC J0856.1+3756 & $0.23\pm0.04$ & $2\pm5$ & 660 (132\arcsec) & $0.28\pm0.01$ & $135\pm10$ \\
RX J0943.0+4659 &  $0.27\pm0.05$ & $48\pm2$ & 470 (95\arcsec) & $0.28\pm0.01$ & $11\pm10$ \\
RXC J1003.0+3254 &  $0.53\pm0.03$ & $22\pm3$ & 450 (89\arcsec) & $0.28\pm0.02$ & $-2\pm10$ \\
RX J1120.1+4318 &  $0.38\pm0.02$ & $95\pm4$ & 720 (119\arcsec) & $0.40\pm0.04$ & $110\pm19$ \\
RXC J1206.2--0848 &  $0.36\pm0.01$ & $86\pm3$ & 680 (130\arcsec) & $0.49\pm0.02$ & $111\pm24$ \\
MS 1241.5+1710 &  $0.10\pm0.03$ & $79\pm4$ & 260 (45\arcsec) & $0.16\pm0.01$ & $5\pm10$ \\
RX J1347.5--1145 &  $0.36\pm0.08$ & $43\pm3$ & 430 (80\arcsec) & $0.31\pm0.03$ & $0\pm10$ \\
MS 1621.5+2640 &  $0.27\pm0.04$ & $22\pm8$ & 380 (74\arcsec) & $0.02\pm0.02$ & $93\pm14$ \\
RX J2228.5+2036 &  $0.30\pm0.10$ & $69\pm5$ & 280 (56\arcsec) & $0.30\pm0.03$ & $93\pm14$ \\
\noalign{\smallskip}\hline
\end{tabular}
\end{center}
\end{table*}

The total luminosity of the clusters was measured by summing the
luminosity of all galaxies in the red sequence interval and located inside
the radius $R_{200}$ estimated with the weak lensing analysis (Paper 1). A
background correction to the total luminosity was included by removing
an average luminosity measured in an annulus defined by $2 R_{200} <
r < 3 R_{200}$ for each cluster and scaled with the adequate area. This
correction is rather small because the total cluster luminosity is
dominated by bright early-type galaxies. Finally a correction for
the magnitude cut of the galaxy catalogs is added \citep{popesso04}. It is
calculated as the integral of a Schechter function up to $0.4 L^\star$
and with a slope $\alpha=1.25$ \citep{blanton03}. A factor
1.6 is then included to get the total luminosity of the clusters
(Table \ref{table:opt_prop}).  We also computed the cluster optical
richness $N_{200}$ which we define as the number of cluster galaxies
within $R_{200}$ after correction from the background \citep{hansen05}.
All values are given in Table \ref{table:opt_prop}.

\subsection{Morphology of the light distribution}
There are many possibilities to map the light distribution in a
galaxy cluster. They depend on the choice of the input catalogs and
the method used to derive a correct mapping (see \citet{okabe10} for
example). 
In the present case we generated the light map trying to take into account more
accurately the ellipticity and the orientation of each individual galaxy for the
building of the cluster light distribution. In practice we selected cluster
galaxies with magnitudes $18<m_r<23$ and colors along the red sequence ($\pm 3
\sigma$). An artificial image was created for each with the package {\sc
Artdata} in IRAF \footnote{IRAF is distributed by the National Optical Astronomy
Observatories, which are operated by the Association of Universities for
Research in Astronomy, Inc., under cooperative agreement with the National
Science Foundation.}, with parameters extracted from the photometric
catalogs of the cluster galaxies 
(position, magnitude, ellipticity $\varepsilon$ and position angle PA). The
resulting image was then smoothed with a Gaussian kernel to generate the cluster
light map. We fixed the FWHM of the kernel to 80\arcsec , a size two times
smaller than the smoothing scale of the dark matter mass map. The maps are
displayed for each cluster together with the mass distributions (Figs.
\ref{fig:0015} to \ref{fig:2228}). The ellipticity $\epsilon = 1 - b/a$ and PA
of the cluster light were measured on a 2D fit of the isophotes with {\sc
Ellipse}. For each cluster we favored the large scale morphology and
we fitted isocontours which in some cases encompass several clumps,
especially in clusters with complex structure. In most cases
this corresponds to a radius 100 to 150\arcsec , {\it i.e.} up to 1 Mpc at the cluster
redshift (Table \ref{table:optMorpho}). Error bars on the elliptical parameters are estimated from the change of the parameters when the radius varies from 90 to 150 \arcsec\ typically.

\subsection{The brightest cluster galaxy (BCG)}
\begin{figure}
\center
\includegraphics[width=0.95\hsize]{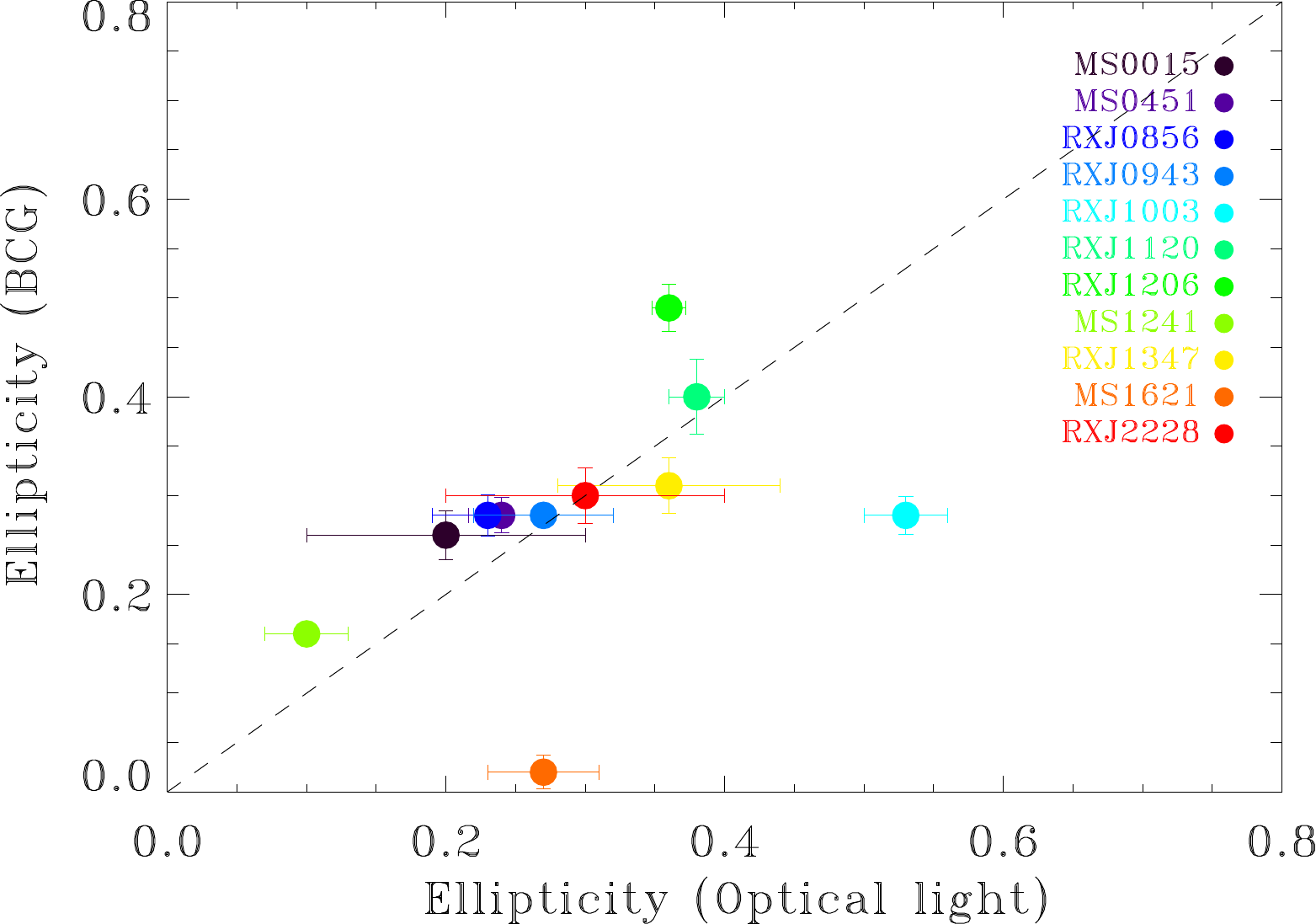}
\caption{Ellipticity of the global light distribution of the clusters
versus the ellipticity of the brightest central galaxy.} 
\label{fig:light_bcg} 
\end{figure}

BCGs are usually located at the very centre of clusters of galaxies.
Numerous studies emphasized their specific properties compared to
lower luminosity cluster members  \citep{lin04,smith10,haarsma10,ascaso11}:
luminosity, size and effective radius, star formation history and
stellar populations \ldots\ Discussions to discriminate between the
role of internal feedback processes and of environment and merging of
satellite galaxies in the formation of the BCG are still open. They
also depend on the scenario of galaxy formation and the $\Lambda$CDM
paradigm seems to favor the importance of galaxy mergers at the centre
of the main halo \citep{delucia07}. On the contrary recent observations
confirm the importance of baryonic feedback in the size evolution of the
BCGs \citep{ascaso11}. It is out of the scope of this paper to produce a
detailed analysis of the structural parameters of the BCGs in our sample,
and we simply compared their ellipticity and orientation (measured with  {\sc Sextractor}) with the large
scale light and dark matter distributions (Table \ref{table:optMorpho}).
The link with the dark matter ellipticity is not obvious but the
correlation between the light distribution and the central BCG is quite
convincing (Fig \ref{fig:light_bcg}), except for two outliers which have
complex sub-structures. Quantitatively we find a weighted mean 
$<\varepsilon_{light}-\varepsilon_{BCG}> = 0.001 \pm 0.12$ and the mean
orthogonal deviation from the 1:1 line ($\varepsilon_{light} =
\varepsilon_{BCG}$) is 0.06. This value decreases to 0.03 if we remove
the two major outliers, well below the uncertainties in the
ellipticity measurements, of the order of 0.05. This result 
is not surprising as the alignment of the BCG with the distribution
of galaxies at large scale was first observed by \citet{lambas88} and
confirmed since then by many studies \citep{panko09,niederste10}.

In a second step we separated the clusters in two classes: those dominated by a
single giant elliptical galaxy, most often embedded in an extended envelope (a
cD-type galaxy) and those for which more than one bright galaxy forms the
cluster centre, or the brightest cluster member does not outshine other
galaxies. 5 out of 11 clusters are dominated by a cD galaxy, namely RXC
J0856.1+3756, RX J1120.1+4318, RXC J1206.2--0848, MS 1241.5+1710 and RX
J2228.5+2036. Surprisingly they are not necessarily the brightest clusters in
terms of total stellar luminosity nor the most massive ones, suggesting that the
formation of a giant cluster galaxy is not only related to the initial halo
conditions but also to the evolution processes in the clusters and the merging
history of the structures \citep{dubinsky98}.


\section{Bi-dimensional weak lensing analysis and dark matter distribution}
\subsection{Mass reconstruction}
We refer to Paper 1 for the details of the weak lensing implementation. In
summary, the galaxies shapes are measured with the {\sc Im2shape}
software \citep{bridle02}. For each object a parametric shape model is
set up with an ellipse. {\sc Im2shape} convolves this model with the
local PSF and subtracts it to the sub-image centered on the galaxy.
A MCMC minimizer applied on the image residuals provides the intrinsic
shape parameters and error estimates. The PSF
is measured directly on the images by averaging the shapes of the 5
closest stars to each galaxy.  All measures are done on the $r'$ images
which were obtained with the highest image quality. Only galaxies for
which the measured ellipticity error is smaller than 0.25 are kept in
the working catalogs. The main results regarding the mass measurements
and the study of the global properties of the clusters are presented in
Paper 1, as well as a careful analysis of the sources of error in the mass
determination. In the present work, we focus on the spatial distribution
of the dark matter traced by the weak lensing map reconstruction.

We used the software {\sc LensEnt2} kindly provided by P. Marshall
\citep{marshall02} to build the weak lensing mass maps. The method is
based on an entropy-regularized maximum-likelihood technique.  It uses
the shape of each background galaxy as an individual estimator of the
local reduced shear. The pixel size on the mass grid is chosen to have
more or less 1 galaxy per pixel leading to a comparable number of data
and free parameters. To consider the fact that clusters have an extended
and smooth mass distribution, the code includes a smoothing via the size
of the Intrinsic Correlation Function (ICF): the physical mass map is
expressed as a convolution of the ``hidden'' distribution (the pixels
grid) with a broad kernel defined by the ICF. The shape and size of this
ICF are the main control parameters of {\sc LensEnt2} and reflect the
spatial resolution of the reconstructed mass map. For simplicity, we used
only a Gaussian ICF with a width of 150\arcsec\ for all clusters. This
represents a good compromise between smoothness and details in  the
mass map. {\sc LensEnt2} also computes error maps which give locally
the width of the probability function in the mass reconstruction.
The average value of the error map within the central area of the CCD
image ($15\arcmin \times 15\arcmin$) is a good estimate of the level of
uncertainty in the mass reconstruction.  This is the value which
determines the signal-to-noise of the detected peaks (Table
\ref{table:resultsWL}). 

{\sc LensEnt2} provides output mass maps in physical units of
surface mass density ($M_\odot \ pc^{-2}$). This is valid provided
that the redshift distribution of the background sources is well known.
In the present case, we worked with source catalogs which were cleaned
for galaxy cluster members, but still contaminated by
foreground galaxies. A rough estimate of such contamination comes from
the redshift distribution of galaxies in the magnitude range selected
for our catalog: if we apply the same selection criteria on the deep
photometric catalog with photometric redshifts built from the CFHTLS-Deep
survey \citep{coupon09}, we find that about 25\%\ of the galaxies are
at redshift smaller than 0.5, {\it i.e.} foreground galaxies. This is
coherent with the number found in deep spectroscopic surveys although
at slightly brighter magnitudes \citep{lefevre05}.  In our case,
the effect of such uniform contamination is mostly a dilution of the
weak lensing signal, so the output mass densities of {\sc LensEnt2}
are not reliable in their absolute values. Moreover, there remains some
additional contamination from cluster galaxies in the very centre of the
clusters (see Figure 3 in Paper 1).  The main effect is to attenuate the
peak intensity and to decrease the S/N ratio of the cluster component.
But we do not expect any significant influence on the shape of the
mass reconstruction, provided the contaminating cluster members are
randomly oriented within the cluster. This assumption is valid in our
case because the galaxy catalogs include cuts in color and magnitude
which eliminate all bright cluster members. The remaining galaxies
are mostly blue and/or faint so they are less sensitive  to intrinsic
alignment effects \citep{mandelbaum11}. Therefore in the rest of the
paper we concentrate our work on the 2D mass distribution.
The mass map reconstructions are displayed for each cluster in the Appendix. 

\subsection{Ellipticity of the mass distribution}
\begin{table*}
\caption{
General properties of the weak-lensing mass maps for the cluster sample:
signal-to-noise ratio of the central peak of the mass map, ellipticity and
position angle (PA), projected distances of the mass density peak from the
brightest cluster galaxy (BCG) and the peak of the X-ray emission, and distance
between the peak of the X-ray emission and the BCG. The elliptical parameters
are fit from a mass isocontour drawn at the $3\sigma$ level, except for 3
clusters (see text for details). Position angles (PA) are given in degrees
counter-clockwise with respect to the NS axis. Shifts between the BCG and the
mass peak smaller than 30\arcsec\ can be considered as non significant and values
larger than this limit are marked in bold. Note that in most cases, the position
of the BCG does not exceed 6\arcsec\ (or about 40 kpc) from the X-ray peak. }
\label{table:resultsWL}
\begin{center}
\begin{tabular}{l c c c c c c}
\hline\hline
\noalign{\smallskip}
Cluster & Peak S/N & Ellipticity &
PA  & Distance & Distance & Distance \\
  & & $\epsilon$ & (deg.) & (Peak, BCG) & (Peak, X-rays) & (X-rays, BCG) \\ 
\noalign{\smallskip}\hline\noalign{\smallskip}
MS 0015.9+1609 & 6.5 & $0.26\pm0.11$ & $+95\pm9$ & {\bf 43}\arcsec & 49\arcsec &
6\arcsec\quad (38 kpc) \\
MS 0451.6--0305 & 4.3 & $0.38\pm0.09$ & $+134\pm6$ & {\bf 33}\arcsec & 27\arcsec &
3\arcsec\quad (19 kpc) \\ 
RXC J0856.1+3756 & 5.5 & $0.20\pm0.07$ & $+54\pm9$ & 6\arcsec & 4\arcsec &
4\arcsec\quad (22 kpc) \\
RX J0943.0+4659 (*) & 6.4 & $0.18\pm0.05$ & $+88\pm12$ & {\bf 72}\arcsec & 7\arcsec &
75\arcsec\quad (407 kpc) \\
RXC J1003.0+3254 (*) & 5.5 & $0.35\pm 0.03$ & $+21\pm9$ & {\bf 82}\arcsec & 94\arcsec
& 151\arcsec\quad (830 kpc) \\
RX J1120.1+4318 & 2.7 & $0.15\pm0.05$ & $+42\pm13$ & {\bf 69}\arcsec & 63\arcsec &
28\arcsec\quad (190 kpc) \\
RXC J1206.2--0848 & 7.0 & $0.25\pm0.04$ & $+74\pm4$ & 5\arcsec & 3\arcsec &
2\arcsec\quad (11 kpc) \\
MS 1241.5+1710 & 4.4 & $0.16\pm0.06$ & $+149\pm21$ & 13\arcsec & 16\arcsec &
1\arcsec\quad (6 kpc) \\
RX J1347.5--1145 & 9.8 & $0.13\pm0.05$ & $-19\pm20$ & 4\arcsec & $<1$\arcsec &
$<1$\arcsec\quad ($<$ 6 kpc) \\
MS 1621.5+2640 & 5.4 & $0.27\pm0.06$ & $+149\pm13$ & 26\arcsec & 42\arcsec &
8\arcsec\quad (45 kpc) \\
RX J2228.5+2036 (*) & 4.7 & $0.40\pm0.07$ & $+104\pm5$ & {\bf 40}\arcsec & 56\arcsec &
4\arcsec\quad (22 kpc) \\
\noalign{\smallskip}\hline
\end{tabular}
\end{center}
(*) double cluster or merger
\end{table*}

The ellipticity of the mass distribution traced by the weak lensing mass
reconstruction has been the focus of several studies. It is a direct evidence of
the triaxiality of the cluster halos and is expected to be a non-negligible
factor in the growth of massive halos in the $\Lambda$CDM paradigm
\citep{limousin13}. \citet{oguri10} studied a sample of 25 massive
X-ray clusters (mostly in the LOCUSS sample) at redshift $\sim 0.2$. They found
an average ellipticity $\left<\epsilon\right> = 1 - b/a = 0.46 \pm 0.04$. More
recently \citet{oguri12} confirmed the trend with another independent sample of
clusters built from the Sloan Giant Arcs Surveys as part of the SDSS
\citep{hennawi08}. These values of the mean ellipticity are in good agreement
with the theoretical predictions based on numerical simulations of cluster dark
matter halos \citep{jing02}. They correspond to what is expected for massive
clusters, contrary to low mass clusters which are expected to be more circular.

We tried to explore this question with the present sample. But because
our sample is at higher redshift than LOCUSS, several difficulties limit
the outcomes of the approach.  
In order to get a estimate of the uncertainties of the elliptical parameters for each mass
reconstruction, we used a ``jackknife'' resampling method to remove
10\%\ of the galaxies in the source catalog, and we repeated the process
10 times. The removed galaxies all differ from one attempt to the
others. 10 new mass maps were computed for each cluster with these
sub-catalogs, as well as the corresponding error maps. The 10 error maps
were averaged and the average level of this frame gives 
the $1\sigma$ level of the mass reconstructions. We then fitted each of
the 10 mass maps with elliptical contours and selected the elliptical
parameters of the $3\sigma$ isocontour. An average of the 10 fits gives
the final values for the elliptical parameters ($\varepsilon$ and PA), as well as their standard
deviation (Table
\ref{table:resultsWL}). This process was acceptable except for the
clusters RX J1120.1+4318 and MS 1241.5+1710 with the lowest S/N maps.
We restricted their fits to the $2\sigma$  and $2.5\sigma$ isocontours
respectively. 

\begin{figure}
\center
\includegraphics[width=0.95\hsize]{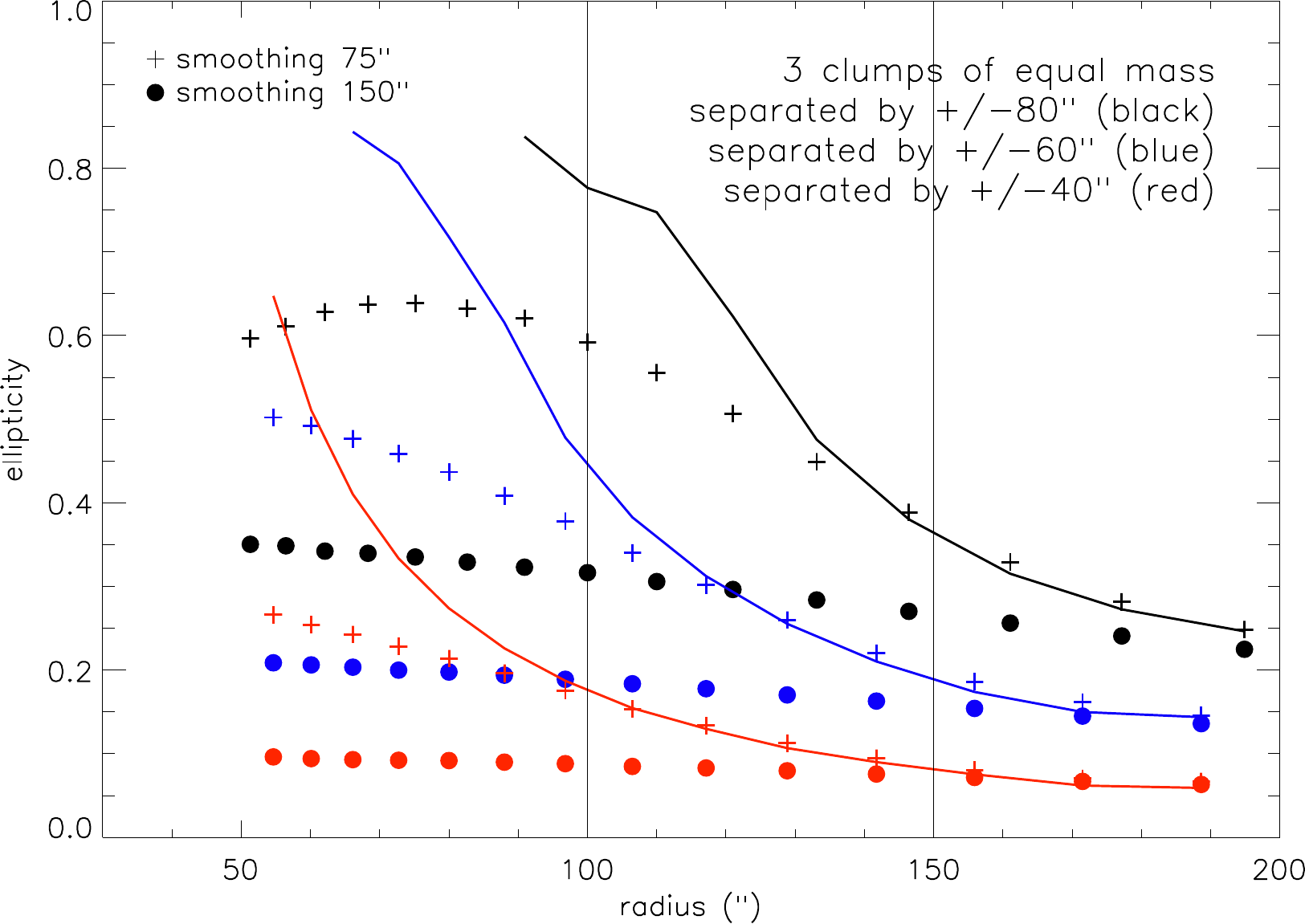}
\caption{Ellipticity of simulated clusters formed by 3 mass clumps with NFW profile, and smoothed with 2
different kernels. The continuum line is the ellipticity measured on
the unsmoothed data. For comparison the ellipticity of
the observed clusters was measured in radii ranging from 100 to 150\arcsec\ approximately.
} 
\label{fig:simulated_ell} 
\end{figure}

In all cases
this corresponds to an isocontour of 100 to 150\arcsec\ in radius (or
600 to 900 $h_{70}^{-1}$ kpc at redshift 0.5). But
because of the limited resolution of the mass reconstruction, the effect
of the central smoothing by the ICF is quite significant and induces
an attenuation in the measure of the mass ellipticity. In practice, an
ICF of 150\arcsec\ corresponds to a Gaussian smoothing with $\sigma \sim
60 \arcsec$.  As a test case we simulated a set of mass maps with multiple clumps of
matter and we tested the effects of the smoothing on the ellipticity
of the mass distribution. In practice each clump was generated with
a NFW profile with $c=4$ and $M_{200}^{2D} = 5.0 \ 10^{15} \ M_\odot$
(associated to $r_s = 60\arcsec = 360$ kpc, $r_{200} =
1.4$ Mpc and $M_{200} = 3.9 \ 10^{14} \ M_\odot$). Three clumps
were aligned along a line and regularly spaced, with separation ranging
from 40\arcsec, 60\arcsec\ and 80\arcsec\ (250 kpc to 500 kpc at $z
\sim 0.5$) between the clumps. The 2D mass maps were then smoothed with
2 different kernels, with $\sigma = 30\arcsec$ and  $\sigma = 60\arcsec$,
corresponding respectively to a ICF of 75\arcsec\ and 150\arcsec\ in the
mass reconstruction. The ellipticity of the simulated mass distributions
was measured with the same method as for the clusters maps, with the
centering fixed on the central mass peak, for both the smoothed and
unsmoothed distributions (Fig \ref{fig:simulated_ell}). The relevant
ellipticiy was measured in radii between 100\arcsec\ and 150\arcsec\
for the observed clusters.  The measures on the simulated clusters show
that the apparent ellipticity is typically decreased by a factor 2 when
applying this severe smoothing. 
We did not attempt to correct more accurately the measured
ellipticities because the ellipticity attenuation should also depend on
the mass profile, which is not constrained enough in this study. A higher
background galaxy density would have allowed sharper mass reconstructions
but this is out of reach with ground-based wide field imaging and requires
data on the quality of HST.

The results of the elliptical fitting of the cluster mass distribution are presented in Table
\ref{table:resultsWL}. Half of the clusters has a low
ellipticity (5 clusters with $\varepsilon_{DM} < 0.2$) while the other
half is clearly elliptical (6 clusters with $\varepsilon_{DM} > 0.2$). 
The ellipticity distribution of the sample shows an weighted mean of  
$\left< \epsilon \right> = 0.25 \pm 0.12$. It appears
narrower than that of other samples like LOCUSS \citep{oguri10}. But
we have a smaller number of clusters and the difficulties to
provide weak lensing maps at redshift 0.5 are stronger. No value 
exceeds 0.40, contrary to what is expected for such a cluster sample
and none of the clusters is really circular ($\varepsilon_{DM} < 0.1$) in
their extended regions. The smoothing process, needed to get an
acceptable mass map at high redshift, is certainly the cause of the
lack of high ellipticity clusters in our sample (Fig \ref{fig:simulated_ell}). In conclusion, even
if we can not draw firm conclusions on the ellipticity distribution of the dark
matter from our sample, we show that we remain compatible with standard
expectations and previous works. A similar attempt to test the evolution of the
ellipticity of the X-ray gas distribution was proposed by \citet{maughan08} with
{\it Chandra} data. They did not find any change in the ellipticity distribution
of the high-z sample ($>0.5$) compared to their low-z sample, contrary to other
morphological parameters like the slope of the surface brightness profile at
large radius. 

We also compared the distance between the main mass peak and the location
of the BCG to check the consistency between the positions of the dark
matter peak and the light peak. However the uncertainty in the centering
of the mass distribution is high and strongly depends on the S/N ratio of
the main mass peak. As demonstrated by \citet{dietrich11} with numerical
simulations, the shape noise in the weak lensing map reconstruction
combined with the smoothing process generate an offset distribution with a
mode as large as 0.3\arcmin\ and median values up to 1\arcmin\ for typical
ground-based observations. In our case, the clusters with the best map
reconstruction and the highest S/N in the central peak ($>5$) are also
those for which the position of the mass peak matches the position of the
BCG as well as the centroid of the gas distribution (Fig. \ref{fig:0015}
to \ref{fig:2228} in the Appendix). This is generally valid, except
for the clusters RX J0943.0+4659 (merging cluster) and MS 0015.9+1609.
This last cluster deserves some comment because previous weak lensing
modeling of the central area, using HST/ACS images, point towards a mass
centre  well centered on the three brightest galaxies \citep{zitrin11}.
Our mass reconstruction suffers from the proximity with a bright star
and its halo which distorts the shape distribution of the faint galaxies.

More interesting is the distribution of the X-ray/BCG offset. It
clearly appears bi-modal with most of the clusters having an offset
smaller than 50 kpc, and three outliers. Among these three clusters,
two are merger systems and the last one, RX J1120.1+4318 is rather
poorly defined in its centre. This is a similar trend as found by
\citet{sanderson09} in the LOCUSS sample of low redshift clusters.
This X-ray/BCG offset is a good indicator of the dynamical state of
the cluster and is highly correlated with the strength of the cooling
core in the centre. The comparison between the X-ray and the mass
peaks is more uncertain due to the limitations mentioned above. But
it shows a similar trend, at least for the clusters with a mass peak
detected with a high enough significance. 

\section{Mass and light distributions}
\subsection{Comparison between light and mass 2D distributions}
\begin{figure}
\center
\includegraphics[width=0.95\hsize]{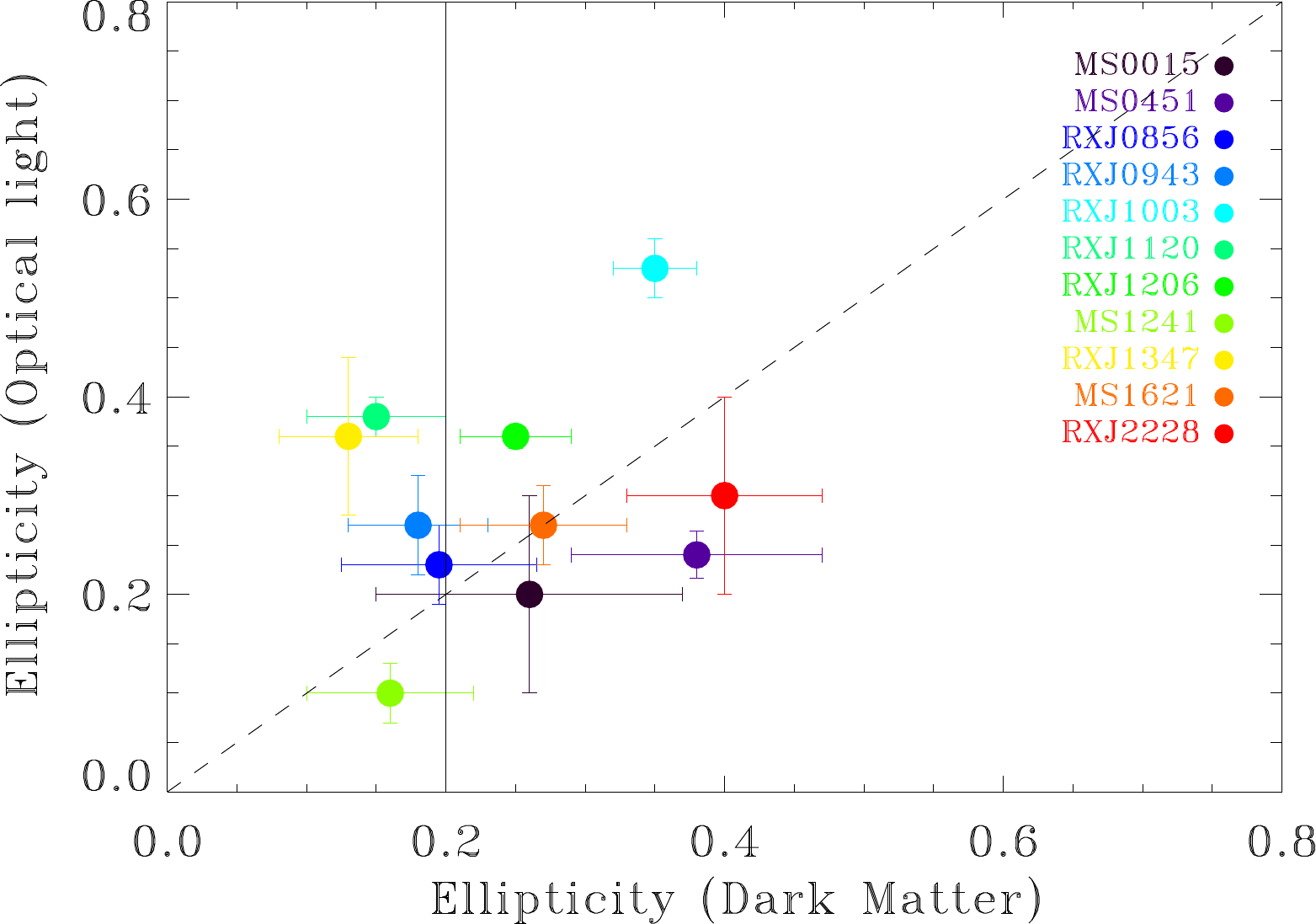}
\caption{Ellipticity of the optical light distribution versus the
ellipticity of the dark matter. The vertical line separates the
cluster sample between the ``circular'' clusters with $\epsilon < 0.2$
and the ``elliptical'' or the irregular ones. } 
\label{fig:ell} 
\end{figure}

Fig. \ref{fig:ell} shows the correlation between the ellipticity of the
dark matter and that of the stellar light distributions. As for the
comparison between the cluster light and the BCG ellipticities, we
measured a weighted mean $<\varepsilon_{DM}-\varepsilon_{light}> = -0.1 \pm 0.02$ and
the mean
orthogonal deviation to the $\varepsilon_{DM} = \varepsilon_{light}$
line is 0.08 for the 11 clusters. Dispersion is slightly higher than
for the light--BCG comparison but again we find a good concordance
between both ellipticities, and a tendency towards a better agreement
for elliptical clusters than for circular ones. We
remind that these measurements correspond to large scale morphologies
so they are more sensitive to substructures which can be found at the Mpc
scale. Weak lensing morphology can also be disturbed by additional mass
halos projected on the line of sight but not physically related to the
clusters. Similar conclusions were reached by  \citet{oguri10} 
from a very similar study. 

\begin{figure}
\center
\includegraphics[width=0.95\hsize]{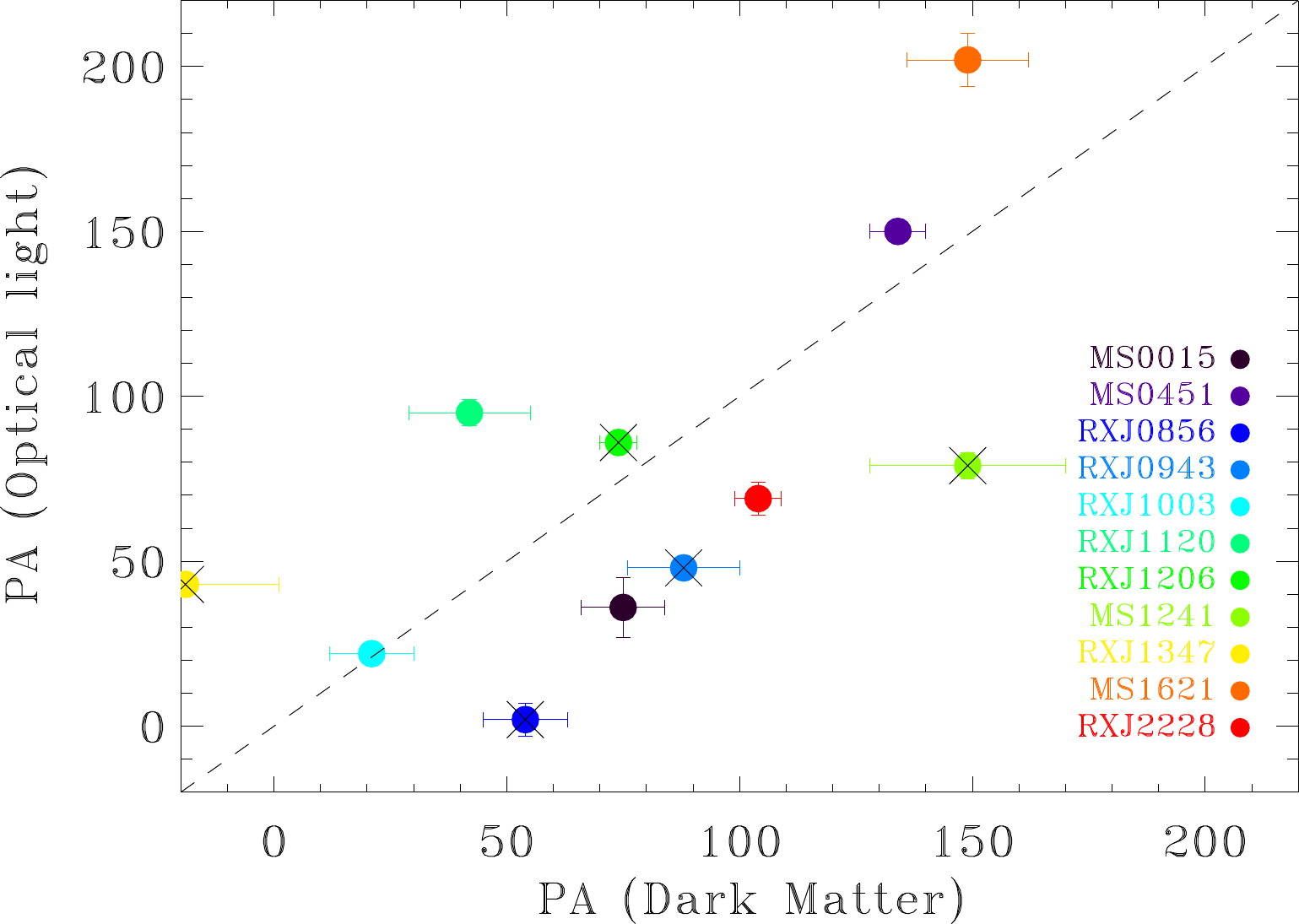}
\caption{
Orientation of the optical light distribution versus the
dark matter. The ``circular'' clusters with an ellipticity $\epsilon <
0.2$ are marked with a cross.} 
\label{fig:pa} 
\end{figure}

\begin{figure}
\center
\includegraphics[width=0.95\hsize]{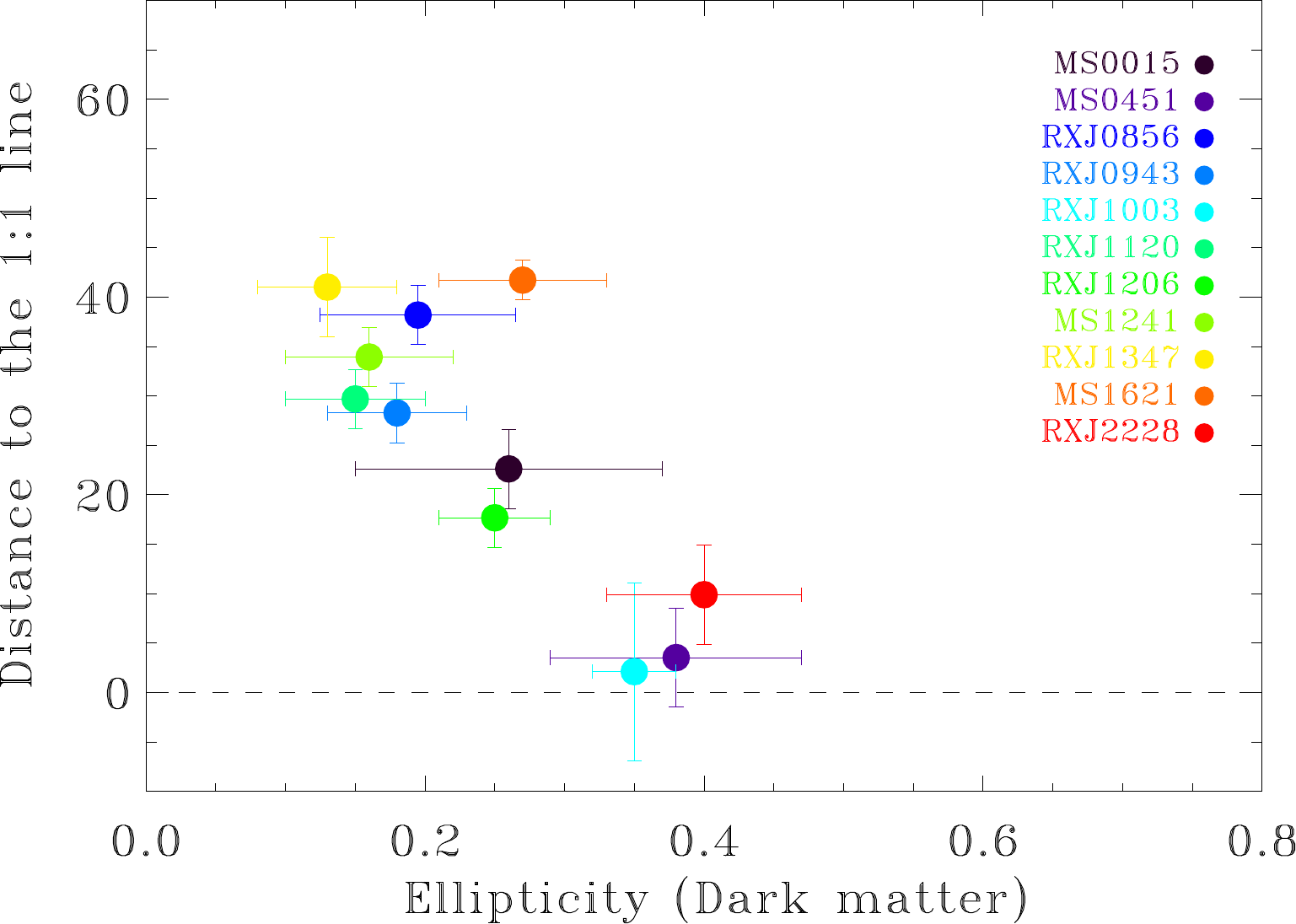}
\caption{
Orthogonal deviation (in degrees) from the 1:1 line for the dark
matter orientation (PA(DM)) compared to the light distribution
orientation (PA(light)) versus the ellipticity of the dark matter
distribution. The more elliptical this distribution is, the better
the alignment between light and dark matter is. 
} 
\label{fig:dist_pa} 
\end{figure}

To study the possible alignment effect between the light and the dark
matter distributions, we represented the position angle of the optical
light versus the position angle of the dark matter distribution (Fig.
\ref{fig:pa}). To better visualize the shift with respect to the $y=x$
line, we also computed the distance between the data points associated
to the clusters and the 1:1 line in that plane.  The position angles
of both distributions are strongly correlated: the mean difference
$<\Delta PA> \ = \ < \textrm{PA(DM) -- PA(light)}> \ = \ +3 \pm 3 ^\circ $
degrees and the average deviation to the 1:1 line is 28$^\circ$. In
addition, as shown in Fig. \ref{fig:dist_pa}, these differences
tend to vanish when the ellipticity increases. We suspect that large
differences at small ellipticities are partly due to biases
in the processing of the elliptical fits or to the influence of further
substructures at large radius. However \citet{oguri10} show in their
detailed study that elliptical fits of weak lensing maps are robust when
they use similar radii of 400 to 800 kpc. So we are confident that in
the present study the ``light traces mass'' assumption is valid
when clusters are in quiescent phases of their evolution. Departures from
this assumption occur when the clusters enter merging processes and
when interactions between large clumps of matter globally perturb their
dynamical equilibrium. 

\subsection{Cluster mass-to-light ratio}
\begin{figure*}
\center
\includegraphics[width=0.8\hsize]{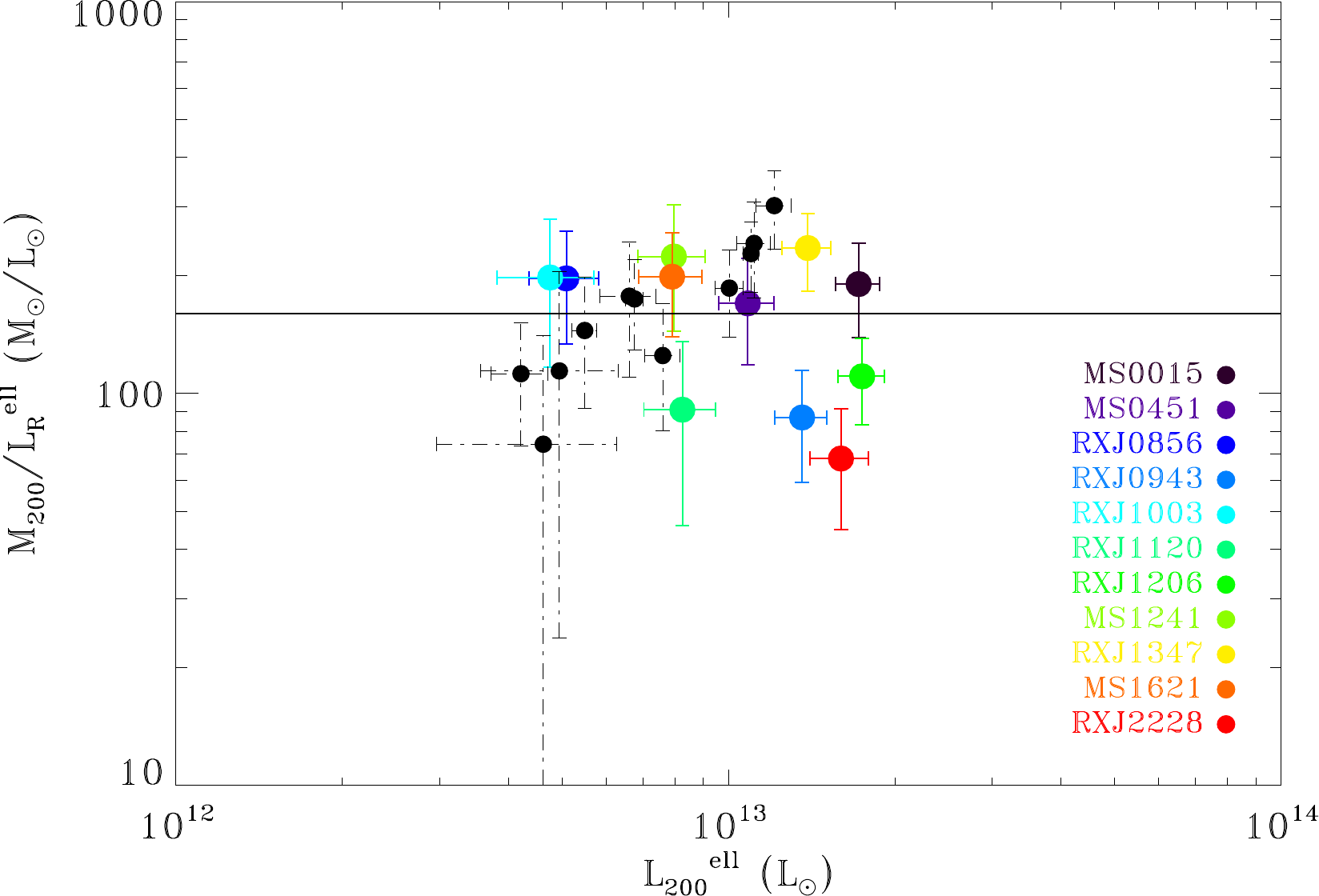}
\caption{Mass-to-light ratio versus the total luminosity $L_{200}^{2D}$ for
the 11 clusters of the sample (color points) and the 11 clusters added from the
Bardeau et al. (2007) sample. The average value $M/L = 166$ in solar units
is drawn as the solid line. }
\label{fig:msurl} 
\end{figure*}

The mass-to-light ratio is a quantity that has been widely studied at
every mass scale, from single galaxies up to rich and massive galaxy
clusters. Its variation across the mass range allows, for instance,
to highlight physical processes that affect the baryonic component of
massive structures, like star formation or galaxy-galaxy interactions
in large dark matter halos \citep{carlberg96,marinoni02,giodini09}. In
Paper 1, we analyzed the correlation between mass and luminosity and
found a logarithmic slope of $0.95 \pm 0.37$ for the mass-luminosity
scaling law, compatible with a constant M/L ratio. In the present paper,
we used 2D projected mass $M_{200}^{2D}$, instead of the 3D mass used
in Paper 1. Both masses differ mostly by a scale factor because all
clusters are assumed to have the same concentration parameter. This
projected mass is the right quantity to be compared to the total 
projected luminosity. So we correlated the 2D mass with the total
luminosity of
early-type galaxies, computed inside the radius $R_{200}$ and corrected
from the non-detected part of the luminosity function. We obtain an
average ratio $\left< M/L \right> = 160 \pm 60 \ h_{70} \ (M/L)_\odot$,
with values ranging from $\sim$ 68 to 235 (Table \ref{table:opt_prop}
and Fig.~\ref{fig:msurl}). Our results are in excellent
agreement with the values obtained by Bardeau et al. (2007) who used a
similar methodology to derive the weak-lensing masses and the optical
luminosities for a sample of clusters at lower redshifts ($z \sim
0.2$). They found an average ratio of $\left< M/L \right> = 170 \pm 67 $
for the same quantities, and their results are very similar to the present
ones. Because the methodology they used is so close to the present one,
we merged the two samples although the clusters differ in redshift. From
the 22 clusters we find an average ratio $\left< M/L \right> = 166 \pm
62 \ h_{70} \ (M/L)_\odot$.  Comparisons with other samples are made
rather difficult due to the several methodologies used to derive optical
luminosities \citep{popesso07} and cannot be discussed further.

In addition, although the mass interval of the sample is rather limited, ranging
from $6 \times 10^{14} \ M_\odot$ to $2.5 \ 10^{15} \ M_\odot$, we tried to consider
the possible variation of the M/L ratio with mass. This is rather speculative
and limited by the fact that the clusters with the lowest mass are also those
with the largest uncertainty in the weak lensing peak detection and
that the 2 quantities are correlated. 
Following the same procedure as described in \citet{foex12}, we fitted the M versus L
relation using a linear regression in the log-log plan with the orthogonal BCES method 
which takes into account errors in both directions and provides a statistical dispersion around the fit $\sigma_{stat}$ as well as the intrinsic dispersion $\sigma-{int}$. We did 
not find any significant departure
from a constant ratio between the 2 quantities, with a slope $\alpha = 0.945
\pm 0.37,  \sigma_{int} = 0.14$ and $\sigma_{stat}=0.11$ for the dispersions of the fit in the log-log space. With the present data we do not find any departure from a constant M/L ratio independent of the total cluster mass, although previous works have found a significant increase of the M/L ratio for massive clusters of galaxies
compared to lower mass clusters and groups  \citep{popesso07,andreon10}. But this was obtained with samples spanning a much larger mass interval than the present one. The physical origin of this
situation is still controversial but it is usually understood in terms of a
decrease of the star formation efficiency with increasing halo mass
\citep{springel03, lin03}.

\section{Summary and conclusions}
The cluster sample presented in this paper is limited to 11 clusters only, at a
redshift $z\sim 0.5$. They were selected by their X-ray emission and are part of
the representative sample \excpres\ which includes 20 clusters in the redshift
range $0.4 < z < 0.6$. Because of additional criteria used to optimize the weak
lensing detection and analysis, the present sample is not any more
representative of the cluster population at intermediate redshift, but it forms
a sub-sample with the brightest X-ray luminosity. We summarize below the
properties of the clusters which have been explored in the paper:
\begin{itemize}
\item We provide for each cluster the total luminosity of the cluster,
after a careful identification of cluster members, and several
morphological parameters of the light distribution. We find good
correlations between the ellipticity of the BCG and the global light
distribution in terms of ellipticity and orientation. 
But whether the BCG has a bright and extended envelope
of cD-type or not does not make significant differences in the general
optical properties of the clusters.
\item The weak lensing mass reconstruction is done for each cluster,
although the peak detection is at low significance in a few cases (one
cluster detected at less than 3$\sigma$). The average ellipticity of
the mass maps is $<\epsilon> = 0.25$, a value compatible with similar
estimates at lower redshift. No evolution of the average ellipticity
of the clusters or the fraction of high ellipticity mass distributions
is detectable in our data. We also explore the distance between the
mass peak, the location of the BCG and the X-ray centre for each
cluster. The position of the mass peak is the most uncertain and is
limited intrinsically by the low density of background galaxies in the
mass reconstruction. On the contrary, we find good agreement between
the location of the BCG and the X-ray centre, especially for regular
clusters. As expected the most discrepant clusters are those with the
most disturbed morphology or clear signs of dynamical perturbations.
\item The mass-to-light ratio distribution shows excellent agreement
with previous measures done with a similar approach and the average
M/L ratio is found: $\left< M/L \right> = 160 \pm 60 \ h_{70} \
(M/L)_\odot$. Previous studies were done at lower redshift and we do not
find significant sign of evolution, as expected for this intermediate
redshift bin. 
\end{itemize}

These properties point towards a general picture of the clusters for which
``the light follows mass'' paradigm is the main driver. This good coherence
is valid both in the central parts of the clusters and at large scale, as
demonstrated with the weak lensing mass reconstructions. Going into more
details, we tried some attempts to separate the sample in two classes as was
done previously for other cluster samples like LOCUSS or the CCCP
\citep{smith05,mahdavi13}. A majority of clusters are regular and follow the
main correlations and 3 or 4 outliers are identified as non-relaxed clusters
or merger systems (namely RX J0943.0+4659, RXC J1003.0+3254, RX J2228.5+2036
and possibly MS 1241.5+1710). To better quantify departures from the
assumption of regularity it will be important to study and better understand
the influence of substructures and 
the role of triaxiality. It is a natural consequence of structure growth
driven by self gravity of Gaussian density fluctuations \citep{limousin13}
but up to now it has been mostly neglected, for simplicity. We now have in
hands a good understanding of the tracers of the distribution of the
different components in clusters so it is timely and appropriate to address
this issue in details. This would allow to improve mass measurements and the
understanding of the mass growth of structures as massive as clusters of
galaxies.


\begin{acknowledgements}
We thank Gabriel Pratt, Marc Huertas and Roser Pello for fruitful
discussions and encouragements. We acknowledge support from the
CNRS/INSU via the Programme National de Cosmologie et Galaxies (PNCG).
EP and GS acknowledge the support of ANR under grant ANR-11--BD56-015.
GF acknowledges support from FONDECYT through grant 3120160.  
ML acknowledges supports from the CNRS and the Dark Cosmology Centre, 
funded by the Danish National Research Foundation. The present work is
based on observations obtained with {\it XMM-Newton}, an ESA science
mission with instruments and contributions directly funded by ESA
Member States and the USA (NASA), and on data products produced at Terapix
and the Canadian Astronomy Data Centre.  The authors wish to recognize
and acknowledge the very significant cultural role and reverence that
the summit of Mauna Kea has always had within the indigenous Hawaiian
community. We are most fortunate to have the opportunity to conduct
observations from this mountain.  
\end{acknowledgements}

\bibliography{references}


\begin{appendix}

\section{Individual properties of clusters}
The 11 clusters presented in the sample are all bright X-ray clusters.
Some of them also present specific optical properties or are already
known as strong gravitational lenses. We review in this section the
properties of the clusters, mostly in the optical. 
Their X-ray properties will be
presented in a companion paper (Arnaud et al., in preparation). 

\subsection{MS 0015.9+1609 $(z=0.541)$}
This very rich cluster has been studied for many years, since the
identification of a high fraction of red galaxies in its population
\citep{koo81}. Included in the CNOC cluster sample, its spectroscopic
survey was presented in \citet{ellingson98} with more than 180
objects observed spectroscopically. The resulting velocity dispersion
$\sigma_{los} = 1127^{+168}_{-112}$ km s$^{-1}$ is a high value
consistent with its galaxy richness \citep{borgani99}.  MS 0015.9+1609
is one of the brightest and most distant X-ray cluster included in the
EMSS sample \citep{gioia94}. It is also part of the highly luminous
X-ray clusters identified in the MACS sample at redshift larger than
0.5 \citep{ebeling07} and is identified as MACS J0018.5+1626.

The weak lensing properties were described by \citet{smail95} and then
by \citet{clowe00}. The authors found a rather low signal and therefore
a total mass not consistent with the optical velocity dispersion of
the galaxies. More recently \citet{hoekstra07} re-analyzed a large
sample of clusters observed in good seeing conditions at CFHT and
found for MS 0015.9+1609 a total mass described by a SIS with $\sigma
= 1164^{+151}_{-173}$ km s$^{-1}$ or by a NFW profile with $M_{200} =
27.0^{+9.0}_{-8.4} \ 10^{14} h^{-1}$ M$_\odot$. Note that despite the high
mass value of the cluster, no strong lensing features were detected in HST
images \citep{sand05}. More recently and thanks to a detailed analysis of
HST/ACS images, \citet{zitrin11} identified 3 systems of multiple images,
though not yet confirmed spectroscopically. They were used to provide
a lensing model of the mass distribution in the centre of the cluster.

\begin{figure*}
  \centering
  \resizebox{0.9\hsize}{!}{
\includegraphics{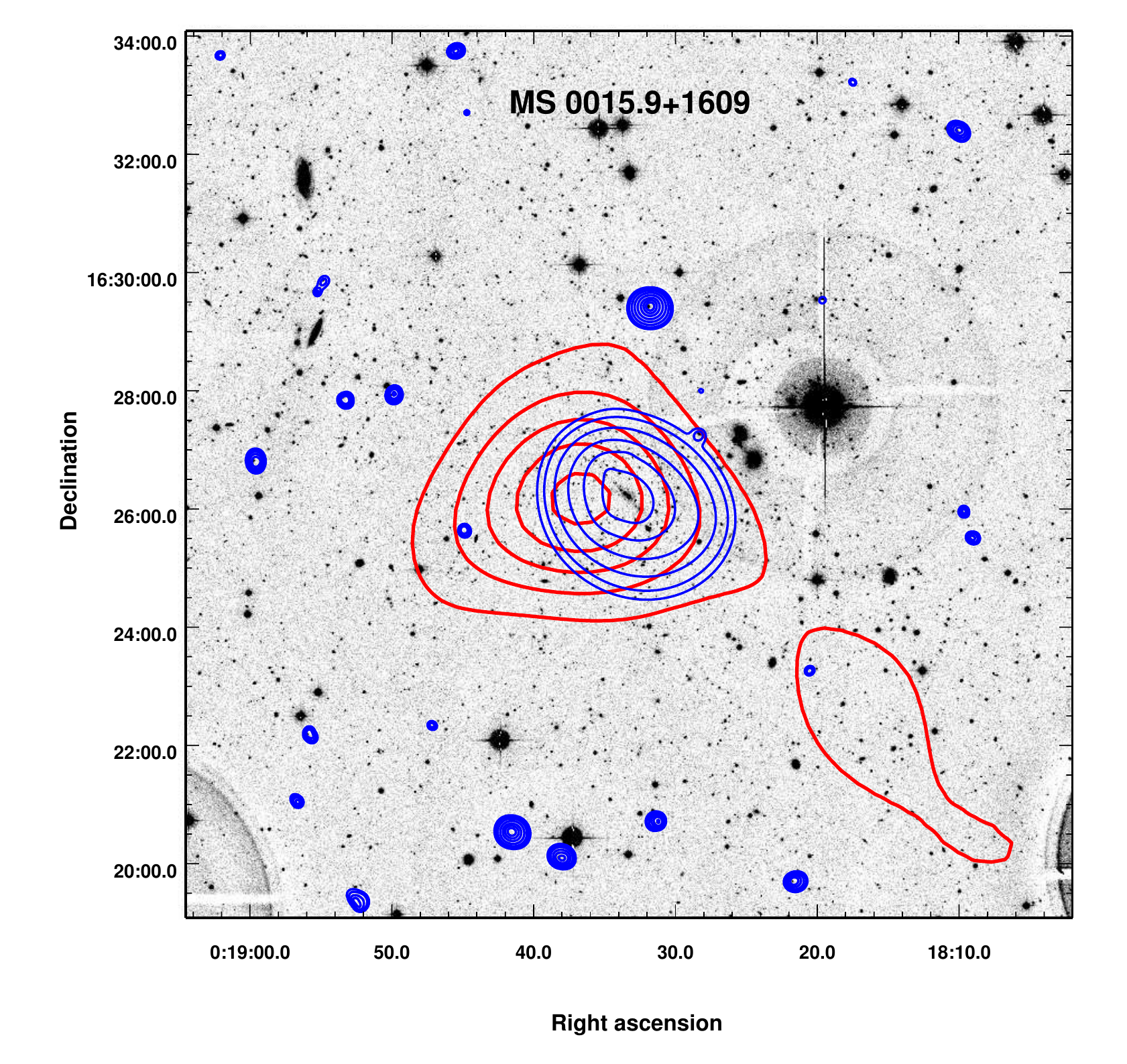}
}
\includegraphics[width=7.5cm]{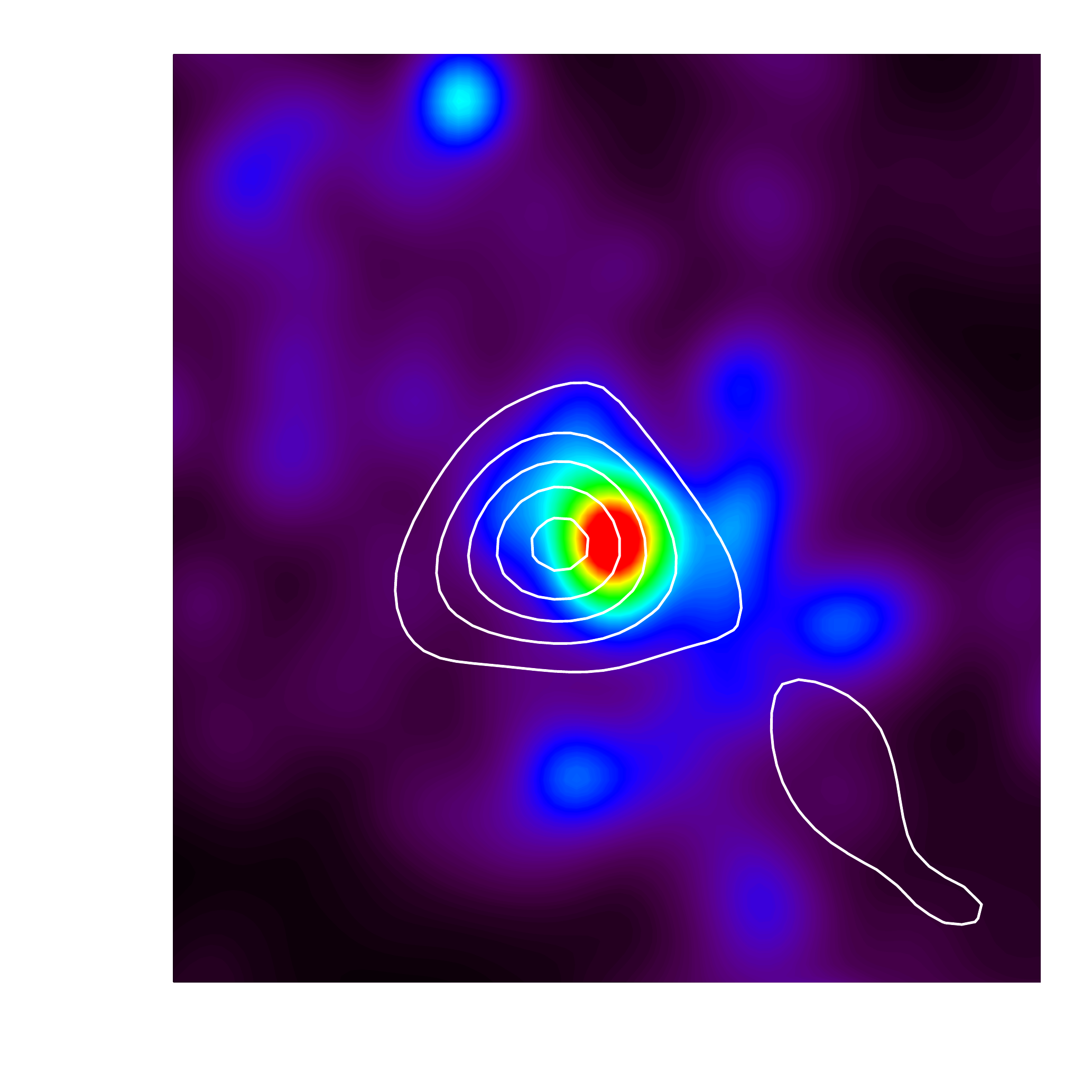}
\includegraphics[width=7.5cm]{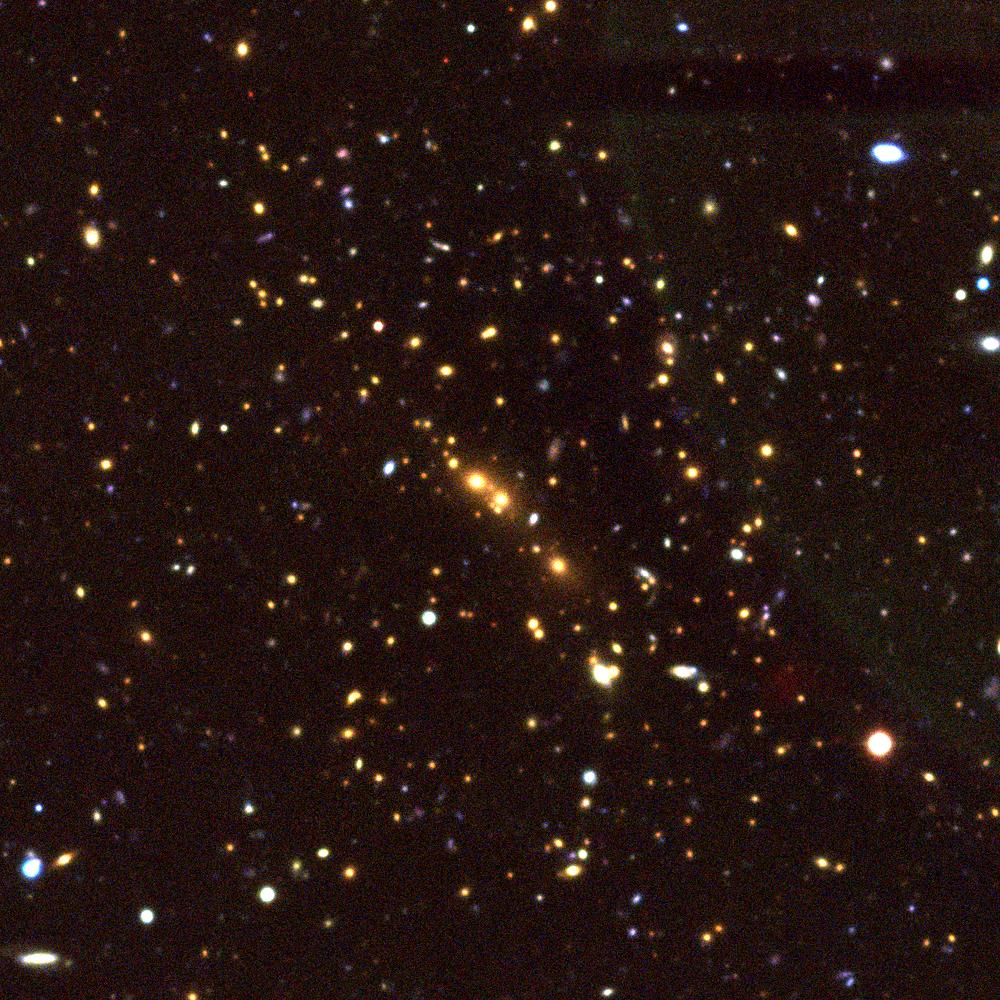}
  \caption{{\bf Up:} $15\arcmin \times 15\arcmin$ insert of the cluster
field extracted from the full $r'$ MegaCam image. The thick red contours
show the mass distribution derived from the 2D weak lensing analysis. The
contour levels are linearly spaced in $\sigma$ of the mass reconstruction,
starting at $2\sigma$. The thin blue contours come from the X-ray
map obtained with \xmm. The image was filtered with wavelets
and the contours are scaled logarithmically.  {\bf Bottom left:} same mass
isocontours overlaid on the galaxy luminosity distribution where cluster
members are selected within the cluster red sequence and $m_r<23$. {\bf
Bottom right:} true color image of the cluster centre from $g'$, $r'$,
$i'$ combination. The field of view is $3\arcmin \times 3\arcmin$.
\label{fig:0015} } 
\end{figure*}

There is no dominant central galaxy in this cluster but a chain of
bright ellipticals, giving a significant elongation in the galaxy
distribution. This elongation was confirmed in the weak lensing map
provided by \citet{zitrin11} on the central area of the cluster. In
our wide field map, the ellipticity of the mass distribution does not
clearly appear (Fig.~\ref{fig:0015}). We we suspect that the bright
star which is close to the cluster centre prevents a correct study of
the cluster mass map obtained from weak lensing reconstruction.

MS 0015.9+1609 is embedded in a large scale structure of the size of a
supercluster, identified spectroscopically by \citet{connolly96}. At least
3 clusters lie within less than 30 Mpc form each other, and
a long and massive filamentary structure crosses the cluster in the
same direction as the galaxies elongation \citep{tanaka07,tanaka09}.
The weak lensing reconstruction presented in the present paper is focused 
on the central area around the cluster only, but we checked that most of
the structures spectroscopically identified by \citet{tanaka07} are also
visible in our global mass map. This may be the case for the South-West
elongation seen in the mass map displayed in Fig.~\ref{fig:0015}.a
Further work is in progress to better quantify these correlations. 

\subsection{MS 0451.6--0305 $(z=0.537)$}
This cluster is the most X-ray luminous cluster in the EMSS catalog
\citep{gioia94} and is also part of the CNOC sample. Intensive
spectroscopic follow-up of the galaxies provided more than 100
spectra of cluster members \citep{ellingson98} and a line-of-sight
velocity dispersion of $\sigma_{los} = 1002^{+72}_{-61}$ km s$^{-1}$
\citep{borgani99}. Weak lensing masses measured by \citet{clowe00}
are roughly compatible with this value as well as  those obtained by
\citet{hoekstra12}. Our own measurements are 50\%\ higher, but they
remain compatible within the uncertainties \citep{foex12}. 
The cluster is also identified as MACS J0454.1--0300.

\begin{figure*}
  \centering
  \resizebox{0.9\hsize}{!}{
\includegraphics{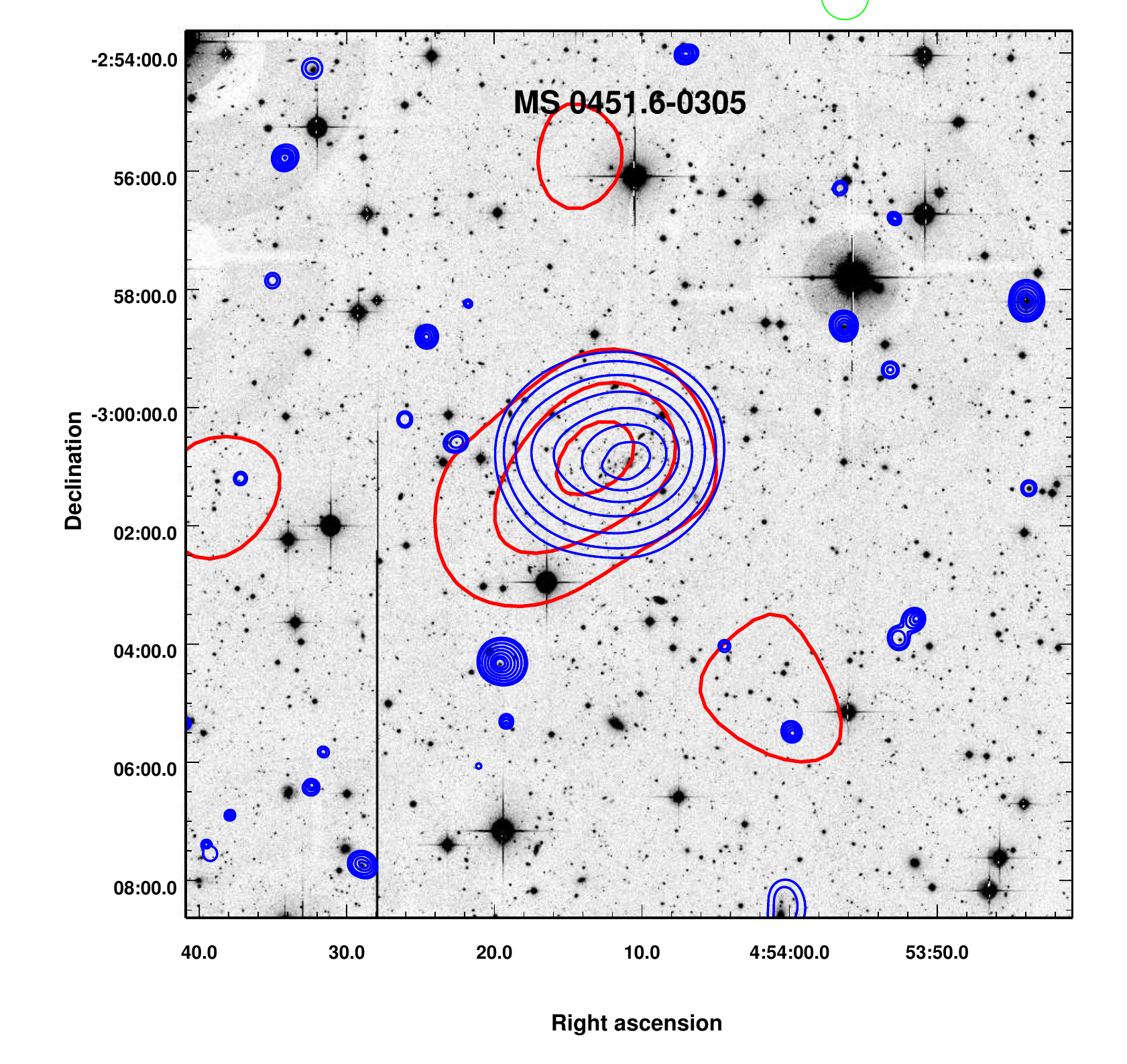}
}
\includegraphics[width=7.5cm]{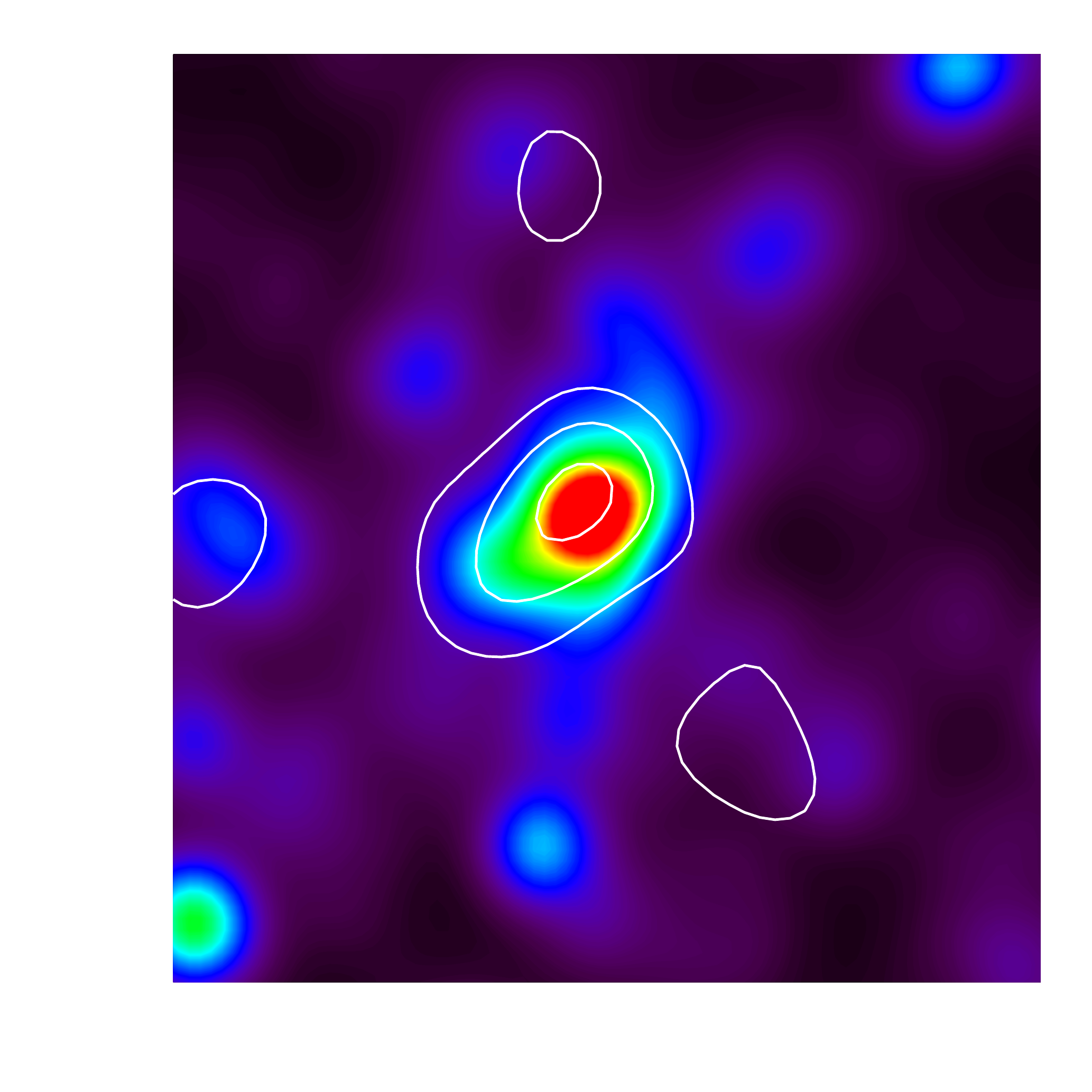}
\includegraphics[width=7.5cm]{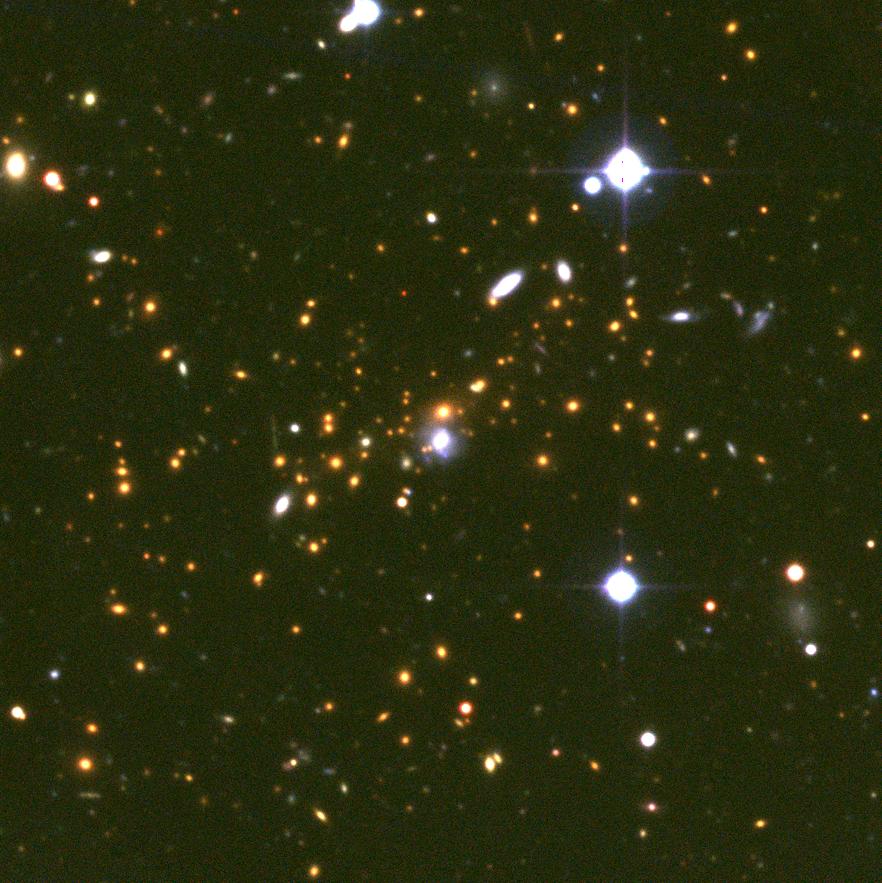}
  \caption{Same as Fig.~\ref{fig:0015}}
\label{fig:0451}
\end{figure*}

A few thin and elongated features were suspected as strong lensing
candidates by \citet{luppino99} and spectroscopically confirmed later
by \citet{borys04}. Interestingly a SCUBA detection of an extended
source in the cluster centre lead to the identification of an ERO
pair, triple imaged \citep{chapman02,takata03,berciano10}. These
features point towards the bright central galaxy as the centre of the
mass distribution. The observed elongation of the weak lensing mass
reconstruction (Fig.~\ref{fig:0451}) is well correlated with the global
elongation of the light distribution in the SE/NW direction. This is
also true for the orientation of the BCG.  The latest strong lens model
presented by \cite{zitrin11} indicates that the central mass distribution
is highly elliptical with an orientation that matches the SE/NW elongation
of the cluster at large scale.

\subsection{RXC J0856.1+3756 $(z=0.411)$} 
\begin{figure*}
  \centering
  \resizebox{0.9\hsize}{!}{
\includegraphics{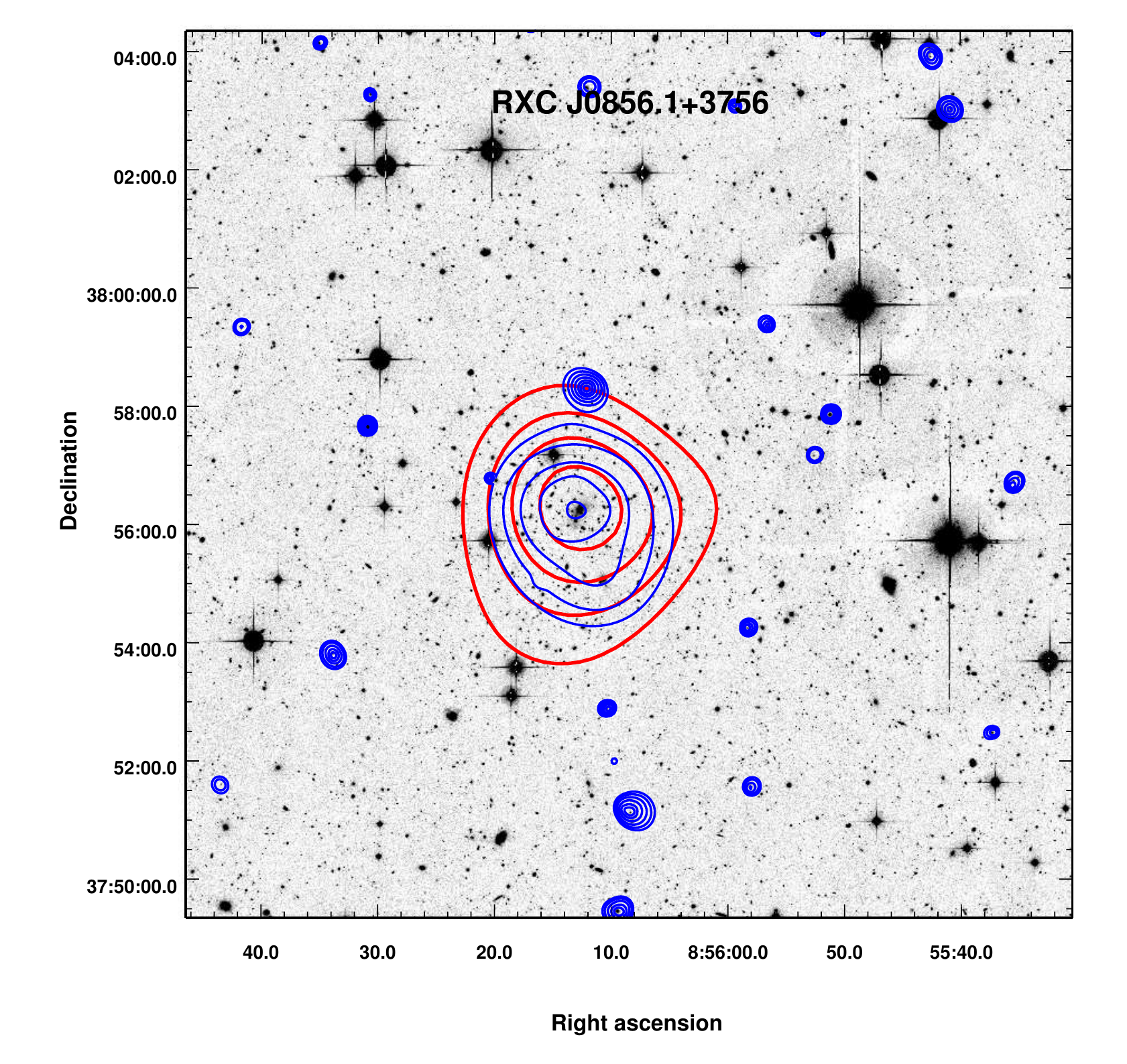}
}
\includegraphics[width=7.5cm]{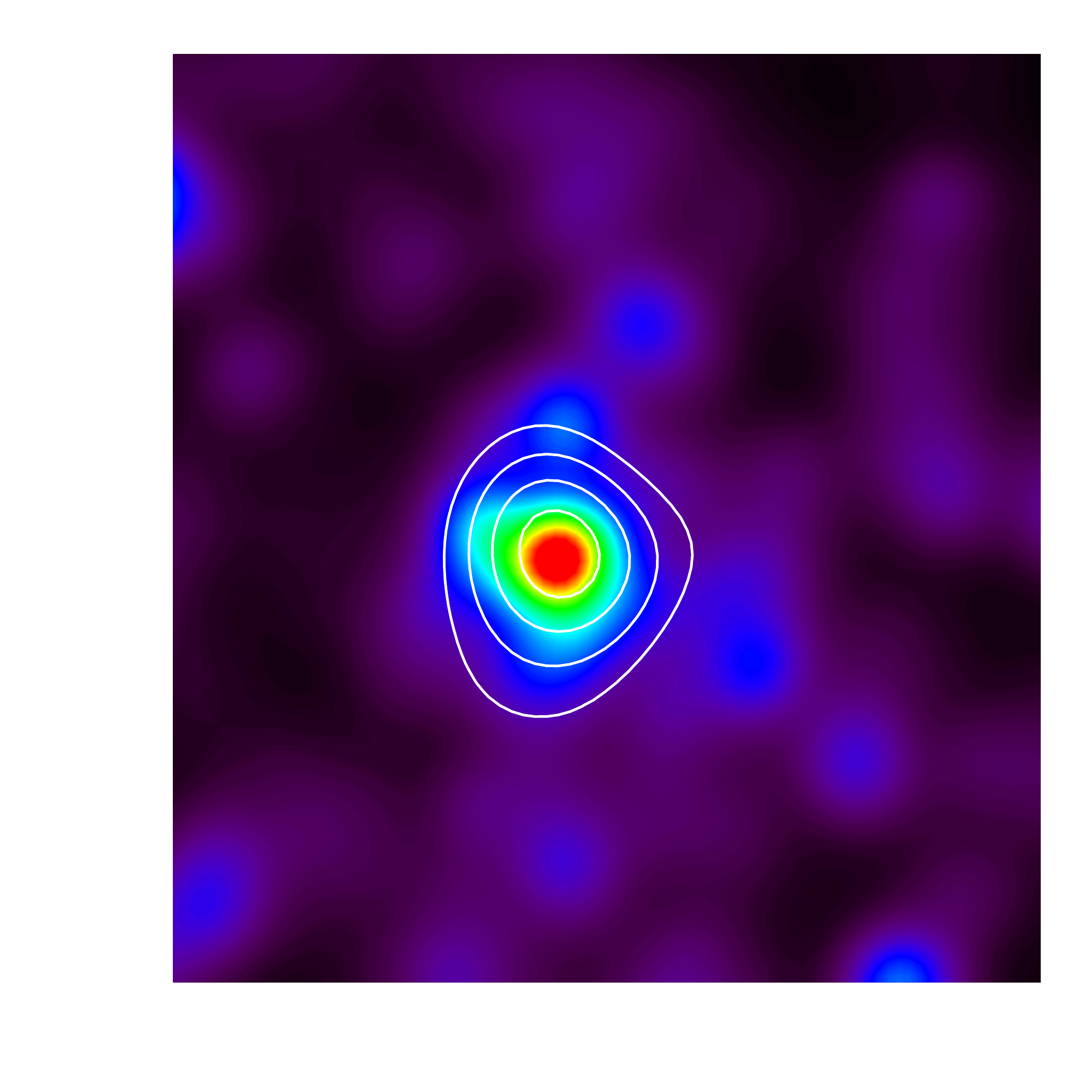}
\includegraphics[width=7.5cm]{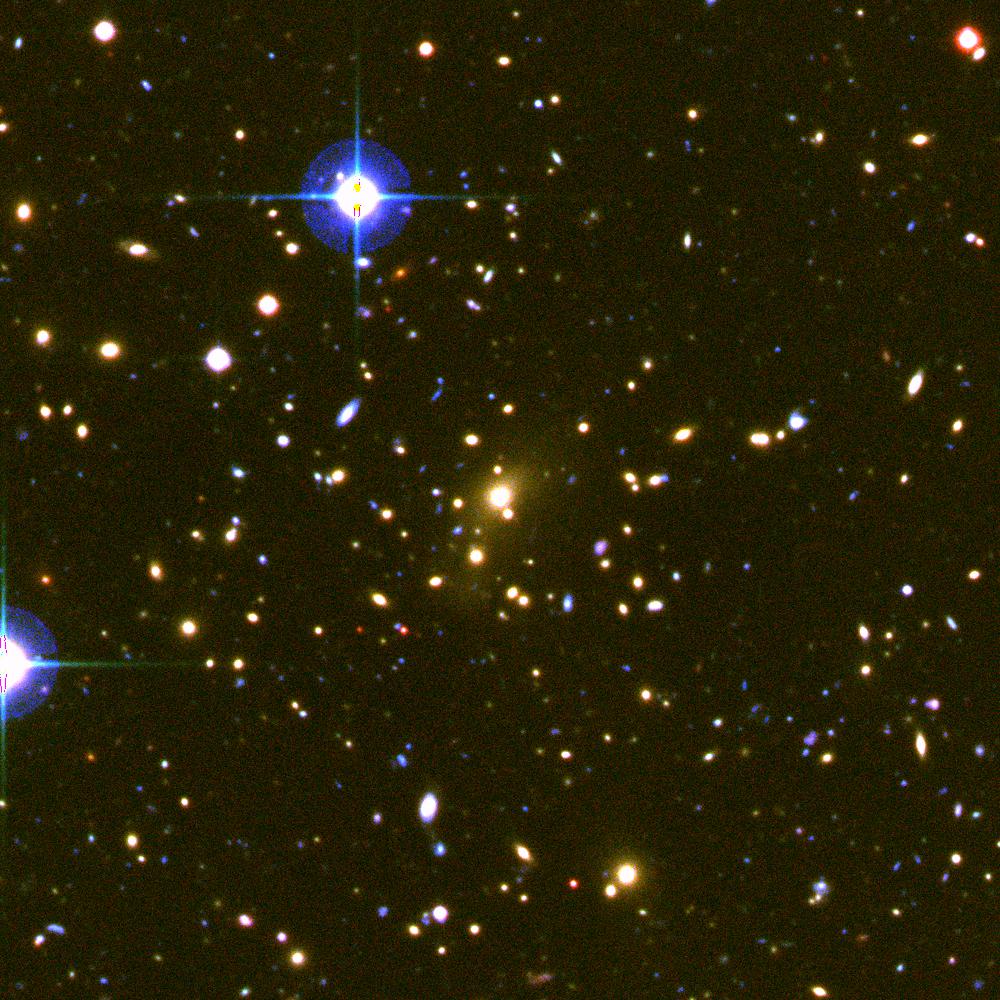}
  \caption{Same as Fig.~\ref{fig:0015}}
\label{fig:0856}
\end{figure*}

This cluster is part of the NORAS sample (Northern ROSAT all-sky galaxy
cluster survey), a pure X-ray selected sample  \citep{bohringer00}. It was
included in the \excpres\ sample because of its high X-ray luminosity. It
is an optical bright cluster identified in the SDSS-DR6 ([WHL2009]
J085612.7+375615, \citep{wen09}), with a redshift measurement of the
BCG at $z=0.411$. The cluster displays a well defined and regular
luminous over-density dominated by a bright and extended cD galaxy
(Fig.~\ref{fig:0856}). The mass map also presents a very regular aspect
around its centre and provides a coherent picture of a relaxed cluster.

\subsection{RX J0943.0+4659 $(z=0.407)$}
This cluster is also known as Abell 851 or Cl 0939+4713. It is the only
Abell cluster of our sample. High resolution HST images of the centre
revealed a large population of blue galaxies and a lot of merging galaxies
\citep{dressler94}. \citet{seitz96} used this deep HST/WFPC2 image to
identify a few lensed objects but no highly magnified gravitational
arcs were detected.  X-ray observations of A851, first with ROSAT
and more recently with \xmm, showed a very perturbed distribution
with pronounced substructures and evidences for a dynamically young
cluster \citep{defilippis03}. Tentative 2D spectro-imaging lead to
the identification of a hot region between the 2 main sub-clusters,
a characteristic of a major merger in an early phase.

\begin{figure*}
  \centering
  \resizebox{0.9\hsize}{!}{
\includegraphics{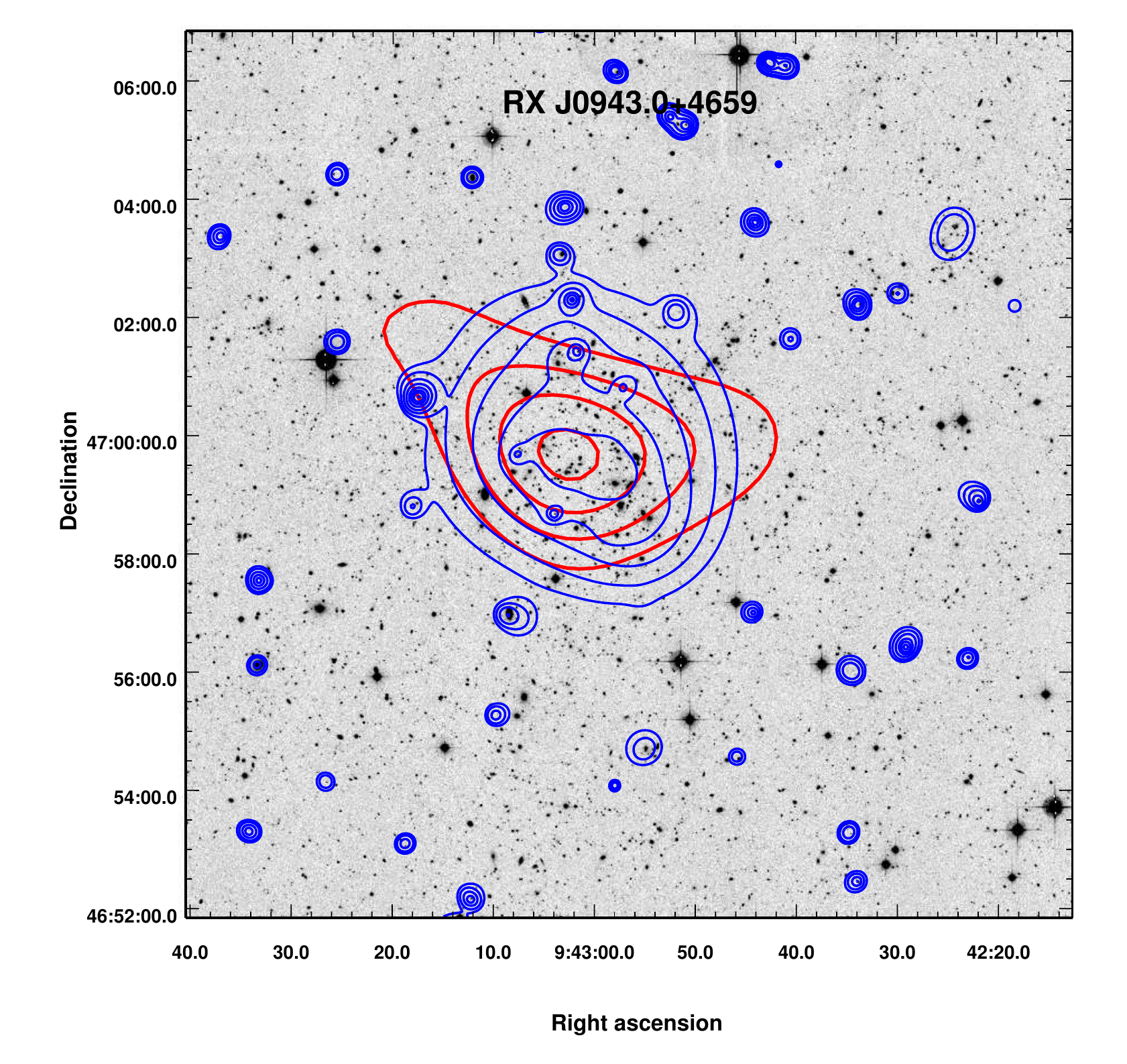}
}
\includegraphics[width=7.5cm]{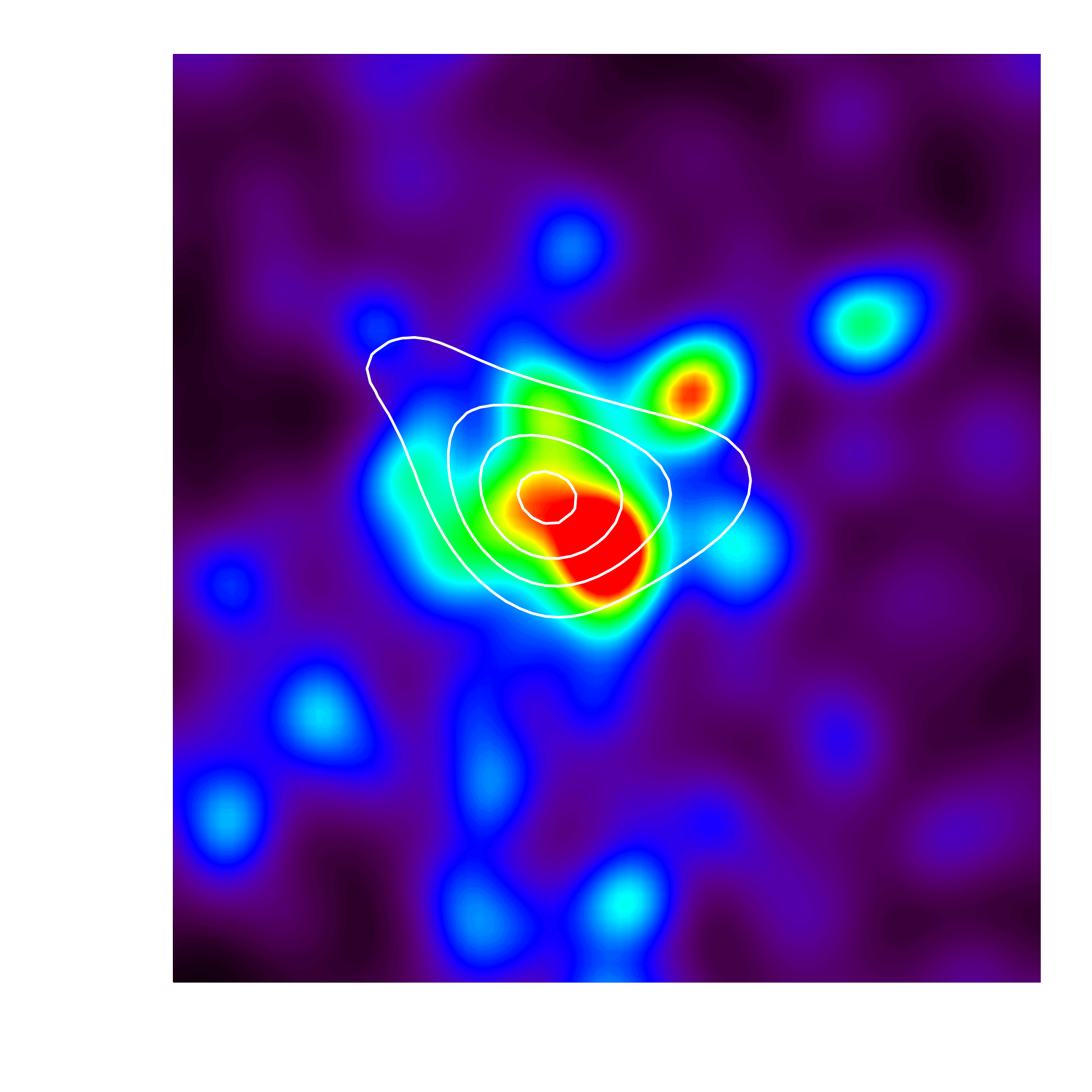}
\includegraphics[width=7.5cm]{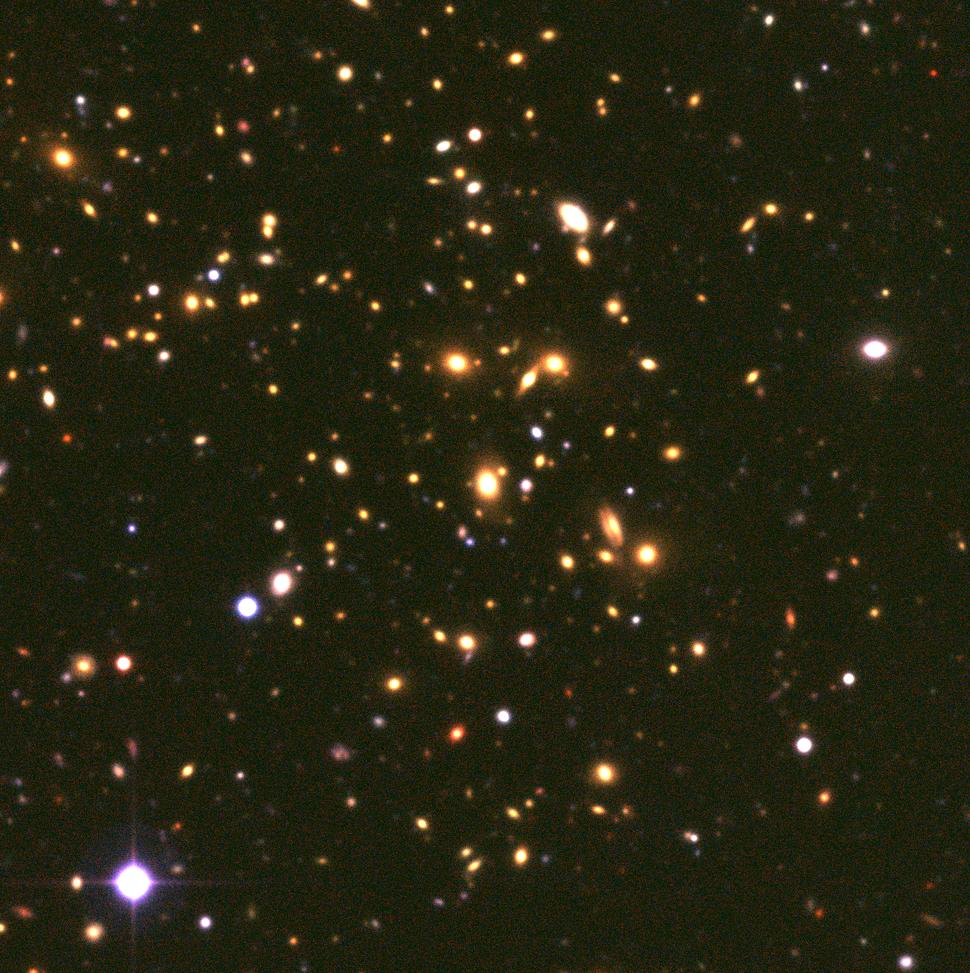}
  \caption{Same as Fig.~\ref{fig:0015}}
\label{fig:0943}
\end{figure*}

The galaxy distribution is complex, with a high galaxy density in the
central area. It can be separated into two clumps which trace the cluster
interaction and are coherent with the gas distribution. Several bright
galaxies dominate the light distribution and are more concentrated in
the South-West extension of the cluster (Fig.~\ref{fig:0943}). On the
contrary the weak lensing mass map is surprisingly regular with only one
main structure but elongated along the direction of the interaction. The
separation between the two X-ray peaks is 50\arcsec , well below the
resolution of the mass map.  So with the present data there is no chance
to have a more detailed view of the mass distribution at a scale where
the physical processes of the cluster merger could be identified. Deeper
imaging should be necessary to go further in this analysis. Due to the
high evidence for merging processes, this cluster was removed later
from the \excpres\ sample, but as optical data were obtained in good
conditions, we kept it in the present sample.

\subsection{RXC J1003.0+3254 $(z=0.416)$}
\begin{figure*}
  \centering
  \resizebox{0.9\hsize}{!}{
\includegraphics{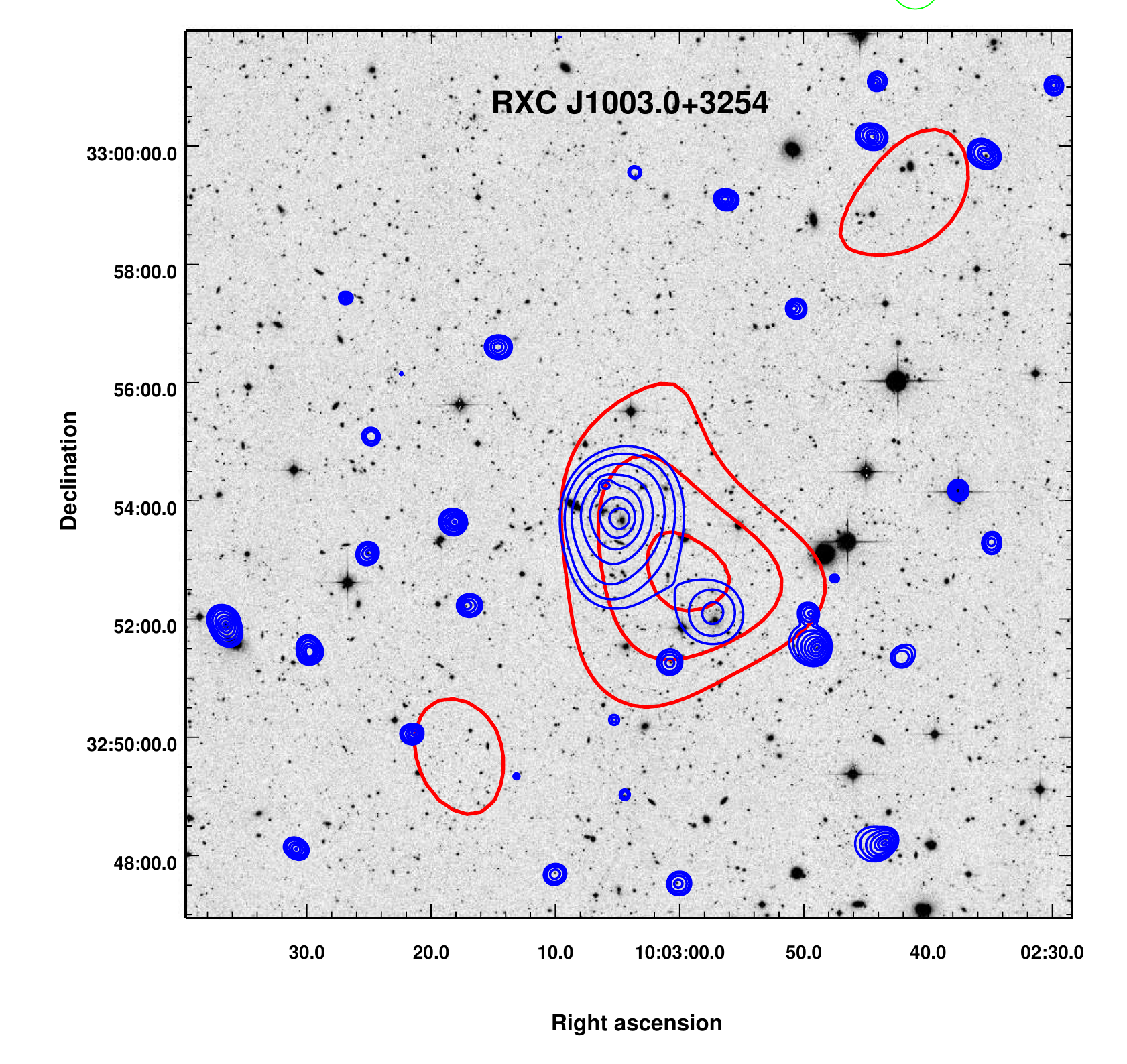}
}
\includegraphics[width=7.5cm]{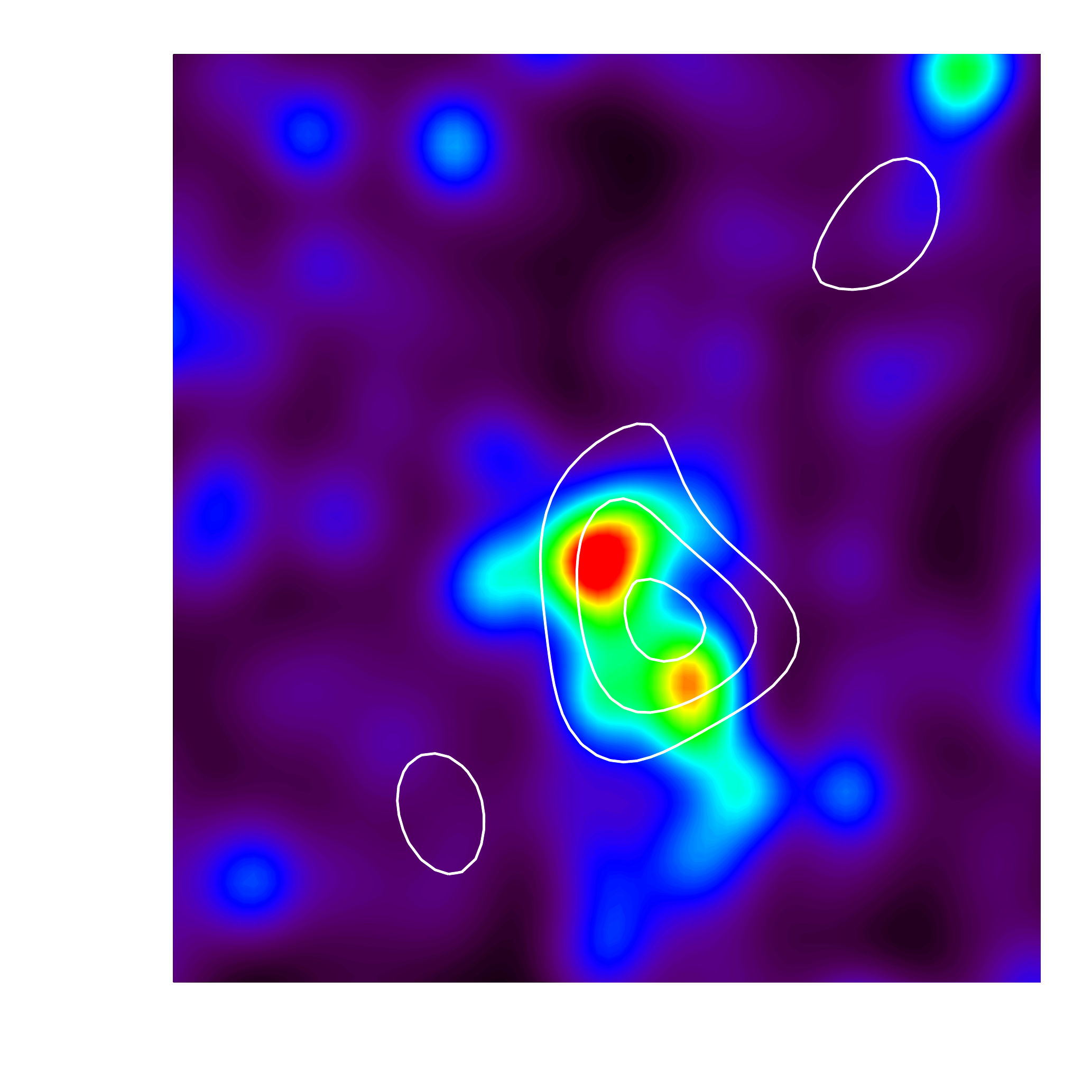}
\includegraphics[width=7.5cm]{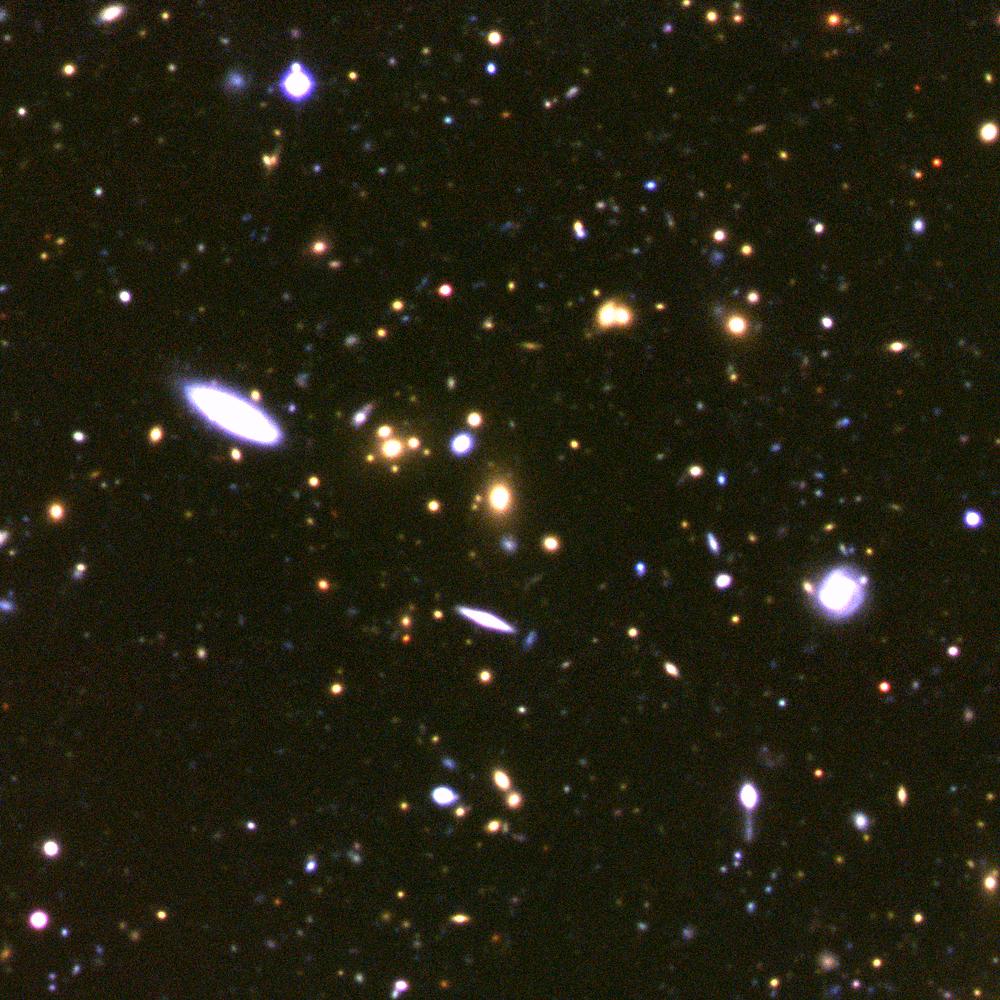}
  \caption{Same as Fig.~\ref{fig:0015}}
\label{fig:1003}
\end{figure*}

The cluster was identified initially by its X-ray extended emission in
the NORAS sample \citep{bohringer00} and it was later re-detected in
the 400d ROSAT sample \citep{burenin07}. Nothing was really known on the
optical properties of this cluster, which displays a bright galaxy in its
centre and a rather loose distribution of cluster members.  Another bright
galaxy is located 2.3\arcmin\ South-West, with similar properties. It is
centered on a secondary peak in the X-ray gas distribution and the mass
map is centered in between these two galaxies. But the bi-modality of the
cluster is more visible on the galaxy distribution than in the mass map
(Fig.~\ref{fig:1003}) which is limited by its spatial resolution. We
suspect that this cluster results from the merging of two sub-clusters
and all conclusions regarding RXC J1003.0+3254 in the global analysis
of the sample must be taken with caution.

\subsection{RX J1120.1+4318 $(z=0.612)$}
\begin{figure*}
  \centering
  \resizebox{0.9\hsize}{!}{
\includegraphics{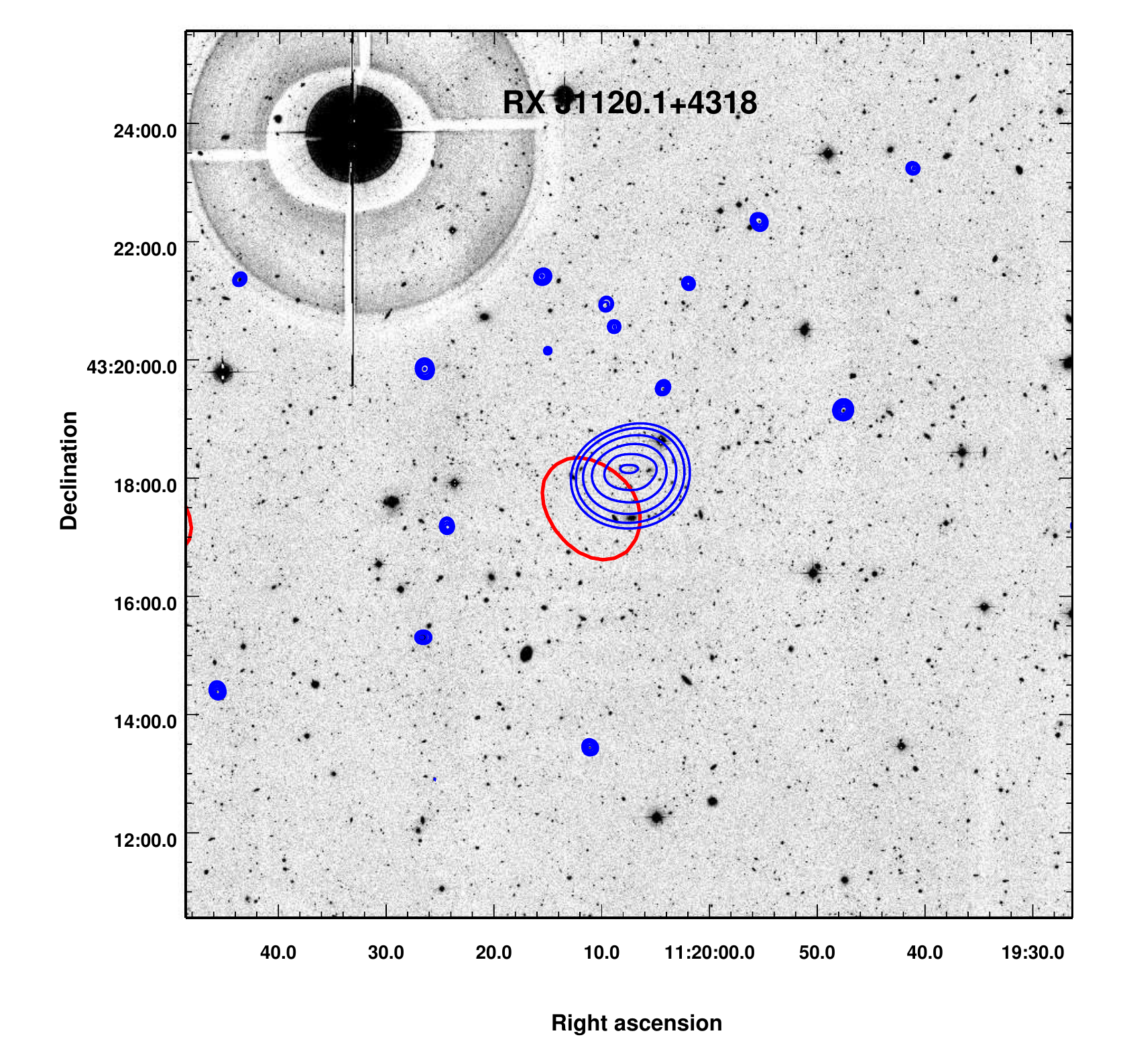}
}
\includegraphics[width=7.5cm]{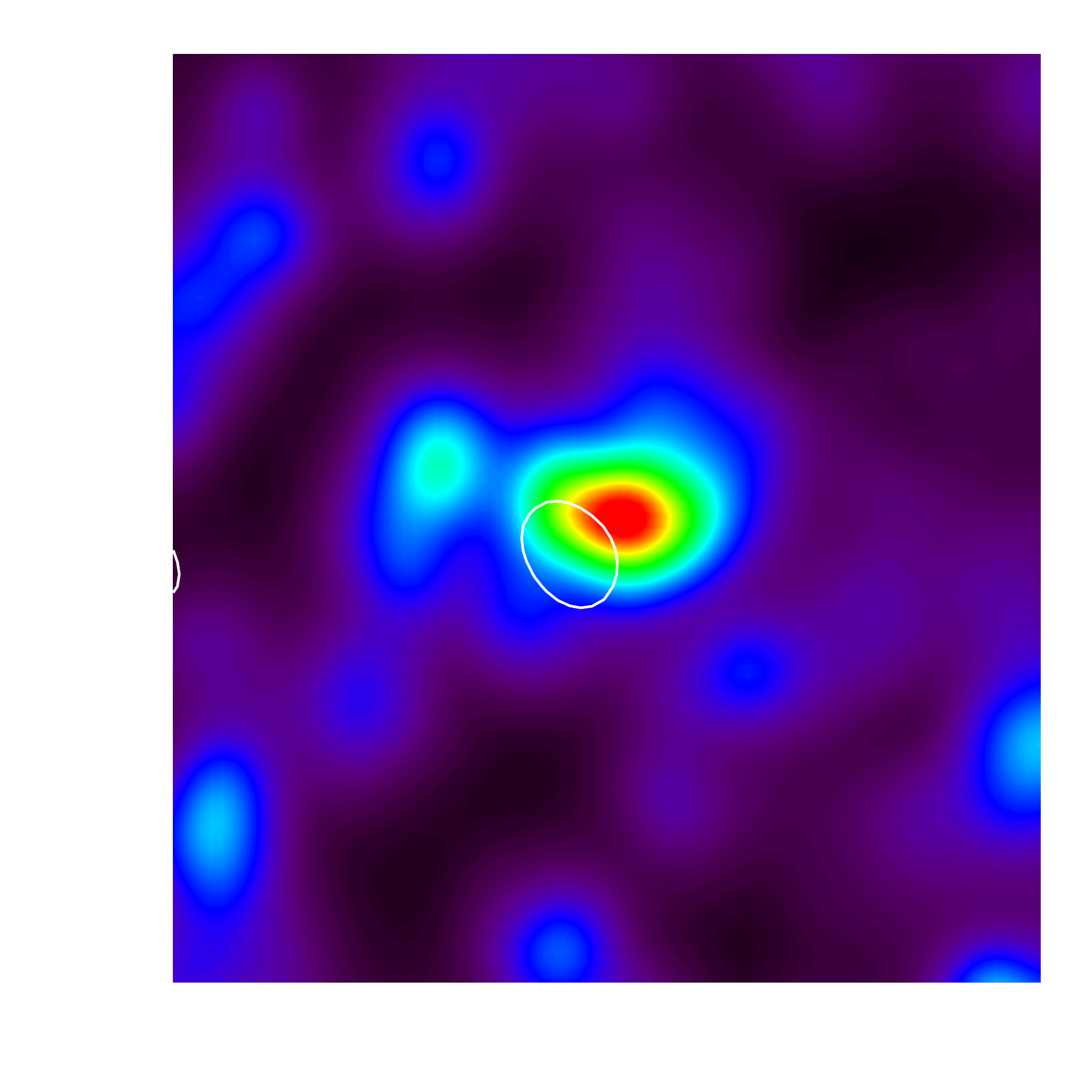}
\includegraphics[width=7.5cm]{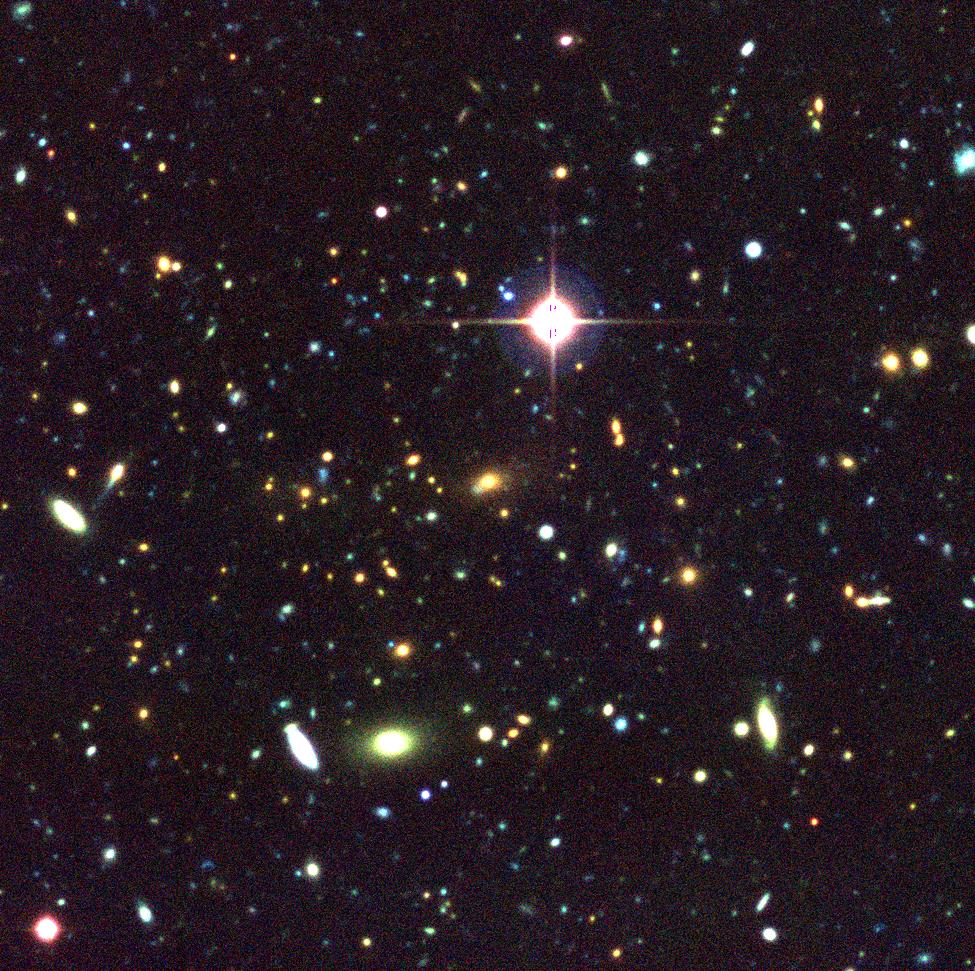}
  \caption{Same as Fig.~\ref{fig:0015}}
\label{fig:1120}
\end{figure*}

This cluster belongs to the Bright SHARC survey \citep{romer00} and 
it was included in the WARPS II catalog (the Wide Angle ROSAT Pointed Survey,
\citet{horner08}). The cluster 
was observed with \xmm\ and analyzed by \citet{arnaud02} who
found a regular X-ray emission with a spherical morphology. They also claim 
that no cooling flow or central gas concentration is
present in this cluster, which is consistent with the cooling time
being larger than the age of the universe at this redshift. Indeed 
with its redshift $z=0.612$, RX J1120.1+4318 is the most distant
cluster of the \excpres\ sample. 
The light distribution of cluster members shows a East-West elongation
which was also been measured in the Chandra X-ray map
\citep{maughan08}. But the ellipticity is rather small and does not
attenuate the regular morphology of the cluster which is clearly in a
relaxed phase. Unfortunately the lensing signal in RX J1120.1+4318 is
barely detected, at less than $3\sigma$ (Fig.~\ref{fig:1120}). So it
is difficult to draw any conclusions on the mass distribution in the
cluster. Even the shift between the mass peak and the light peak
can not be considered as significant. 

\subsection{RXC J1206.2--0848 $(z=0.441)$}
This cluster is one of the brightest clusters of the REFLEX sample
\citep{bohringer04}. It belongs to the MACS sample
\citep{ebeling10} as MACS J1206.2--0847 and is part of the CLASH
sample \citep{postman12}. It displays a
bright and spectacular arc system, initially spectroscopically observed
by \citet{sand04} and confirmed more recently by \citet{ebeling09} at a
redshift $z=1.036$. A detailed analysis of the central mass distribution
was done both with strong lensing and X-ray data, giving a discrepancy
of a factor 2 between the two mass estimates. But the X-ray distribution
of the gas shows some signs of merging processes in the centre which
could explain this discrepancy. Similar trends have already been noticed
in other clusters like A1689 for example \citep{limousin07}.

\begin{figure*}
  \centering
  \resizebox{0.9\hsize}{!}{
\includegraphics{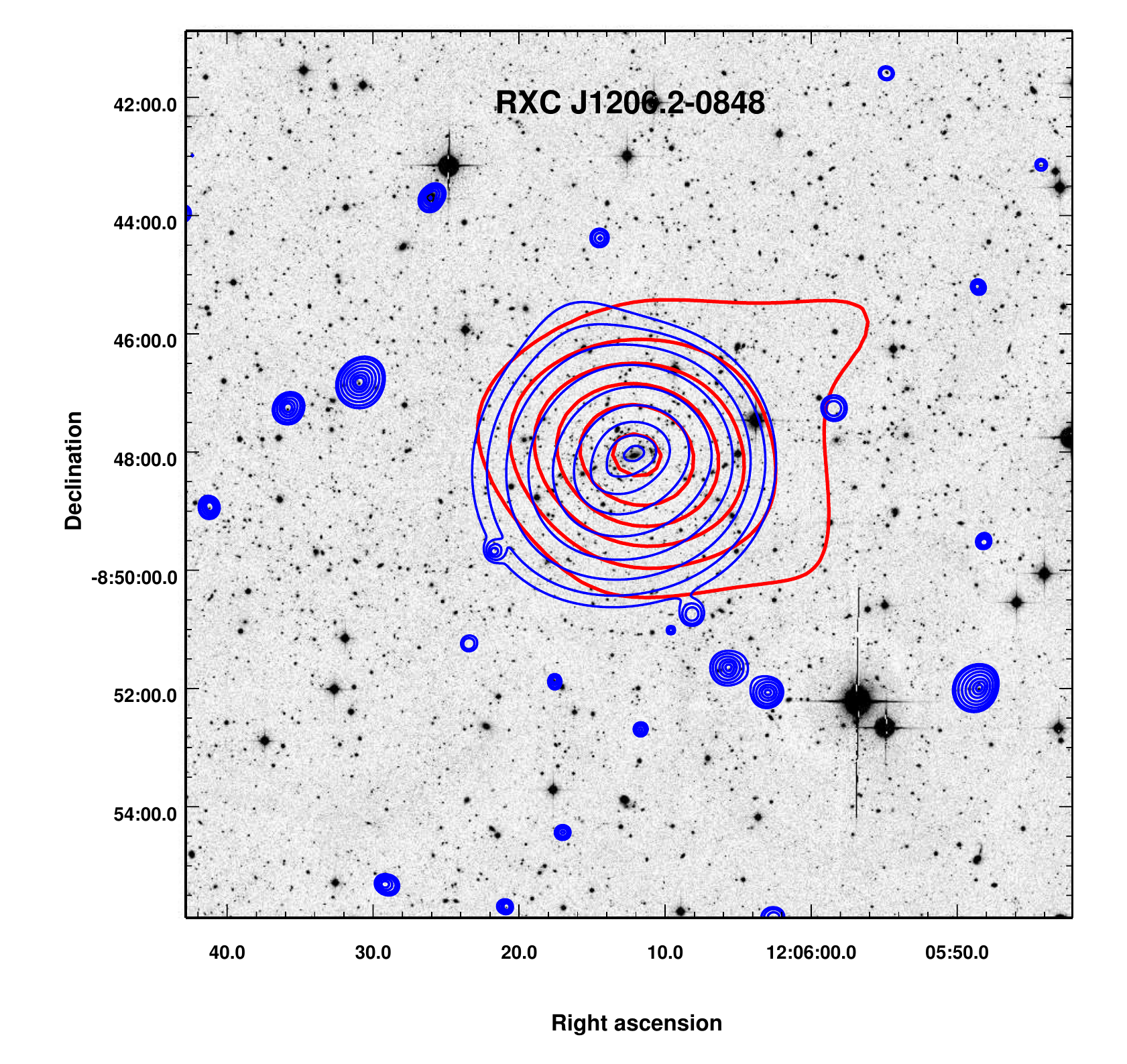}
}
\includegraphics[width=7.5cm]{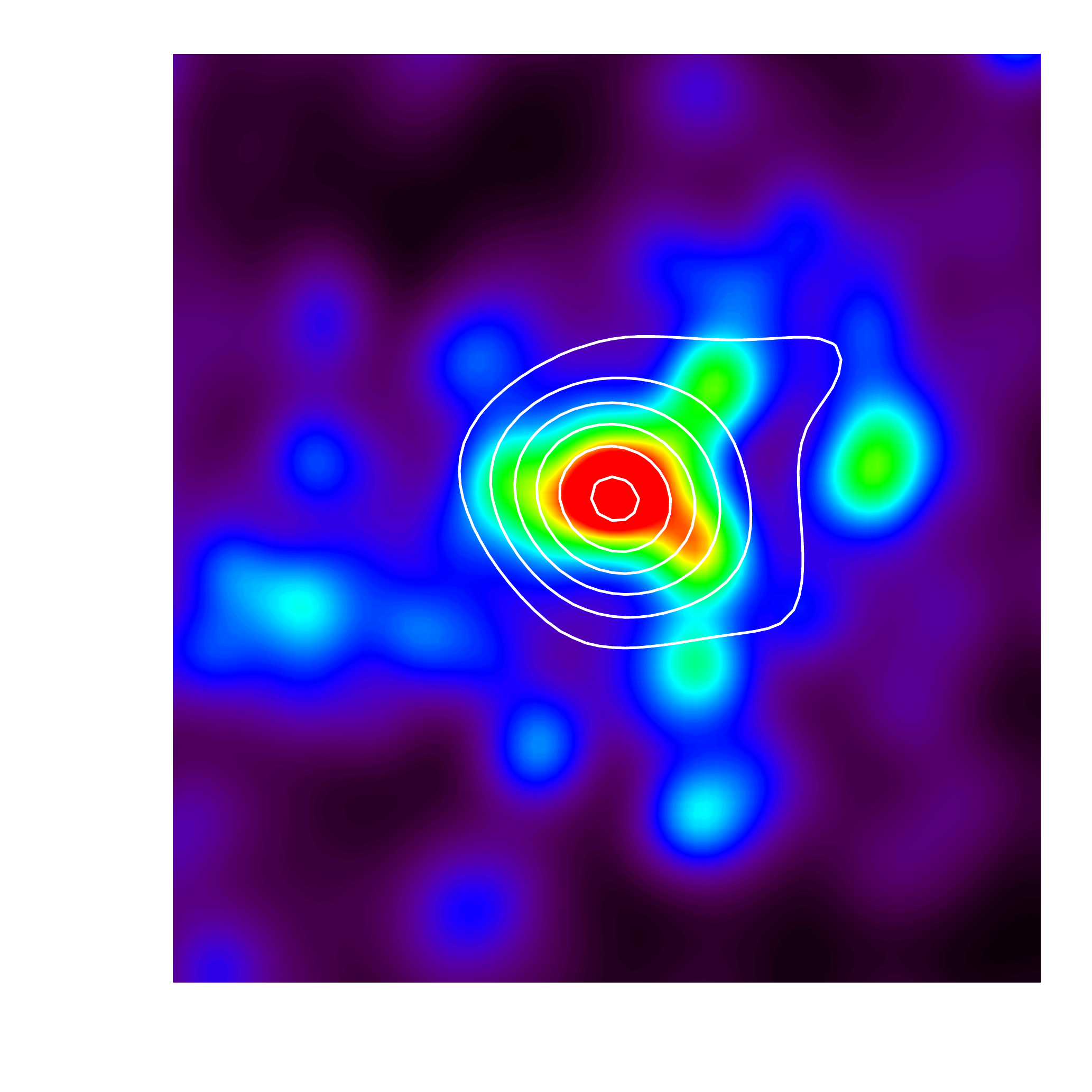}
\includegraphics[width=7.5cm]{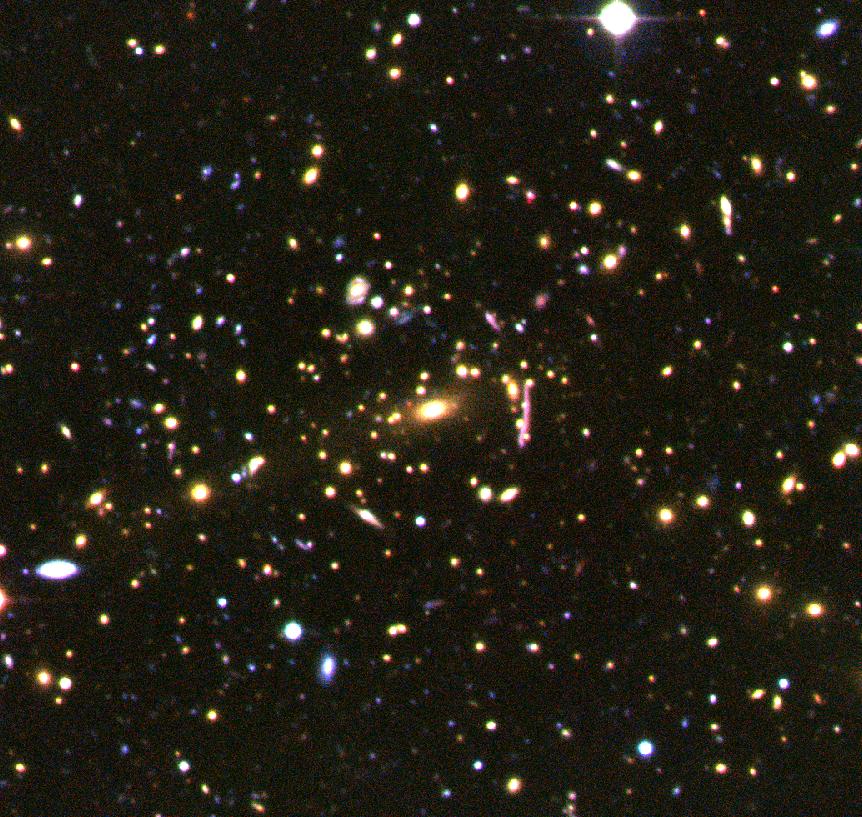}
  \caption{Same as Fig.~\ref{fig:0015}}
\label{fig:1206}
\end{figure*}

The weak lensing mass distribution is well peaked, with a regular shape
and a central concentration which fits the luminous mass as well as the
X-ray mass (Fig.~\ref{fig:1206}).  Note that the central galaxy is also
a bright radio source with a steep spectrum \citep{ebeling10}.

\cite{umetsu12} recently did a comprehensive
analysis of this cluster combining weak and strong lensing derived
from wide-field Subaru imaging and HST observations. Their morphological
analysis of both the reconstructed mass map and light distribution reveals
the presence of a large-scale structure around RXC J1206.2--0848. The
orientation of this structure matches the position angle of the BCG
and that of the cluster light distribution and projected mass map. The
ellipticity they derive for the latter is somehow larger than ours, but
we obtain consistent results for the light distribution. The overall
shape of RXC J1206.2--0848 indicates that light follow mass up to the
large scales of the cosmic web.

\subsection{MS 1241.5+1710 $(z=0.549)$}
\begin{figure*}
  \centering
  \resizebox{0.9\hsize}{!}{
\includegraphics{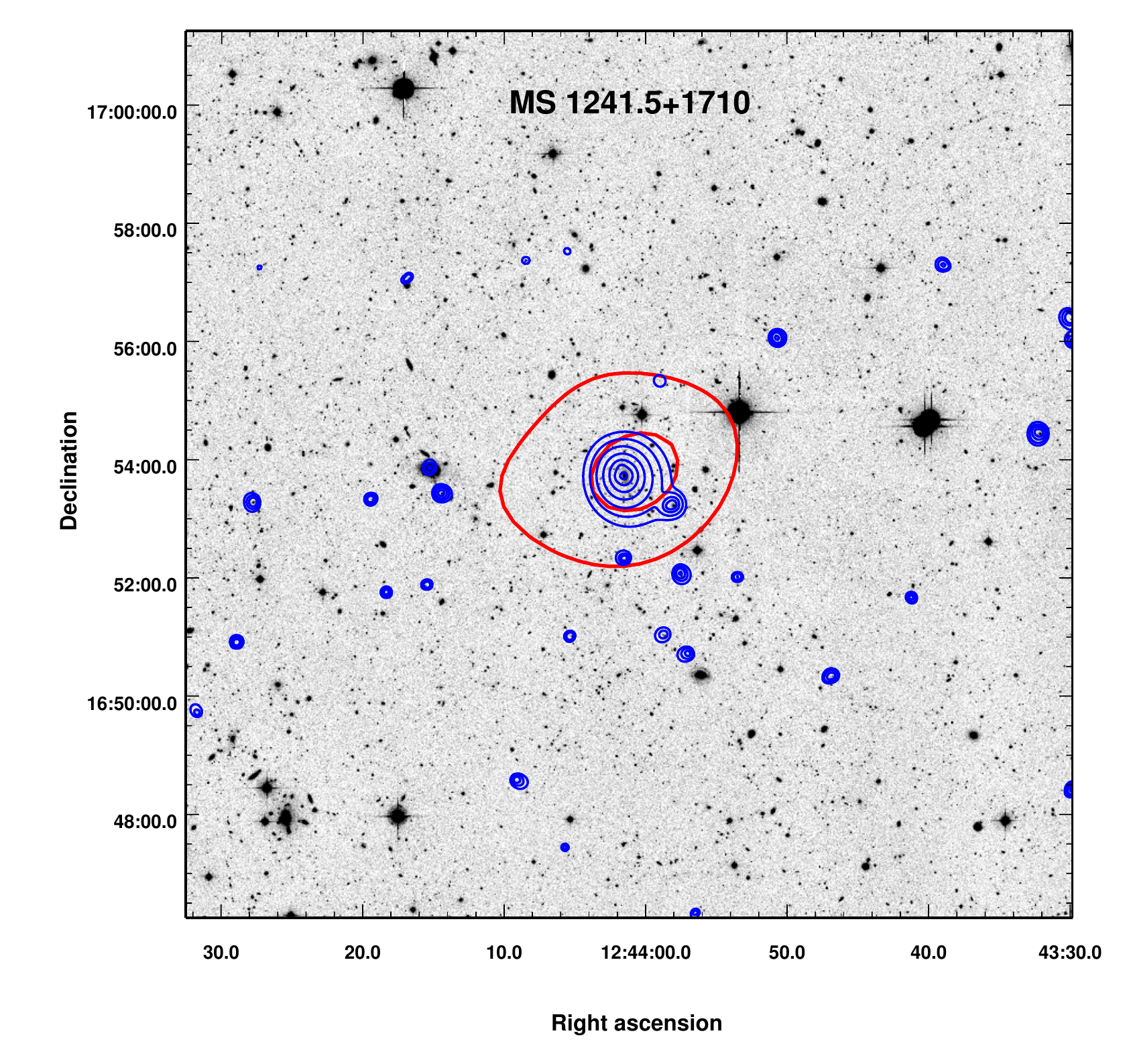}
}
\includegraphics[width=7.5cm]{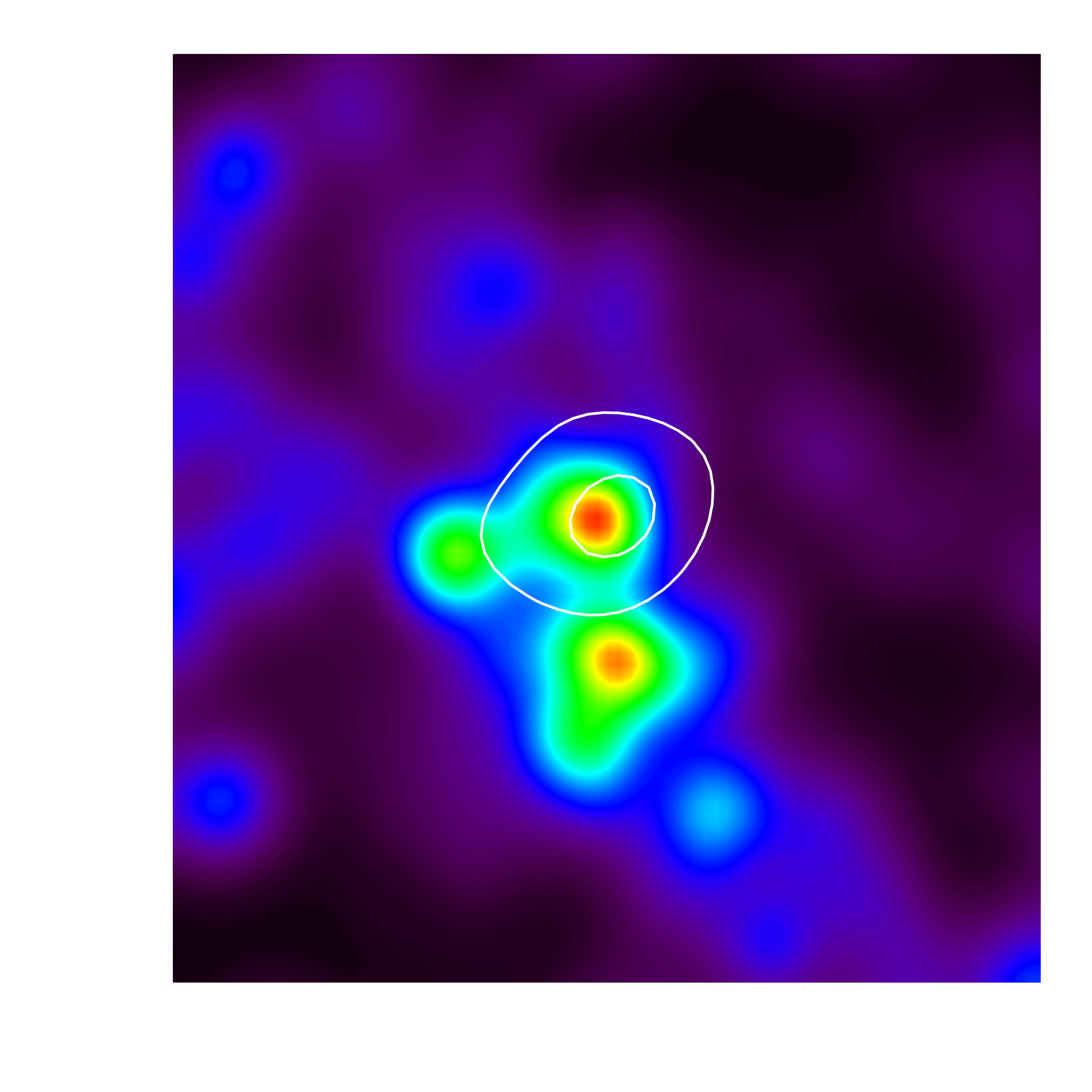}
\includegraphics[width=7.5cm]{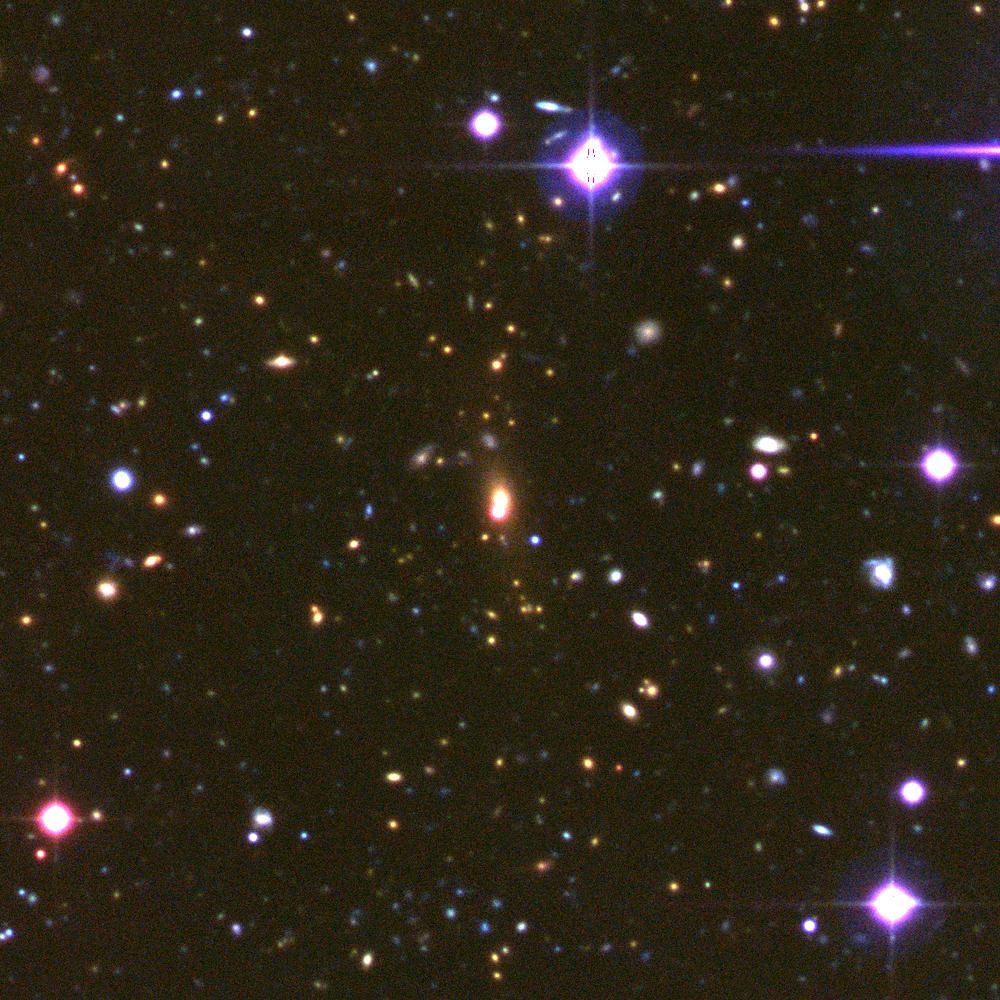}
  \caption{Same as Fig.~\ref{fig:0015}}
\label{fig:1241}
\end{figure*}

As part of the EMMS sample \citep{gioia94}, this cluster was also
observed in optical, but no significant strong lensing feature was
detected \citep{luppino99}. The luminosity distribution is complex, with
a southern extension possibly related to the main cluster. However the
mass distribution does not show a similar trend nor the X-ray gas. Both
are regular and centered in the BCG embedded in a bright and extended
envelope. So firm conclusions are difficult to draw because of the low
S/N ratio of the mass map (Fig.~\ref{fig:1241}). The second over-density of
galaxies could also be due to some contamination along the line of sight.
Deeper and  multi-color images would be necessary to confirm the reality
of an in-falling substructure on the main cluster.

\subsection{RX J1347.5--1145 $(z=0.451)$}
\begin{figure*}
  \centering
  \resizebox{0.9\hsize}{!}{
\includegraphics{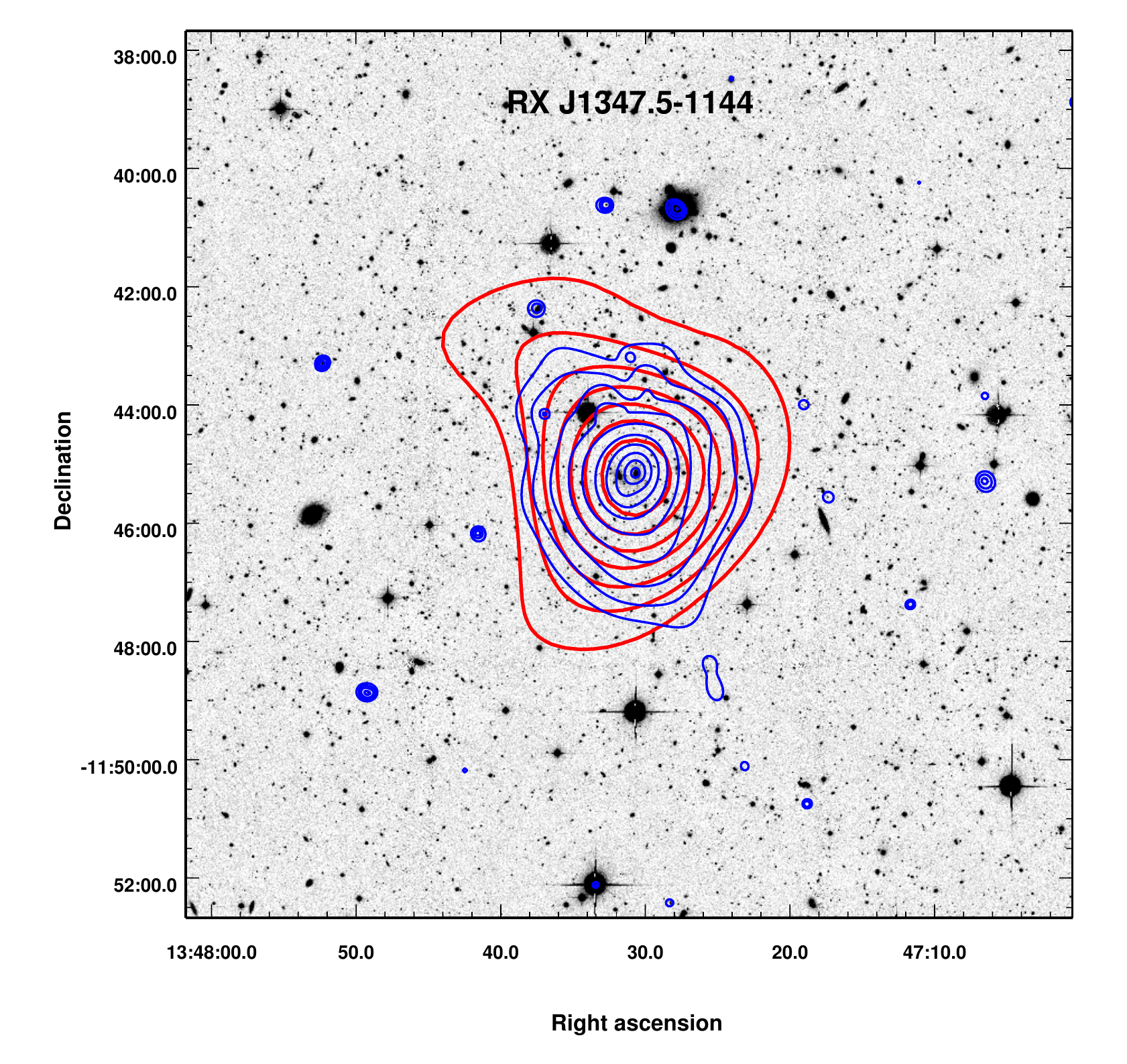}
}
\includegraphics[width=7.5cm]{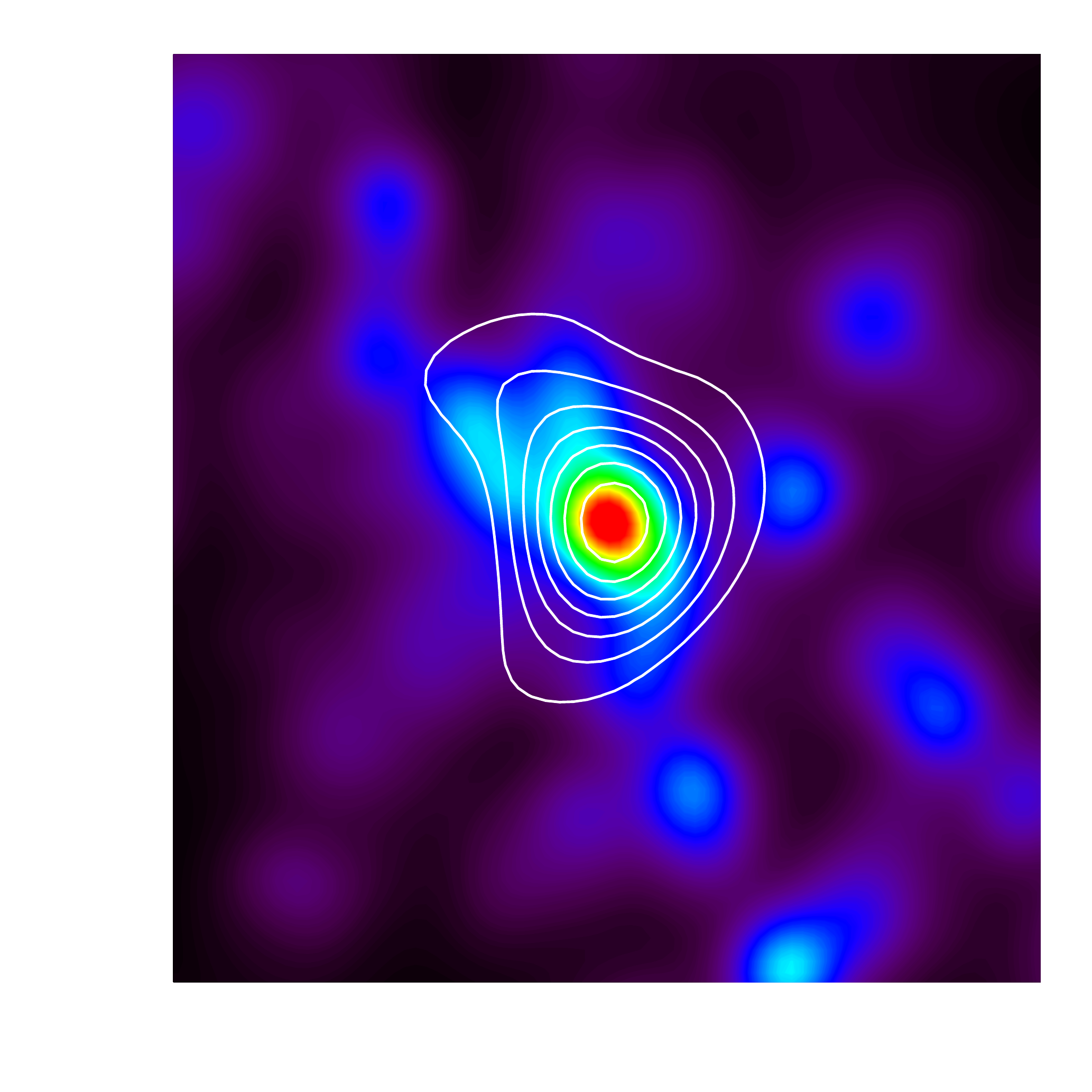}
\includegraphics[width=7.5cm]{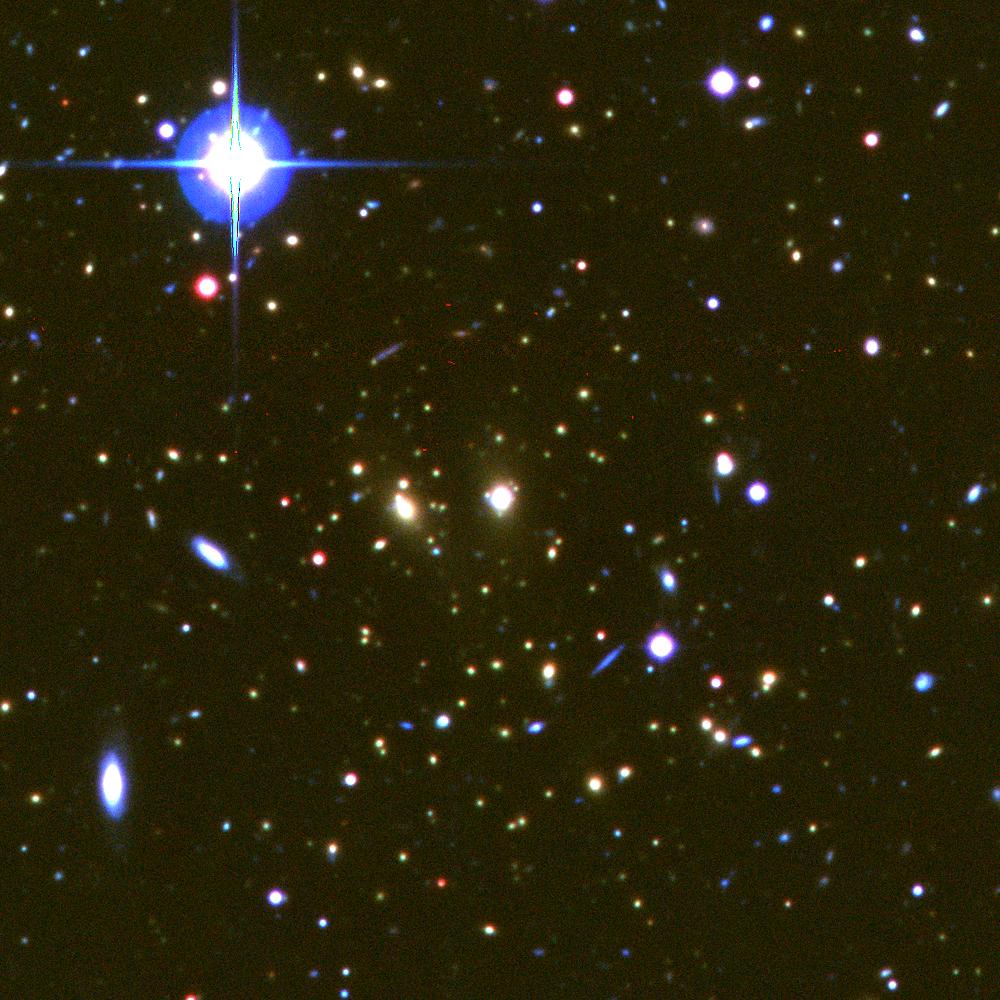}
  \caption{Same as Fig.~\ref{fig:0015}}
\label{fig:1347}
\end{figure*}

This is the brightest cluster of the REFLEX sample \citep{bohringer04}
and is part of the CLASH sample \citep{postman12}. It presents a 
spectacular strong lensing system detected by
\citet{schindler95}.  It was also identified as a cluster with a strong
central cooling flow \citep{allen00}, feeding a powerful radio source
\citep{pointecouteau01}. A detailed combined analysis of the strong
and weak lensing effects \citep{bradac05b,bradac05a} lead to a very
accurate view of the dynamical status of the cluster in the inner
regions: the cluster presents a mass concentration centered on the
BCG with some extension to the SW and many evidences of sub-cluster
merging. But RX J1347.5--1145 is definitely not a major merger.
After some controversy, the different mass estimates seemed to
converge, especially those measured close to the centre using the
strong lensing features \citep{halkola08,bradac08}. But a factor
of 2 remains between the X-ray and the weak lensing masses at large
radius \citep{fisher97,kling05,gitti07}.  In this context, our mass
map confirms the previous results and does not bring new evidences on
the mass distribution (Fig.~\ref{fig:1347}). It is mostly used in order
to check and validate our weak lensing procedure before applying it to
other clusters. Similar results were published on RX J1347.5--1145 by
\citet{hoekstra12} with the same CFHT data. Fortunately they obtain
very similar mass measures.

Several studies \citep{lu10,verdugo12} revealed that RX J1347.5--1145
is embedded in a large-scale structure, extending up to 20 Mpc in the
NE-SW direction. Our luminosity map confirms the existence of several
over-densities on a large scale, aligned along this direction. The main
orientation of the cluster light distribution also follows the same
direction. As for RXC J1206.2--0848, this cluster supports the picture
of the cosmic web where massive clusters are fed by filaments whose
orientation matches the global morphology of the central node.

\subsection{MS 1621.5+2640 $(z=0.426)$}
\begin{figure*}
  \centering
  \resizebox{0.9\hsize}{!}{
\includegraphics{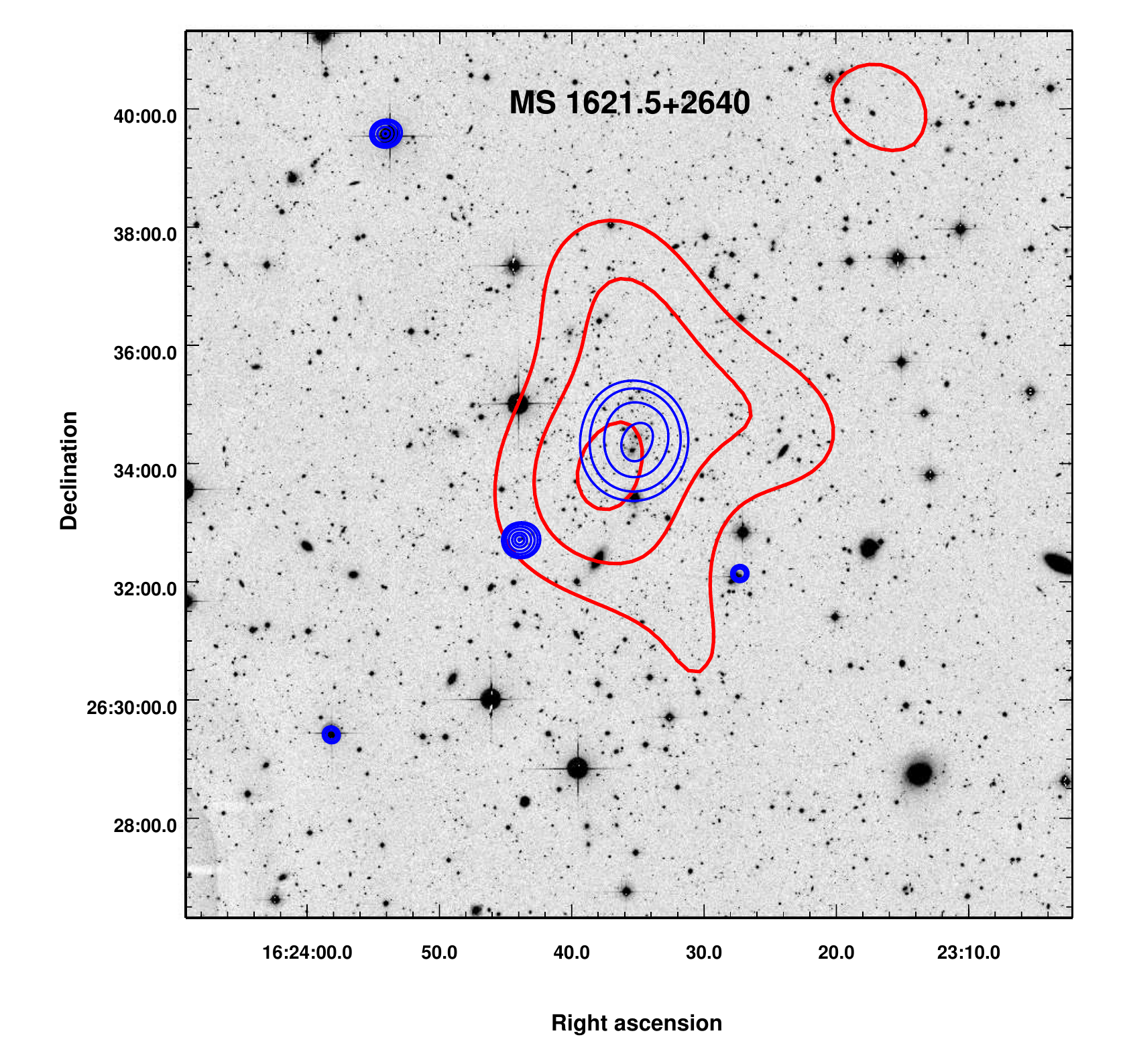}
}
\includegraphics[width=7.5cm]{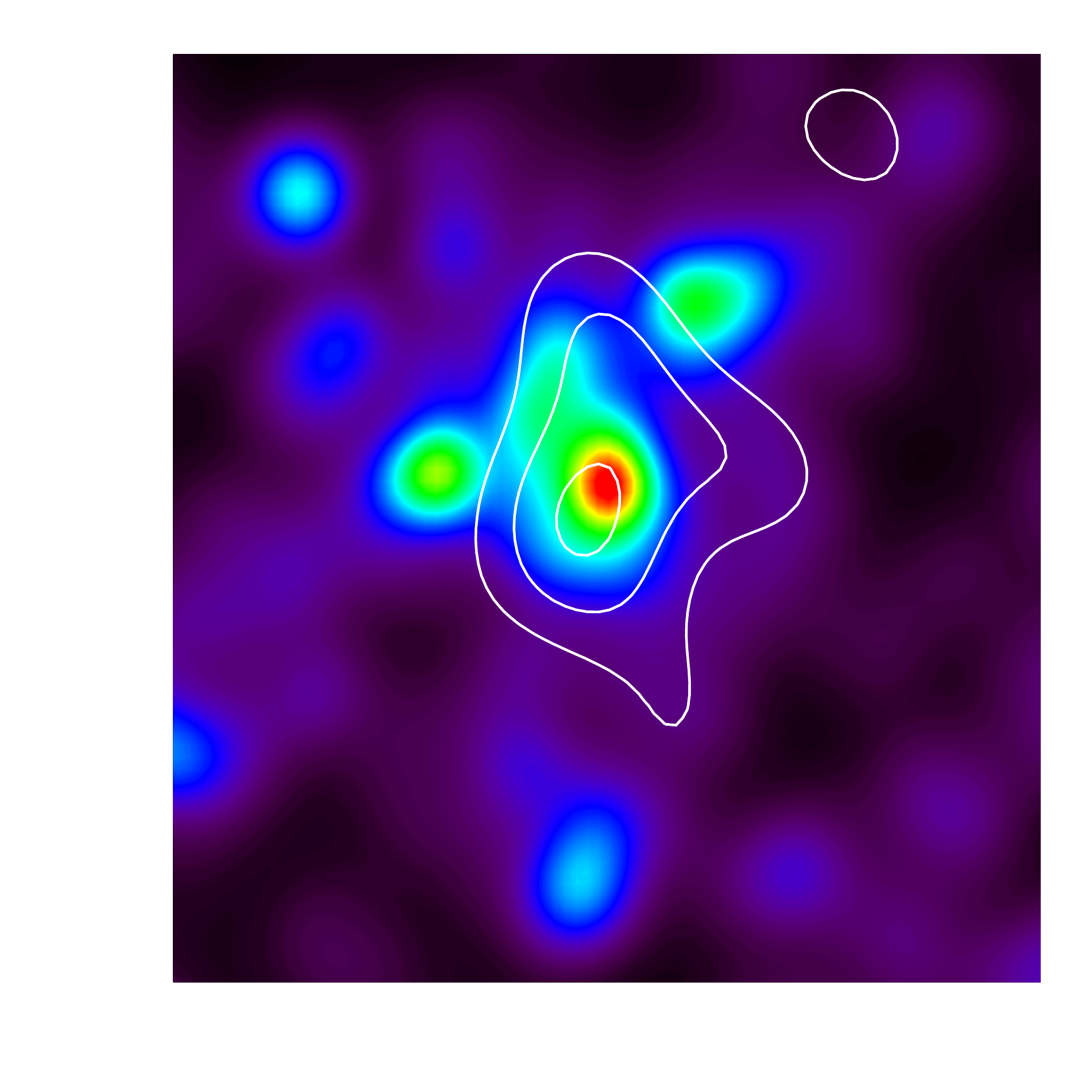}
\includegraphics[width=7.5cm]{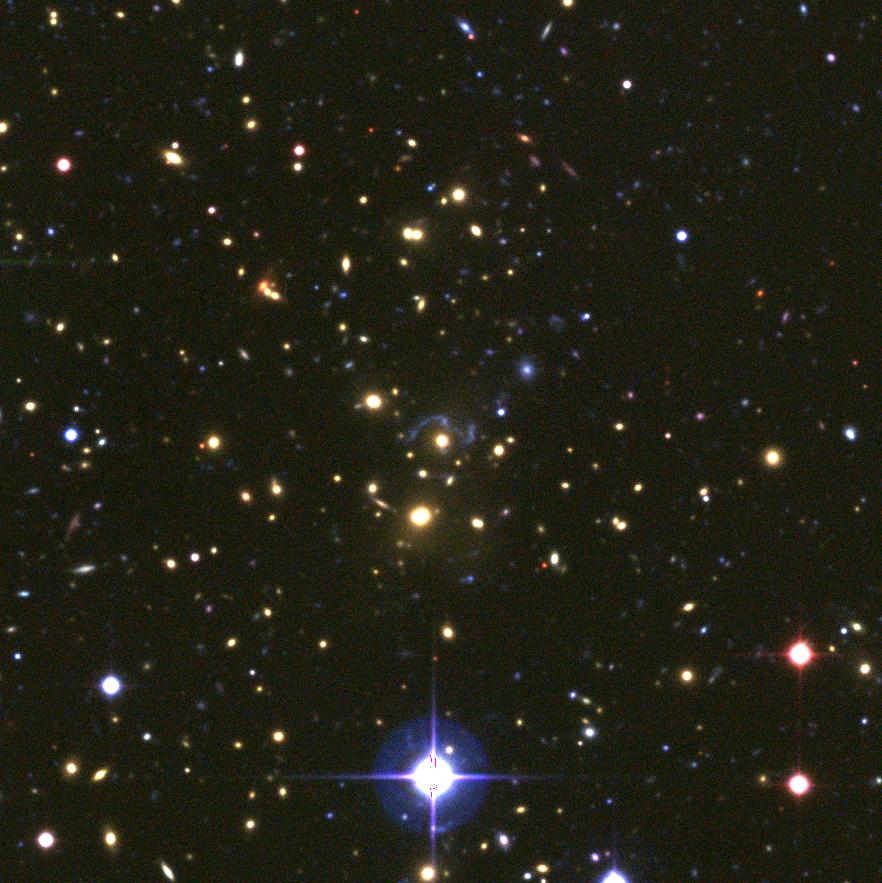}
  \caption{Same as Fig.~\ref{fig:0015}}
\label{fig:1621}
\end{figure*}

As part of the EMSS cluster sample \citep{gioia94} the cluster was
rapidly identified as a strong lens with a nice gravitational arc
located around a radio galaxy which is not the brightest cluster
galaxy \citep{luppino99}. No spectroscopic redshift is presently
available for the arc, although its lensed nature is not in doubt
\citep{sand05}.  MS 1621.5+2640 is also part of the CNOC sample and
was observed spectroscopically with more than 100 cluster redshifts
available \citep{ellingson97}. The velocity dispersion is not very high
($\sigma_{los} = 839 ^{+67}_{-53}$ km/s, \citet{borgani99}). More recently
\citet{hoekstra07} did a very accurate weak lensing analysis and his
results are in good agreement with the dynamical mass estimate. It is
also consistent with the X-ray mass obtained with ROSAT \citep{hicks06}.
With our weak lensing mass reconstruction, we find a mass distribution
rather elongated and coherent with the light distribution. The large
shift between the mass and light peaks is more probably an artifact than
a real one (Fig.~\ref{fig:1621}).

\subsection{RX J2228.5+2036 $(z=0.412)$}
\begin{figure*}
  \centering
  \resizebox{0.9\hsize}{!}{
\includegraphics{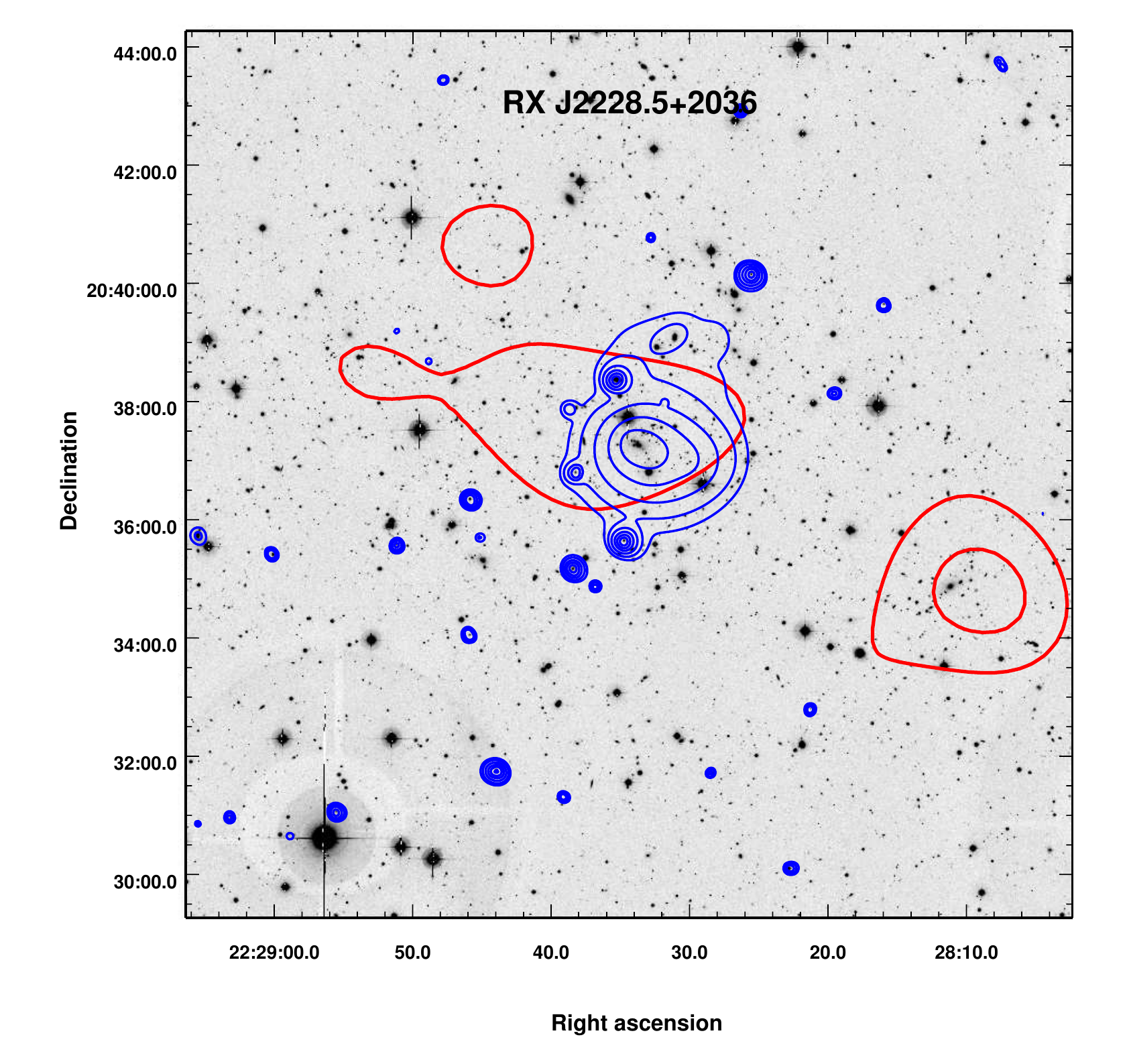}
}
\includegraphics[width=7.5cm]{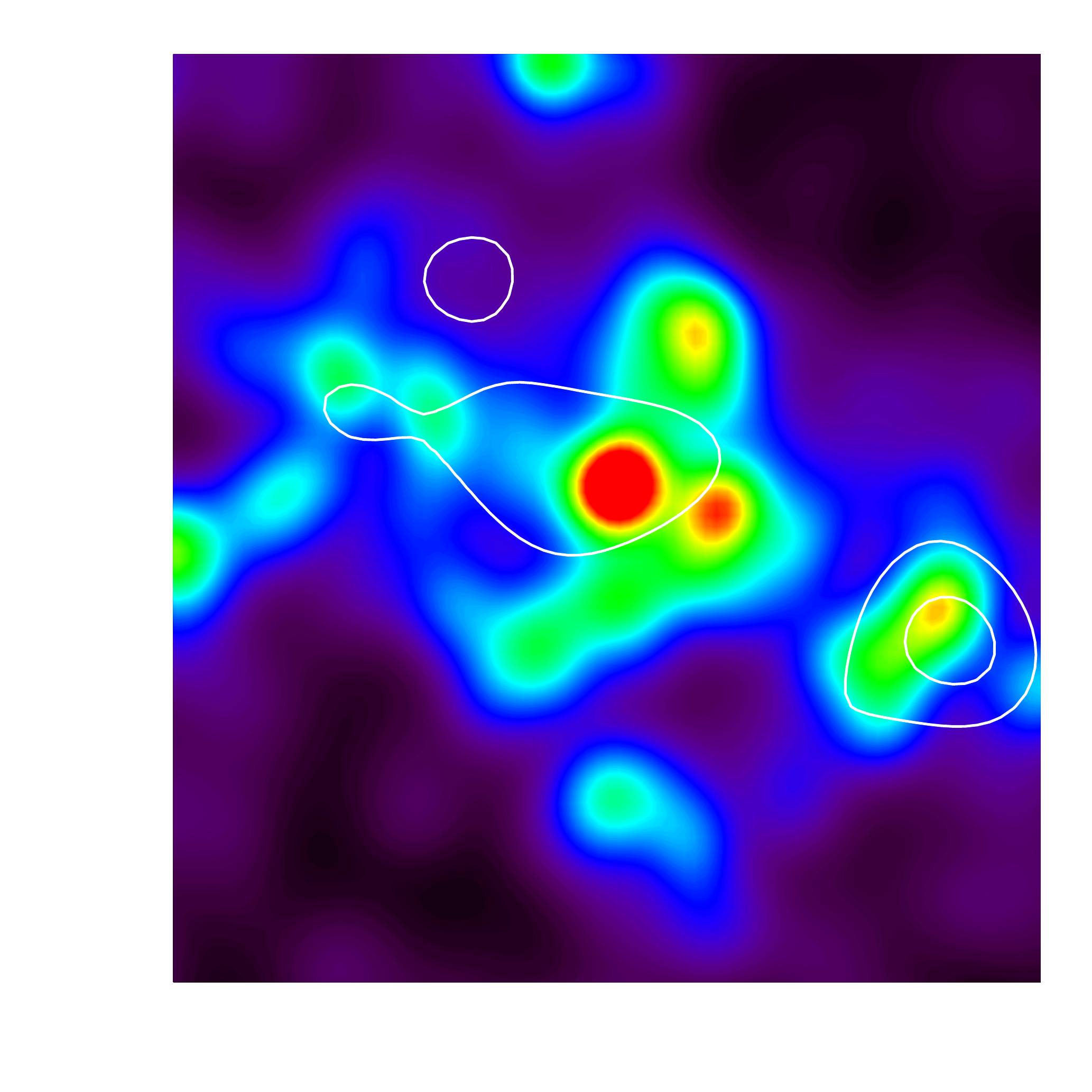}
\includegraphics[width=7.5cm]{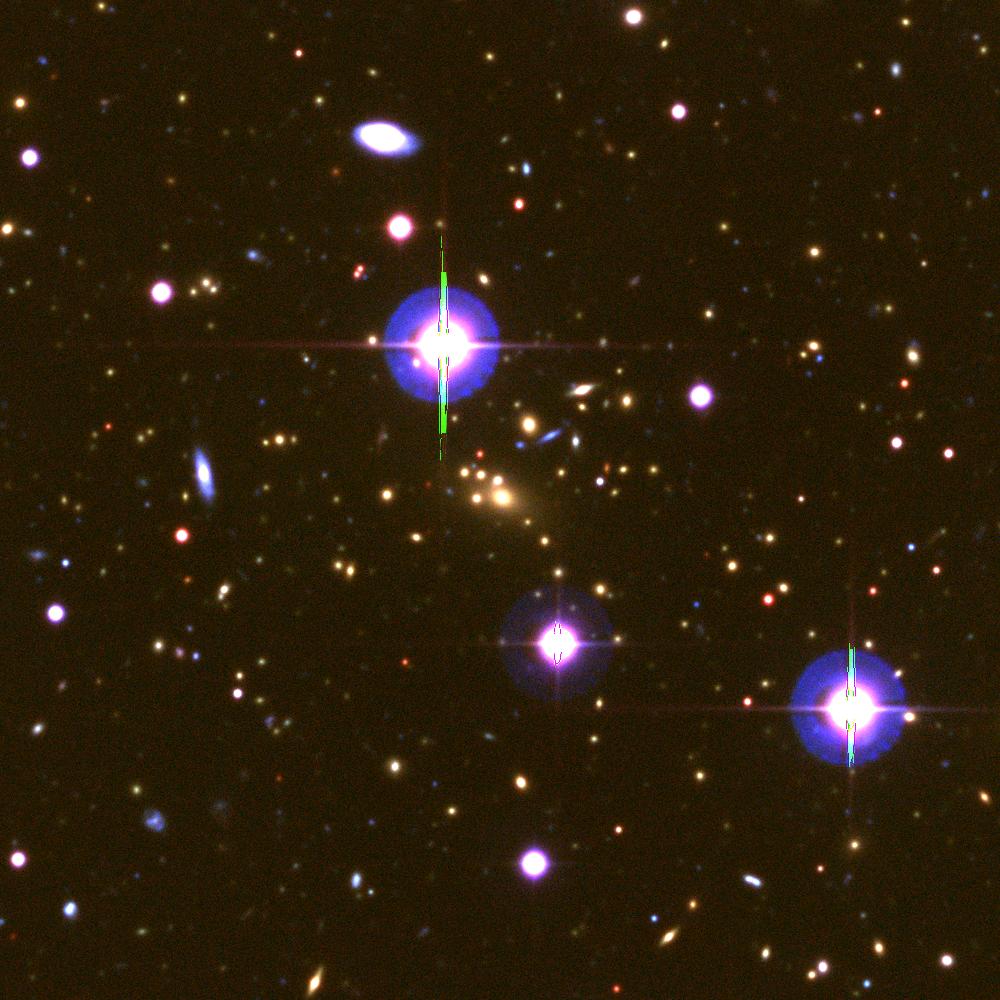}
  \caption{Same as Fig.~\ref{fig:0015}}
\label{fig:2228}
\end{figure*}

This cluster is part of the NORAS sample \citep{bohringer00} and also
belongs to the MACS sample (MACS J2228.5+2036, \citet{ebeling07}).
Because it is at low galactic latitude, very few optical observations
are available.  Our weak lensing reconstruction is rather uncertain
(Fig.~\ref{fig:2228}) and possibly flawed because of the large number
of bright stars in the field of view. However, in addition to its
strong X-ray emission, this cluster was detected for its SZ signal,
allowing one of the first combined analysis between the X-ray and the
SZ signals \citep{pointecouteau02,jia08}. Both confirm that the cluster
is quite massive and dynamically perturbed. 
The weak lensing map 
shows a poor signal close to the cluster centre but suggests that the
cluster has an elongated shape. This is also valid for the complex light
distribution. The most convincing feature is a galaxy clump located in the
South-West direction, detected on the mass map with higher significance
than the main cluster. It is associated with a galaxy excess centered on
a bright elliptical galaxy with similar magnitude as the cluster BCG.
We suspect that this clump is at a similar redshift as RX J2228.5+2036
and may be the cause of a future major merger with RX J2228.5+2036.
Surprisingly there is no X-ray counter-part to this clump. 


\end{appendix}

\end{document}